\documentclass[reprint,amsmath,amssymb,aps,nofootinbib]{revtex4-1}
\usepackage{graphicx}
\usepackage{amsmath}
\usepackage{dcolumn}
\usepackage{bm}

\usepackage{mathtools}
\usepackage{blkarray}
\usepackage{footnote}
\usepackage{hyperref}		
\hypersetup{
    colorlinks,
    citecolor=blue,
    filecolor=blue,
    linkcolor=blue,
    urlcolor=blue,
    pdfdisplaydoctitle=true}
\usepackage[toc,page]{appendix}
\usepackage{adjustbox}
\usepackage{float}
\usepackage{color}
\usepackage{xcolor}
\usepackage{soul}
\usepackage[normalem]{ulem}
\usepackage{amsmath}

\usepackage{scalerel}
\usepackage{tikz, braket}
\usetikzlibrary{svg.path}

\definecolor{orcidlogocol}{HTML}{A6CE39}
\tikzset{
  orcidlogo/.pic={
    \fill[orcidlogocol] svg{M256,128c0,70.7-57.3,128-128,128C57.3,256,0,198.7,0,128C0,57.3,57.3,0,128,0C198.7,0,256,57.3,256,128z};
    \fill[white] svg{M86.3,186.2H70.9V79.1h15.4v48.4V186.2z}
                 svg{M108.9,79.1h41.6c39.6,0,57,28.3,57,53.6c0,27.5-21.5,53.6-56.8,53.6h-41.8V79.1z M124.3,172.4h24.5c34.9,0,42.9-26.5,42.9-39.7c0-21.5-13.7-39.7-43.7-39.7h-23.7V172.4z}
                 svg{M88.7,56.8c0,5.5-4.5,10.1-10.1,10.1c-5.6,0-10.1-4.6-10.1-10.1c0-5.6,4.5-10.1,10.1-10.1C84.2,46.7,88.7,51.3,88.7,56.8z};
  }
}

\newcommand\orcidicon[1]{\href{https://orcid.org/#1}{\mbox{\scalerel*{
\begin{tikzpicture}[yscale=-1,transform shape]
\pic{orcidlogo};
\end{tikzpicture}
}{|}}}}

\definecolor{myred}{rgb}{1,0,0}
\definecolor{myblue}{rgb}{0,0,1}
\colorlet{lightred}{myred!60!white}
\colorlet{lightblue}{myblue!60!white}
\colorlet{lightgreen}{green!60!white}
\colorlet{darkgreen}{green!60!black}
\colorlet{darkred}{myred!60!black}

\definecolor{mypurple}{rgb}{1,0,1}
\colorlet{lavender}{mypurple!50!white}

\newcommand{\todohidden}[1]{\textcolor{lightred}{}}

\newcommand{\oldjbm}[1]{#1}
\newcommand{\oldmi}[1]{#1}

\newcommand{\oldyy}[1]{#1}

\newcommand{\headline}[1]{\textcolor{black}{#1}} 
\newcommand{\mi}[1]{\textcolor{black}{#1}}   
 \newcommand{\mib}[1]{\textcolor{black}{#1}}
\newcommand{\jbm}[1]{\textcolor{black}{#1}}
\newcommand{\jbmb}[1]{\textcolor{black}{#1}}
\newcommand{\jbmc}[1]{\textcolor{black}{#1}} 
\newcommand{\jbmd}[1]{\textcolor{black}{#1}} 
\newcommand{\jbme}[1]{\textcolor{black}{#1}} 

\newcommand{\jbmf}[1]{\textcolor{black}{#1}}
\newcommand{\jbmfmain}[2]{\textcolor{black}{#2}}
\newcommand{\jbmg}[1]{\textcolor{black}{#1}}

\newcommand{\yy}[2]{\textcolor{black}{#2}}

\begin{document}

\title{Dome structure in pressure dependence of superconducting transition temperature for HgBa$_2$Ca$_2$Cu$_3$O$_8$ --- Studies by \textit{ab initio} low-energy effective Hamiltonian}
\author{
Jean-Baptiste Mor\'ee$^{1,2}$ \orcidicon{0000-0002-0710-9880},
Youhei Yamaji$^{3}$ \orcidicon{0000-0002-4055-8792}, 
and Masatoshi Imada$^{1,4}$ \orcidicon{0000-0002-5511-2056}
}

\affiliation{
$^{1}$ Research Institute for Science and Engineering, Waseda University, 3-4-1, Okubo, Shinjuku, Tokyo 169-8555, Japan\\
$^{2}$ RIKEN Center for Emergent Matter Science, 2-1 Hirosawa, Wako, Saitama 351-0198, Japan\\
$^{3}$ \oldyy{Research Center for Materials Nanoarchitectonics (MANA) and Center for Green Research on Energy and Environmental Materials (GREEN), National Institute for Materials Science (NIMS)}, Namiki, Tsukuba-shi, Ibaraki, 305-0044, Japan\\
$^{4}$ Physics Division, Sophia University, Kioi-cho, Chiyoda-ku, Tokyo 102-8554, Japan
}


\begin{abstract}
The superconducting (SC) cuprate HgBa$_2$Ca$_2$Cu$_3$O$_8$ (Hg1223)
has the highest SC transition temperature $T_{c}$ among cuprates at ambient pressure $P_{\rm amb}$, 
namely, $T_{c}^{\rm opt}\simeq 138$ K experimentally at the optimal hole doping concentration.
$T_{c}^{\rm opt}$ further increases under pressure $P$ and reaches $164$ K at optimal pressure $P_{\rm opt}\simeq 30$ GPa, then $T_{c}^{\rm opt}$ decreases with increasing $P > P_{\rm opt}$ generating a dome structure [Gao \textit{et al.}, Phys. Rev. B 50, 4260(R) (1994)].
This nontrivial and nonmonotonic $P$ dependence of $T_{c}^{\rm opt}$ calls for theoretical understanding and mechanism.
To answer this open question, we consider the {\it ab initio} low-energy effective Hamiltonian (LEH) for the antibonding (AB) Cu$3d_{x^2-y^2}$/O$2p_{\sigma}$ band derived generally for the cuprates. 
In the AB LEH for cuprates with $N_\ell \leq 2$ laminated CuO$_2$ planes between block layers, 
it was proposed that $T_c^{\rm opt}$ is determined by a universal scaling $T_{c}^{\rm opt}\simeq 0.16|t_1|F_{\rm SC}$ [Schmid \textit{et al.}, Phys. Rev. X 13, 041036 (2023)], 
where $t_1$ is the nearest neighbor hopping, and the SC order parameter at optimal hole doping $F_{\rm SC}$
mainly depends on the ratio $u=U/|t_1|$ where $U$ is the onsite effective Coulomb repulsion:
The $u$ dependence of $F_{\rm SC}$ has a peak at $u_{\rm opt} \simeq 8.5$ and a steep decrease with decreasing $u$ in the region $u < u_{\rm opt}$ irrespective of materials dependent other {\it ab initio} parameters.
In this paper, we show that (I) $|t_1|$ increases with $P$, whereas (II) $u$ decreases with $P$ in the {\it ab initio} Hamiltonian of Hg1223.
Based on these facts, we show that the dome-like $P$ dependence of $T_{c}^{\rm opt}$ can emerge at least qualitatively if we assume
(A) Hg1223 with $N_\ell = 3$ follows the same universal scaling for $T_{c}^{\rm opt}$,
and (B) Hg1223 is located at the slightly strong coupling region $u \gtrsim u_{\rm opt}$ at $P_{\rm amb}$
and $u \simeq u_{\rm opt}$ at $P_{\rm opt}$ by taking account of expected corrections to our {\it ab initio} calculation. 
The consequence of (A) and (B) is the following:
With increasing $P$ within the range $P < P_{\rm opt}$, the increase in $T_{c}^{\rm opt}$ is accounted for by the increase in $|t_1|$, whereas $F_{\rm SC}$ is insensitive to the decrease in $u$ around $\simeq u_{\rm opt}$ and hence to $P$ as well. 
At $P > P_{\rm opt}$, the decrease in $T_{c}^{\rm opt}$ is accounted for by the decrease in $u$ below $u_{\rm opt}$, which causes a rapid decrease in $F_{\rm SC}$ dominating over the increase in $|t_1|$.
We further argue the appropriateness of the assumptions (A) and (B) based on the insight from studies in other cuprate compounds in the literature.
In addition, we discuss the dependencies of $u$ and $|t_1|$ on each crystal parameter (CP), which provides hints for designing of even higher $T_{c}^{\rm opt}$ materials.
\end{abstract}

\maketitle

\renewcommand{\theequation}{\arabic{equation}}

\section{Introduction}
\label{sec:introduction}

\begin{figure}[ht]
\centering
\includegraphics[scale=0.42]{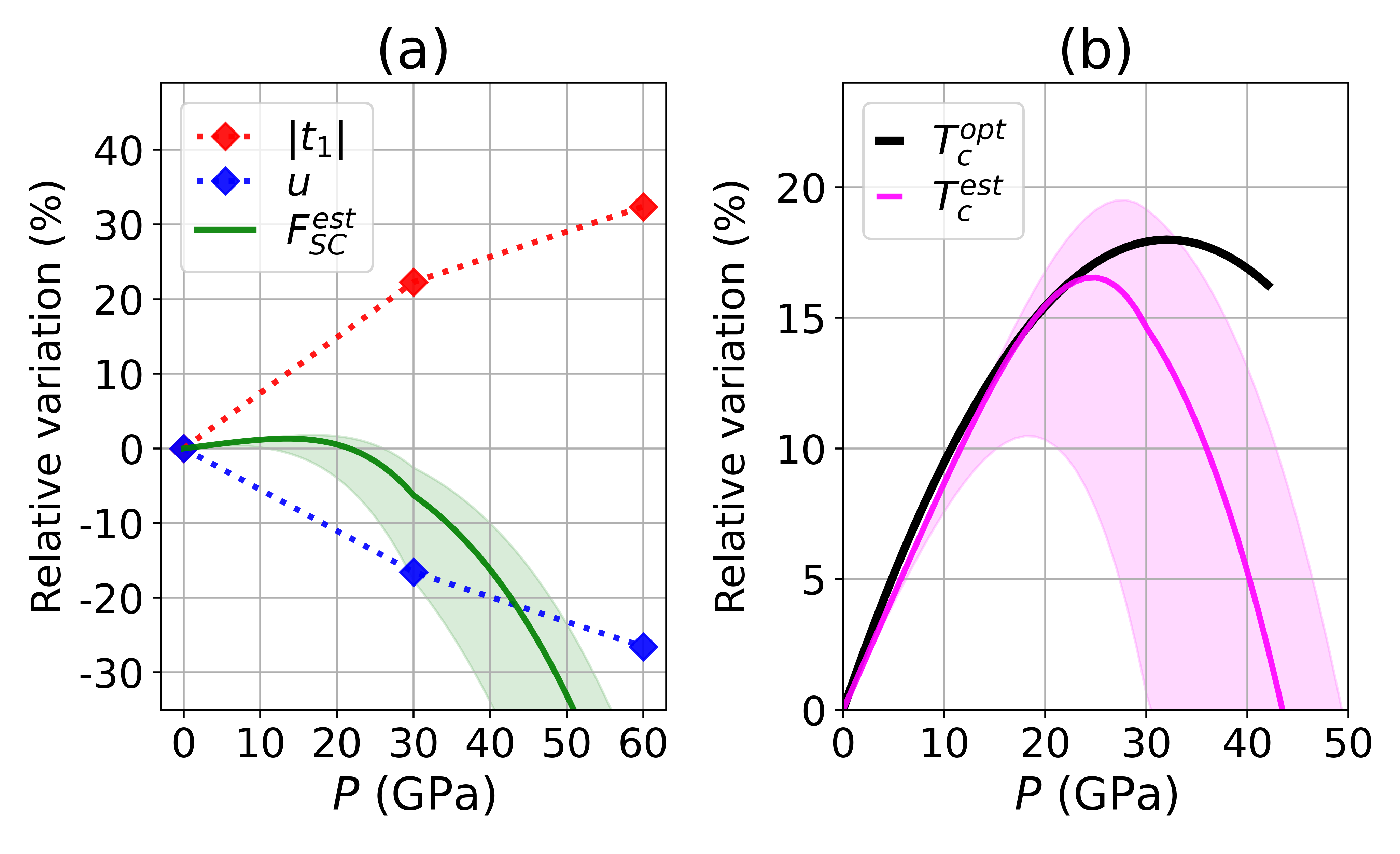}
\caption{
Summary of the main results obtained in this paper, and theoretical prediction of the dome structure in the $P$ dependence of $T_{c}$.
Panel (a): 
\mib{$P$ dependence of $|t_1|$ and $u$ deduced at the most sophisticated c$GW$-SIC+LRFB level. Here, these values are estimated from the calculated results at the c$GW$-SIC level by supplementing the correction from the c$GW$-SIC to the c$GW$-SIC+LRFB levels defined in Eqs.~(\ref{xLRFB}) and (\ref{yLRFB}). The pressure $P$ is measured from $P_{\rm amb}$.
$P$ dependence of $F_{\rm SC}$ estimated at the c$GW$-SIC+LRFB level denoted as $F_{\rm SC}^{\rm est}$ is also plotted, where $F_{\rm SC}^{\rm est}$ is deduced from the universal $u$ dependence found in Ref.~\cite{Schmid2023} and by using $u$ for Hg1223 shown here. See the main text below Eqs.~(\ref{xLRFB}) and (\ref{yLRFB}) for the detailed corrections of c$GW$-SIC and c$GW$-SIC+LRFB levels, where the notations for the quantities improved in such ways are denoted as $u_{{\rm c}GW{\rm -SIC+``LRFB"}}$ ($u_{{\rm c}GW{\rm -SIC}}$) and $|t_1^{}|_{{\rm c}GW{\rm -SIC+``LRFB"}}$ ($|t_1^{}|_{{\rm c}GW{\rm -SIC}}$) instead of $u$ and $|t_1|$, respectively to indicate the c$GW$-SIC+LRFB (c$GW$-SIC) levels explicitly.}
Diamond symbols show results at $P=P_{\rm amb}$, $30$ GPa and $60$ GPa, and dashed lines show linear interpolations between diamonds.
\mib{Panel (b): Experimental $T_{c}^{\rm opt}$~\cite{Gao1994} (black curve) and the present theoretical optimum $T_c$ denoted as $T_{c}^{\rm est}$ deduced from Eq.~(\ref{Tcscaling}) 
proposed in Ref.~\cite{Schmid2023} by replacing $F_{\rm SC}$ with $F_{\rm SC}^{\rm est}$. Shaded areas in (a) and (b) indicate the uncertainty described below Eqs.~(\ref{xLRFB}) and (\ref{yLRFB}) in the main text. Qualitative dome structure of $T_c^{\rm opt}$ is reproduced in the present prediction, $T_c^{\rm est}$.}
}
\label{fig:tcexp}
\end{figure}

\headline{Unconventional SC occurs in cuprates \cite{Bednorz1986} with a diverse distribution of $T_{c}^{\rm opt}$.}
At $P_{\rm amb}$, known values of $T_{c}^{\rm opt}$ range from $T_{c}^{\rm opt} \simeq 6$ K in Bi$_2$Sr$_2$CuO$_6$ (Bi2201) \cite{Torrance1988} to $T_{c}^{\rm opt} \simeq 138$ K in HgBa$_2$Ca$_2$Cu$_3$O$_8$ (Hg1223) \cite{Gao1994,Dai1995}.
\oldmi{$T_{c}^{\rm opt}$ further increases under pressure. 
In the case of Hg1223 and other Hg-based cuprates,}
$T_{c}^{\rm opt}$ has a dome-like structure as a function of $P$ \jbmd{\cite{Gao1994,Yamamoto2015}}. 
An example is shown in \mib{Fig.~\ref{fig:tcexp}(b)} for Hg1223:
$T_{c}^{\rm opt}$ increases \oldmi{with pressure and shows the maximum
164 K at $P_{\rm opt}\simeq 30$ GPa \cite{Nunez1993,Gao1994},}
which is the highest known value of $T_{c}^{\rm opt}$ in the cuprates.

\headline{
A wide range of $T_{c}^{\rm opt}\simeq6$-164 K in the cuprates 
has inspired studies on chemical substitution and pressure application to gain insights into the microscopic mechanism of the diversity in $T_{c}^{\rm opt}$.
}
\oldmi{For example,} for Y-based \cite{Crommie1989,Meingast1991,Belenky1991,Welp1992,Meingast1993,Mito2012,Mito2014,Mito2014meissner,Mito2016} 
and Hg-based \cite{Hardy2010,Mito2017} high-$T_{c}$ cuprates,
the uniaxial pressures $P_a$ and $P_c$ were applied.
(In this paper, $P_a$ refers to the simultaneous compression along axes ${\bf a}$ and ${\bf b}$ 
in Fig.~\ref{fig:cryst}, while keeping $|{\bf a}|=|{\bf b}|$, and $P_c$ refers to the compression along axis ${\bf c}$. 
The axes are represented in Fig.~\ref{fig:cryst} for the tetragonal cell in Hg1223.)
This decomposition of pressure revealed, in the case of HgBa$_2$CuO$_4$ (Hg1201,  $T_{c}^{\rm opt}\simeq 94$ K \cite{Putilin1993}), that $T_{c}^{\rm opt}$ decreases with out-of-CuO$_2$ plane contraction caused by $P_{c}$ ($\partial T_c^{\rm opt} / \partial P_c \simeq -3$ K/GPa) but increases with in-plane contraction caused by $P_{a}$ 
($\partial T_c^{\rm opt} / \partial P_a \simeq \oldmi{5}$ K/GPa) \cite{Hardy2010}.

\headline{
However, microscopic mechanisms leading to the 
\jbme{$P$} dependence of $T_{c}^{\rm opt}$
are not well understood yet, while understanding the mechanism of them certainly helps future materials design.
}
Since it is difficult to isolate these hidden mechanisms by experiments only,
further theoretical studies of cuprates under $P$ are desirable.

\headline{
In this paper, we propose a microscopic mechanism for the \jbme{$P$} dependence of $T_{c}^{\rm opt}$ for the carrier doped Hg1223 based on an {\it ab initio} study.
}
\oldmi{For {\it ab initio} studies, the density functional theory (DFT) has been widely applied in the history~\cite{Hohenberg1964,Kohn1965}. However, its insufficiency in strongly correlated electron systems is also well known. Instead,} 
\oldmi{we apply} the multiscale \textit{ab initio} scheme for correlated electrons (MACE) \cite{Aryasetiawan2004,Aryasetiawan2006,Imada2010,Hirayama2013,Hirayama2015,Hirayama2018,Hirayama2019},
which 
\oldmi{has succeeded in correctly reproducing the \jbme{SC} properties of the cuprates~\cite{Hirayama2019,Ohgoe2020,Moree2022,Schmid2023} at ambient pressure}
and has motivated further studies on hypothetical Ag-based compounds \cite{Hirayama2022silverarxiv}.

\jbmd{
MACE consists of a three-step procedure that determines the LEH parameters for the single-band AB Hamiltonian;
}
\jbmd{this procedure has several different accuracy levels, which are defined below and whose details are given in Appendix~\ref{app:method}.}
\jbmd{At the earliest stage of the MACE, the simplest level denoted as LDA+cRPA~\cite{Aryasetiawan2004,Aryasetiawan2006} or GGA+cRPA was employed; at this level, we start from the electronic structure at the Local Density Approximation (LDA) or Generalized Gradient Approximation (GGA) level, and the effective interaction parameters are calculated on the level of the constrained random phase approximation (cRPA)~\cite{Aryasetiawan2004}.}
\jbmd{The next level is denoted as c$GW$-SIC~\cite{Hirayama2018}, in which the starting electronic structure is preprocessed from the LDA or GGA level to the one-shot $GW$ level, and the one-particle part is improved by using the constrained $GW$ (c$GW$)~\cite{Hirayama2013} and the self-interaction correction (SIC)~\cite{Hirayama2015}.}
\jbmd{The most recent and accurate level is denoted as c$GW$-SIC+LRFB~\cite{Hirayama2019},
which is essentially the same as the c$GW$-SIC, except that the $GW$ electronic structure is further improved: The level renormalization feedback (LRFB) \cite{Hirayama2019} is used to correct the onsite Cu$3d_{x^2-y^2}$ and O$2p_{\sigma}$ energy levels.}

\jbmd{
Although the c$GW$-SIC+LRFB level is the most accurate and was used to reproduce the \jbme{SC} properties of the cuprates~\cite{Hirayama2019,Ohgoe2020,Moree2022,Schmid2023},
 we mainly employ the simplest GGA+cRPA version  for the purpose of the present paper, because the qualitative trend of the parameters can be captured by this simplest framework. \jbmd{(See Appendix~\ref{app:method} for a more detailed discussion.)}}
\jbmd{\mi{We also reinforce the analysis by \mib{deducing} more refined \jbme{c$GW$-SIC+LRFB} level in a limited case \mib{from the explicit cGW-SIC level calculations} to remove the known drawback of \jbmc{GGA+cRPA} as we detail later.}}

\headline{
We derive and analyze the pressure dependence of AB LEH parameters including various intersite hoppings and interactions;
however, we restrict the main discussion to $|t_1|$ and $u$, since they are the principal parameters \mi{that} control $T_{c}$ in the proposal~\cite{Schmid2023}.
}
Other LEH parameters are given in the Supplemental Material (S1)~\cite{hg1223supplemental}.
\jbmd{In the following, we mainly discuss $|t_1^{\rm avg}|$ and $u^{\rm avg}$, which are the \textit{ab initio} values of $|t_1|$ and $u$ at GGA+cRPA level, averaged over the inner and outer CuO$_2$ planes. (See Fig.~\ref{fig:cryst} for a representation of the CuO$_2$ planes.)
}
\\



This paper is organized as follows.
\mib{In Sec.~\ref{Overview}, the central results of the present paper \yy{is}{are} outlined to capture the essence of the results before detailed presentation.}
In Sec.~\ref{sec:meth}, we give \oldjbm{the crystal structure of Hg1223, the hole concentration and a reminder of the \jbmc{GGA+cRPA} scheme.}
In Sec.~\ref{sec:band}, we give \oldmi{the DFT electronic structure} at the GGA level as a function of $P$.
In Sec.~\ref{sec:leh}, we show the \oldjbm{pressure dependence of AB LEH parameters \jbmd{at the GGA+cRPA level.}}
In Sec.~\ref{sec:disc}, we discuss the \mi{adequacy of the assumptions made in Sec.~\ref{Overview}.} 
\mi{We also discuss} the consistency of our results with the experimental $P$ dependence of $T_{c}^{\rm opt}$ in Fig.~\ref{fig:tcexp}.
Summary and Conclusion are given in Sec.~\ref{sec:summary}.
In Appendix~\ref{app:method}, methodological details of MACE scheme are summarized.
In Appendix~\ref{app:comp_details}, computational details used in this paper are described.
\jbme{In Appendices~\ref{app:corr-c} and \ref{app:corr-d}, we detail the corrections used in Secs.~\ref{Overview} and \ref{sec:disc}.}
\oldjbm{
\jbme{In Appendix~\ref{app:pdepinterm}, we discuss in detail 
the $P$ dependence \mib{at the intermediate stage of the present procedure.}
}
In Appendix~\ref{app:unc}, we detail the dependence of AB LEH parameters on crystal parameters (CP) around optimal pressure.
}

\section{Overview }
\label{Overview}

\jbmd{The main results obtained in this paper are summarized \mib{as (I) and (II) below.}}

\noindent
\jbm{(I) $|\jbmd{t_1^{\rm avg}}|$ increases with $P$.
This increase in $|t_1|$ is caused specifically by the uniaxial pressure $P_{a}$,
in agreement with previous experimental studies on e.g. Hg1201 \cite{Hardy2010}.
}

\noindent
\jbm{(II) $u\jbmd{^{\rm avg}}$ decreases with $P$.
The decrease in $u$ is caused mainly by (I), \mib{namely by} the increase in $|t_1|$, but is slowed down by the increase in $U$ at $P<P_{\rm opt}$.
The increase in $U$ is also caused by $P_{a}$.
}
\\

\jbm{The \mi{nontrivial} pressure dependence of $T_{c}^{\rm opt}$ can be understood \mi{from (I,II), which is
derived from our \textit{ab initio} Hamiltonian  even} at the preliminary level \jbmc{GGA+cRPA}, if we assume the following (A) and (B)\yy{}{.}}
\jbme{\yy{[The reality of (A) and (B) will be discussed later in Sec.~\ref{sec:disc}, and details of (C) and (D) are given in Appendices~\ref{app:corr-c} and \ref{app:corr-d}, respectively.]}{[The reality of (A) and (B) will be discussed later in Sec.~\ref{sec:disc}.]}
}

\begin{description}
\item[\yy{}{(A)}] The universal scaling for $T_c^{\rm opt}$ \jbmfmain{is given}{given} theoretically as
\mib{
\begin{equation}
T_{c}^{\rm est}\simeq 0.16|t_1|F_{\rm SC}
\label{Tcscaling}
\end{equation} 
recently proposed for the cuprates with \jbmfmain{$N_{\ell} \leq 2$}{$N_{\ell}=$ 1, 2 and $\infty$}~\cite{Moree2022,Schmid2023} is also valid for Hg1223 with $N_{\ell} = 3$.}
\item[\yy{}{(B)}] \mi{ \mib{$F_{\rm SC}$ follows a universal $u$ dependence revealed in Ref.~\cite{Schmid2023}, where $F_{\rm SC}$ has a peak at $u=u_{\rm opt} \simeq \jbme{8.0-}8.5$.} In addition, at $P_{\rm amb}$, Hg1223 is located at slightly strong coupling side $u \gtrsim u_{\rm opt}$, while 
the highest pressure $P=60$ GPa applied so far is in the weak coupling side $u < u_{\rm opt}$.
\mib{In fact, we justify later} $u \jbmfmain{\sim}{ \ \simeq \ } u_{\rm opt}$ at optimal pressure $P_{\rm opt} \simeq 30$ GPa for Hg1223.
}
\end{description}

\mib{To understand the consequences of the assumptions (A) and (B) appropriately \yy{}{and to complement the consequences quantitatively}, we correct the errors anticipated in our {\it ab initio} GGA+cRPA calculation by using the following (C) and (D).}
\yy{}{[Details of (C) and (D) are given in Appendices~\ref{app:corr-c} and \ref{app:corr-d}, respectively.]}
\begin{description}
\item[\yy{}{(C)}] We correct the values of $u^{\rm avg}$ and $|t_1^{\rm avg}|$ \mib{obtained at the GGA+cRPA level by deducing the most sophisticated c$GW$-SIC+LRFB level}.
Since GGA+cRPA is known to underestimate $u$ in Bi2201 and Bi$_2$Sr$_2$CaCu$_2$O$_8$ (Bi2212), it is desirable to improve the AB LEH to the more accurate c$GW$-SIC+LRFB level.
However, \mib{the explicit calculation at the c$GW$-SIC+LRFB level is computationally demanding, while the corrections from the explicitly calculated c$GW$-SIC to the c$GW$-SIC+LRFB levels are known to be small and are relatively materials insensitive. Thus we represent the correction by a universal constant with admitted uncertainty.
The estimates of $u$ and $|t_1|$ improved in such ways are denoted as $u_{{\rm c}GW{\rm -SIC+``LRFB"}}$ and $|t_1^{}|_{{\rm c}GW{\rm -SIC+``LRFB"}}$
The 
procedure consists in the two steps (C1) and (C2):}
\jbme{
\item[\yy{}{(C1)}] c$GW$-SIC calculation: Starting from \mib{the whole and detailed pressure dependence of $u^{\rm avg}$ and $|t_1^{\rm avg}|$ for Hg1223 calculated at the GGA+cRPA level,
we calculate explicitly \mib{the level of the c$GW$-SIC denoted as} $u_{{\rm c}GW{\rm -SIC}}$ and $|t_1^{}|_{{\rm c}GW{\rm -SIC}}$ in \mib{limited cases of pressure choices of Hg1223 to reduce the computational cost.}}}
\jbme{
\item[\yy{}{(C2)}] \mib{Estimate at the c$GW$-SIC+LRFB level: We use
\begin{equation}
u_{{\rm c}GW{\rm -SIC+``LRFB"}} = x_{\rm LRFB} u_{{\rm c}GW{\rm -SIC}}
\label{xLRFB}
\end{equation}
 and 
 \begin{equation}
 |t_1|_{{\rm c}GW{\rm -SIC+``LRFB"}} = y_{\rm LRFB}|t_1|_{{\rm c}GW{\rm -SIC}}
\label{yLRFB}
\end{equation} 
and estimate constants $x_{\rm LRFB}$ and $y_{\rm LRFB}$ \jbmfmain{from}{in Hg1223 from} the already explicitly calculated results\jbmfmain{.}{ \ for other compounds Hg1201, CaCuO$_2$, Bi2201 and Bi2212.}
The estimated values are $x_{\rm LRFB} = 0.95$ (with the range of uncertainty $0.91-0.97$) and $y_{\rm LRFB} = 1.0$.
See Appendix~\ref{app:corr-c} for detailed procedure to estimate $x_{\rm LRFB}$ and $y_{\rm LRFB}$ for the case of Hg1223. 
}}
\jbme{
The concrete effect of (C) \mib{for Hg1223} is to increase $u$ \mib{from the cRPA level} by the ratio 
$u_{{\rm c}GW{\rm -SIC+``LRFB"}}/u^{\rm avg} \simeq 1.29$ at $P_{\rm amb}$, 
$\simeq 1.15$ at 30 GPa, and $\simeq 1.08$ at 60 GPa;
also, the $\simeq 13-14\%$ increase in $|t_1^{\rm avg}|$ from $P_{\rm amb}$ to 30 GPa becomes $\simeq 17\%$ by this correction. 
}
\noindent
\jbme{
\item[\yy{}{(D)}] After applying (C), we further correct the value of 
\jbmfmain{\mib{$|t_1|_{{\rm c}GW{\rm -SIC+``LRFB"}}$ and $u_{{\rm c}GW{\rm -SIC+``LRFB"}}$}}{$|t_1|_{{\rm c}GW{\rm -SIC+``LRFB"}}$} by considering the plausible error in crystal parameters at high pressure.}
Structural optimization by {\it ab initio} calculation is known to show quantitative error and it is preferable to correct it if experimental value is known. We compare our structural optimization and the experimental \jbme{cell parameter $a$} if it is available (this is the case at $P<8.5$ GPa) and assume that this trend of the deviation continues for $P > 8.5$ GPa, where experimental data are missing.
\jbme{Namely, at $P > 8.5$ GPa, we assume that our calculation overestimates the experimental $a$ by $\simeq 0.05$ \AA,
and we correct $a$ \jbme{by $\Delta a = -0.05$ \AA} \ accordingly.
The concrete effect of (D) is that the increase in $|t_1|_{{\rm c}GW{\rm -SIC+``LRFB"}}$ from $P_{\rm amb}$ to 30 GPa is now $\simeq 22\%$.
}
\end{description}

\jbme{The final estimates of
$u_{{\rm c}GW{\rm -SIC+``LRFB"}}$ and $|t_1|_{{\rm c}GW{\rm -SIC+``LRFB"}}$
are shown in Fig.~\ref{fig:tcexp}(a).
Since (C1) is computationally demanding, we perform (C) and (D) only at $P_{\rm amb}$, $30$ GPa and $60$ GPa, and infer the correction at other pressures by linear interpolation for the pressure dependence.
}

\mi{
\mib{Even by considering only (A) and (B) above, the present mechanism} qualitatively accounts for the microscopic trend of the dome structure:}
\jbm{At $P<P_{\rm opt}$, (I), namely the increase in $|t_1|$, plays the role to increase $T_{c}$, whereas the decrease in $u$ does not appreciably affect $F_{\rm SC}$ and thus $T_{c}$, because $F_{\rm SC}$ passes through the broad peak region in the $u$ dependence.
At $P>P_{\rm opt}$, (II), namely the decrease in $u$, drives the decrease in $F_{\rm SC}$ and thus $T_{c}$ \mi{surpassing} the increase in $|t_1|$,
which generates a dome structure.
If we \mi{take into account (C) and (D)} in addition to (A) and (B), 
the \jbme{dome structure in the $P$ dependence of experimental $T_{c}^{\rm opt}$ is \mib{more} quantitatively reproduced 
(see Fig.~\ref{fig:tcexp})}.} 
\jbm{
In addition to the above results, we discuss the dependence of AB LEH parameters on each CP,
which provides \mi{us with} hints for future designing of even higher $T_{c}^{\rm opt}$ materials.}
\\

\section{Framework of Method}
\label{sec:meth}

\begin{table*}[!htb]
\caption{Irreducible Cartesian atomic coordinates $(x,y,z)$ within the unit cell given by the $({\bf a},{\bf b},{\bf c})$ frame in Fig.~\ref{fig:cryst}.
The atom index $l$ is either Cu($i$) (Cu atom in the IP), O($i$) (O atom in the IP), Cu($o$) (Cu atom in the OP), O($o$) (O atom in the OP), 
O(ap) (apical O atom), Ca, Ba or Hg.
The coordinates of other atoms in the unit cell may be deduced by applying the transformations $(y,x,z)$ to O($i$) and O($o$) and $(x,y,-z)$ to Cu($o$), O($o$), Ca, Ba and O(ap).
The atomic coordinates are entirely determined by the seven CPs $a$, $c$, $d^{z}_{\rm Ca}$, $d^{z}_{\rm Cu}$, $d^{z}_{\rm buck}$, $d^{z}_{\rm Ba}$ and $d^{z}_{\rm O(ap)}$.
Note that $d^{z}_{\rm buck}$ [the displacement of O($o$) due to the Cu($o$)-O($o$)-Cu($o$) bond buckling] may be either positive or negative.
The CP values are listed in Fig.~\ref{fig:cryst} as a function of $P$. 
}
\begin{ruledtabular}
\begin{tabular}{m{2cm}m{1.8cm}m{1.8cm}m{1.8cm}m{1.8cm}m{1.8cm}m{1.8cm}m{1.8cm}m{1.8cm}}
Atom index $l$      & Cu($i$) & O($i$) & Ca & Cu($o$) & O($o$) & Ba & O(ap) & Hg\\
$x$ & 0 & $a/2$ & $a/2$ & 0 & $a/2$& $a/2$ & 0 & 0\\
$y$ & 0 & 0 & $a/2$ & 0 & 0 & $a/2$ & 0 & 0\\
$z$ & 0 & 0 & $d^{z}_{\rm Ca}$ & $d^{z}_{\rm Cu}$ & $d^{z}_{\rm Cu}-d^{z}_{\rm buck}$ & $d^{z}_{\rm Cu} +  d^{z}_{\rm Ba}$ &  $d^{z}_{\rm Cu} +  d^{z}_{\rm O(ap)}$ & $c/2$\\
\end{tabular}
\end{ruledtabular}
\label{tab:cryst}
\end{table*}
\begin{figure*}[!htb]
\includegraphics[scale=0.57]{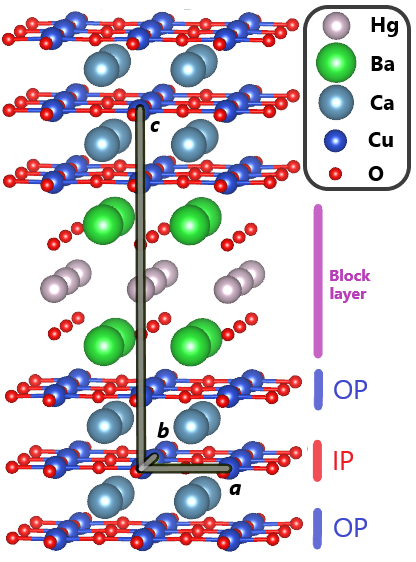}
\includegraphics[scale=0.29]{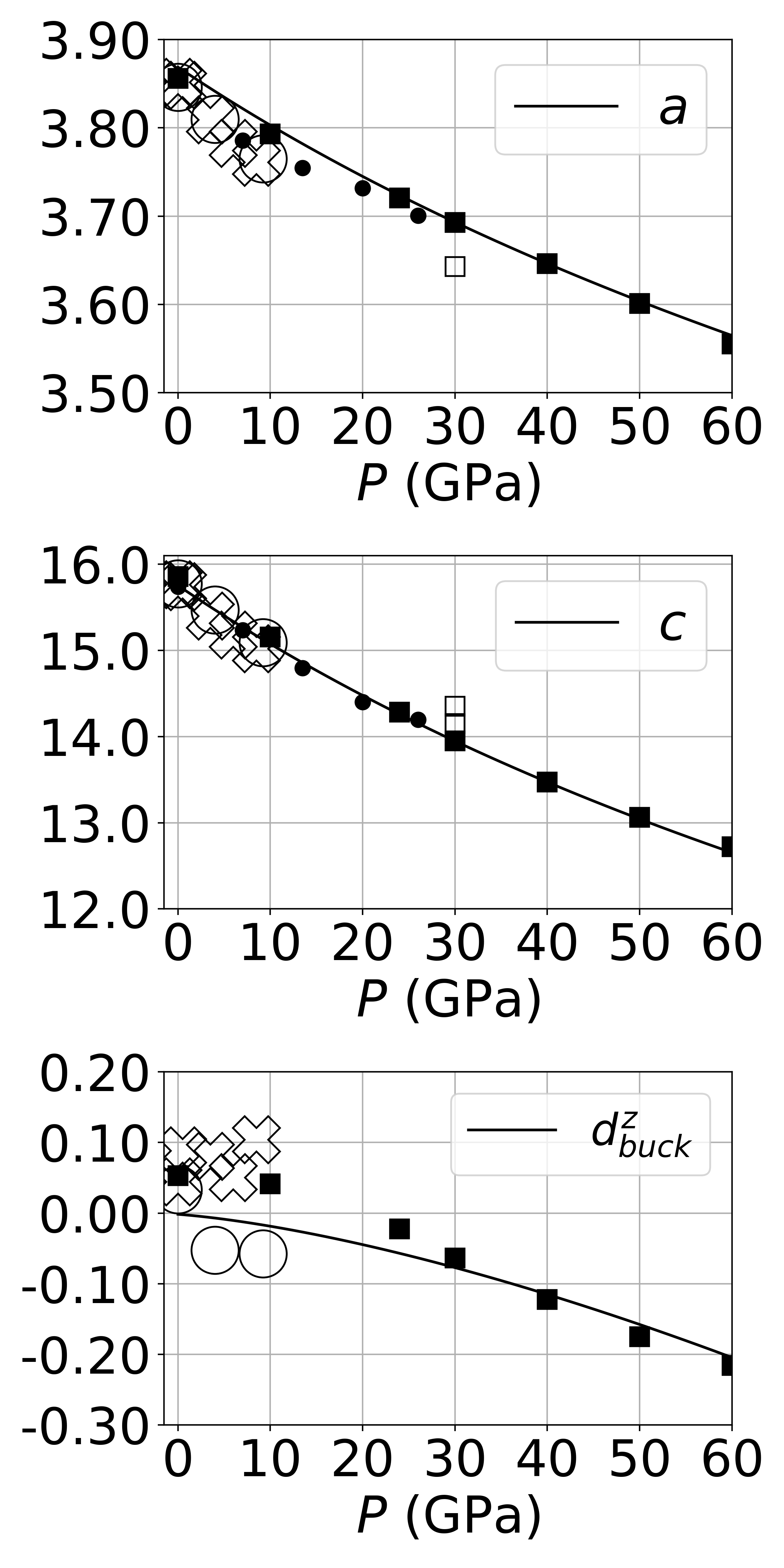} 
\includegraphics[scale=0.29]{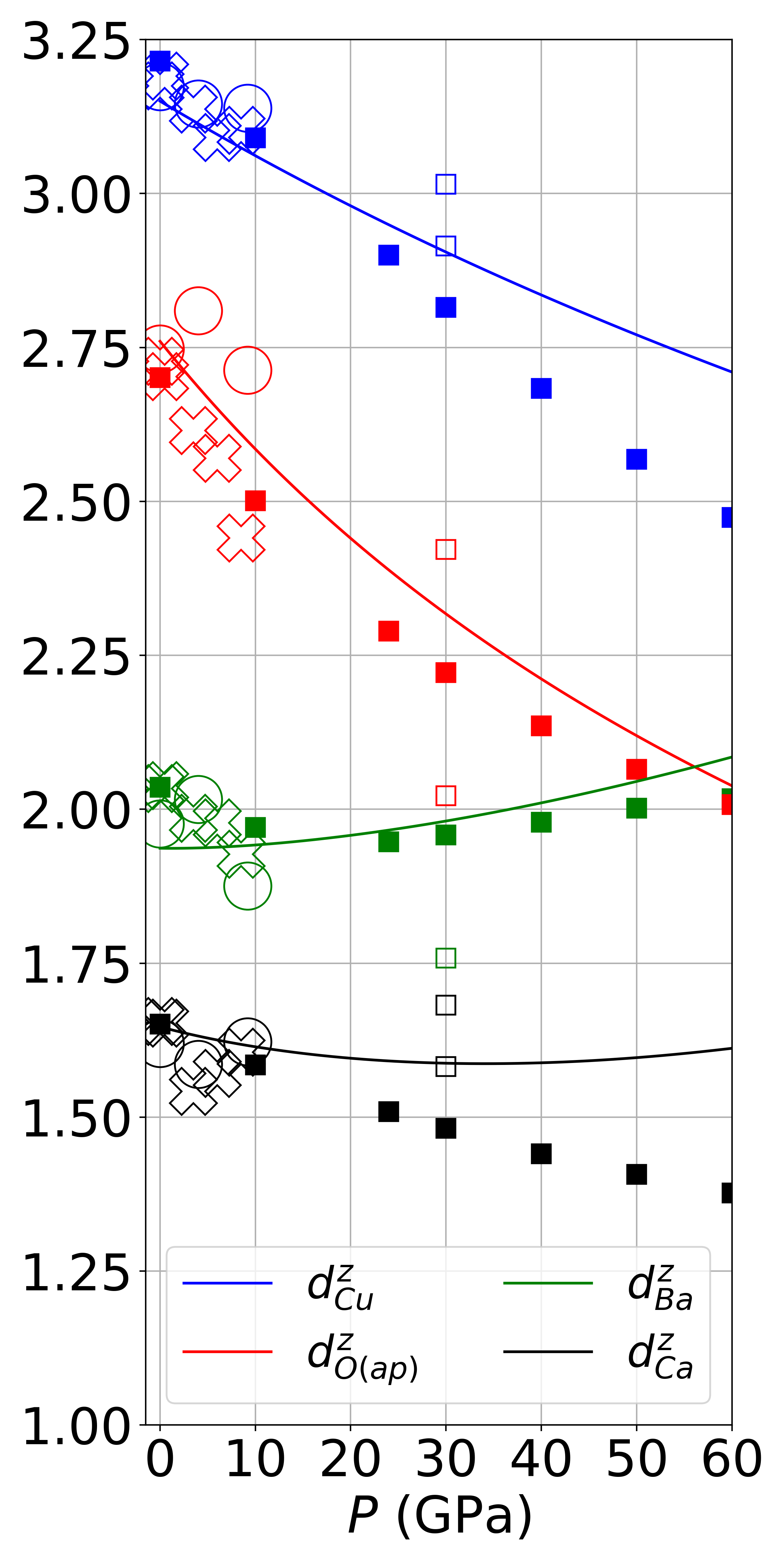} 
\caption{
Left panel: Crystal structure of Hg1223. 
We show the block layer, the inner CuO$_2$ plane (IP), the outer CuO$_2$ plane (OP), and the interstitial Ca atoms.
The thick \yy{black}{gray} lines represent the primitive lattice vectors ${\bf a},{\bf b},{\bf c}$.
The cell parameters are $a=|{\bf a}|=|{\bf b}|$ and $c=|{\bf c}|$; other CPs are defined in Table \ref{tab:cryst}.
Middle and right panels: Pressure dependence of the CP values in \AA. 
We show the optimized CP values (squares) and the extrapolated CP values from Zhang et al. \cite{Zhang1997} (solid lines); for details, see Appendix~\ref{app:choicecryst}. 
The open squares show the modifications of the optimized CP values at $P_{\rm opt} = 30$ GPa that are considered in Appendix~\ref{app:unc}.
For comparison, we also show the experimental CP values from Armstrong \textit{et al.} \cite{Armstrong1995} (\oldmi{open crosses}) and Hunter \textit{et al.} \cite{Hunter1994} (\oldmi{open circles}), and the values of $a$ and $c$ from Eggert \textit{et al.} \cite{Eggert1994}  (\oldmi{dots}).
}
\label{fig:cryst}
\end{figure*}
\begin{table}[!htb]
\caption{Definitions of the uniform pressure $P$ and uniaxial pressures $P_a$, $P_c$, $P_a^{\rm buck}$ and $P_{c}^{\overline{\rm buck}}$ considered in this paper.
Each CP is marked with a checkmark if its value is modified by the application of the pressure, and with a cross if not.
If the CP value is modified, the value is that in the $P$ dependence in Fig.~\ref{fig:cryst}.
If not, the value is that at $P_{\rm amb}$ in Fig.~\ref{fig:cryst}.
}
\label{tab:pu}
\begin{ruledtabular}
\begin{tabular}{cccccc}
&  $P$ & $P_a$ & $P_a^{\rm buck}$ & $P_c$ & $P_{c}^{\overline{\rm buck}}$ \\
$a$ & $\checkmark$ & $\checkmark$ & $\checkmark$ & $\times$ & $\times$ \\
$d^{z}_{\rm buck}$  & $\checkmark$ & $\times$ & $\checkmark$ & $\checkmark$ & $\times$ \\
$c$, $d^{z}_{\rm Ca}$, $d^{z}_{\rm Cu}$, $d^{z}_{\rm Ba}$, $d^{z}_{\rm O(ap)}$ & $\checkmark$ & $\times$ & $\times$ & $\checkmark$ & $\checkmark$ \\
\end{tabular}
\end{ruledtabular}
\end{table}

\headline{
We start from the crystal structure of Hg1223 and the pressure dependence of the CP values in Fig.~\ref{fig:cryst}.
}
\oldjbm{We abbreviate the inner and outer CuO$_2$ planes shown in Fig.~\ref{fig:cryst} as IP and OP, respectively.}
The crystal structure is entirely determined by \oldjbm{the seven CPs defined in Table~\ref{tab:cryst},
which consist of the two cell parameters $a$ and $c$ and the five characteristic distances $d^{z}_{l}$.}
\oldjbm{The CP values} considered in this paper are \oldmi{listed} in Fig.~\ref{fig:cryst}, as a function of $P$.
\oldjbm{In the main analyses of this paper, we consider (i) CP values obtained by a structural optimization, which are denoted as optimized CP values.
For comparison, we also consider (ii) the theoretical calculation of the CP values in Zhang \textit{et al.} \cite{Zhang1997} for the region between $P_{\rm amb}$ and $20$ GPa, and extrapolate the pressure dependence up to $60$ GPa.
Details about (i,ii) are given in Appendix \ref{app:choicecryst}.
We also consider (iii) the experimental CP values from Armstrong \textit{et al.} \cite{Armstrong1995} between $P_{\rm amb}$ and $8.5$ GPa. (The values at $P_{\rm amb}$ correspond to the SC phase with \jbm{the experimental SC transition temperature} $T_{c}^{\rm exp} \simeq 135$ K close to $T_{c}^{\rm opt} \simeq 138$ K.)
} \mi{It is known that the optimized CP values slightly deviate from the experimental values and it is indeed seen in Fig.~\ref{fig:cryst}. From the comparison of the optimized and experimental CPs, we take into account the correction \jbme{(D)} addressed in Sec.~\ref{sec:introduction}.}  

\headline{We simulate at the experimental optimal hole concentration $p$, 
which allows a reliable comparison with the $P$ dependence of $T_{c}^{\rm opt}$~\cite{Gao1994}.
}
We use the same procedure as \oldmi{that in Ref.~\onlinecite{Moree2022} employed for Hg1201}: We partially substitute Hg by Au.
We consider the chemical formula Hg$_{1-x_{\rm s}}$Au$_{x_{\rm s}}$Ba$_2$Ca$_2$Cu$_3$O$_{8}$ with $x_{\rm s}=0.6$ 
in order to realize the average hole concentration per CuO$_2$ plane
$p_{\rm av} = 0.2$ \cite{Bordet1996,Kotegawa2001,Yamamoto2015}.
This choice is discussed and justified in Appendix~\ref{app:hole}.

\headline{In addition, we examine the distinct effects of the uniaxial pressures along axis ${\bf a}$ and axis ${\bf c}$, 
whose definitions are given in Table~\ref{tab:pu} and discussed below.}
\oldjbm{
The nontrivial point is: Experimentally, what are the variations in CP values when the crystal structure is compressed along ${\bf a}$ (${\bf c}$) ?
First, the compression along ${\bf a}$ obviously modifies the cell parameter $a$ 
as well as the amplitude $|d^{z}_{\rm buck}|$ of the Cu-O-Cu bond buckling in the OP, but it should not affect the other CPs $d^{z}_{\rm Cu}$, $d^{z}_{\rm Ca}$, $d^{z}_{\rm Ba}$, and $d^{z}_{\rm O(ap)}$.
Thus, we define the uniaxial pressure $P_a^{\rm buck}$ along ${\bf a}$ as follows: 
The compression along ${\bf a}$ modifies the values of $a$ and $d^{z}_{\rm buck}$, 
and all other CP values are those at $P_{\rm amb}$.
We also consider a simplified definition, denoted as $P_a$: 
The compression along ${\bf a}$ modifies only the value of $a$, and all other CP values are those at $P_{\rm amb}$.
As we will see, $P_a$ is sufficient to describe the main effect of the compression along ${\bf a}$.
Second, the compression along ${\bf c}$ modifies the values of $d^{z}_{l}$, that is, all CP values except that of $a$.
This uniaxial pressure is denoted as $P_c$.
For completeness, we also consider a second definition, denoted as $P_{c}^{\overline{\rm buck}}$: 
The compression along ${\bf c}$ modifies all CP values except those of $a$ and $d^{z}_{\rm buck}$.
This allows to discuss the effect of the relatively large value of $|d^{z}_{\rm buck}|$ at $P > P_{\rm opt}$.
In the main analyses of this paper, we consider $P_a$ ($P_c$) to simulate the compression along ${\bf a}$ (${\bf c}$).
We also give complementary results by considering $P_a^{\rm buck}$ and $P_{c}^{\overline{\rm buck}}$.
}

\headline{We first compute the electronic structure at the DFT level.}
The \jbme{$P$} dependence of the GGA band structure is demonstrated in Fig.~\ref{fig:bandp}, from which we derive the LEH spanned by the Cu$3d_{x^2-y^2}$/O2$p_{\sigma}$ AB bands by employing the \jbmc{GGA+cRPA} scheme sketched in Appendix~\ref{app:method}. 
Computational details of DFT and \jbmc{GGA+cRPA} scheme are described in \jbme{Appendix~\ref{app:comp_details}.}

\headline{Then, we define the AB LEH as follows.}
In the AB LEH for multi-layer cuprates \cite{Moree2022},
there is only one AB orbital centered on each Cu atom. 
Then the AB LEH reads
\begin{equation}
\mathcal{H}^{}_{} = \sum_{l,l'} \mathcal{H}^{l,l'}_{} = \sum_{l,l'} \Big[ \mathcal{H}_{\rm hop}^{l,l'} + \mathcal{H}_{\rm int}^{l,l'} \Big],
\label{eq:h}
\end{equation}
where $l,l'=\{i,o,o'\}$ with $i$ \oldmi{being an IP site, and $o,o'$ belonging to the two equivalent OPs.}
in which we distinguish the hopping and interaction parts between planes $l$ and $l'$, as, respectively,
\begin{align}
\mathcal{H}_{\rm hop}^{l,l'} &  = \sum_{(\sigma{\bf R}),(\sigma'{\bf R'})} t^{l,l'}_{}({\bf{R'-R}}) \hat{c}^{\dagger}_{l\sigma{\bf R}} \hat{c}^{}_{l'\sigma' {\bf R'}}, \label{htll} \\
\mathcal{H}_{\rm int}^{l,l'} &  = \sum_{(\sigma {\bf R}),(\sigma' {\bf R'})} U^{l,l'}_{}({\bf{R'-R}}) \hat{n}^{}_{l \sigma {\bf R}} \hat{n}^{}_{l' \sigma' {\bf R'}}, \label{hrll}
\end{align}
where $\sigma,\sigma'$ are the spin indices.
By using these notations, ($l \sigma {\bf R}$) is the AB spin-orbital in the plane $l$ and in the unit cell at ${\bf R}$, with spin $\sigma$.
$c^{\dagger}_{l\sigma{\bf R}}$, $c^{}_{l\sigma{\bf R}}$ and $\hat{n}^{}_{l\sigma{\bf R}}$ are respectively the creation, annihilation and number operators in ($l \sigma {\bf R}$), 
and $t^{l,l'}_{}({\bf R'-R})$ and $U^{l,l'}_{}({\bf R'-R})$ are respectively the hopping and direct interaction parameters between ($l \sigma {\bf R}$) and ($l' \sigma' {\bf R'}$)\oldjbm{.} 
The translational symmetry allows to restrict the calculation of LEH parameters to 
$t^{l,l'}_{\sigma,\sigma'}({\bf R})$ and $U^{l,l'}_{\sigma,\sigma'}({\bf R})$ between ($l \sigma {\bf 0}$) and ($l' \sigma' {\bf R}$).

\headline{
In this paper, we focus on the intraplane LEH $\mathcal{H}_{\rm}^{l} = \mathcal{H}_{\rm}^{l,l}$ 
within the plane $l$ and analyze only the
first nearest-neighbor hopping $t_{1}^{l}=t^{l,l}_{}([100])$ 
and the onsite effective interaction $U^{l}=U^{l,l}_{}([000])$,
because these two parameters were
proposed to essentially determine $T_{c}^{\rm opt}$ at least for single- and two-layer cuprates~\cite{Schmid2023}.
}
\oldmi{(Other LEH parameters are given in the Supplemental Material 
 (S1)~\cite{hg1223supplemental}.) Then within this restricted range,} $\mathcal{H}_{}^{l}$ is rewritten as
\begin{equation}
\mathcal{H}_{}^{l} = |t_1^{l}| \Big[ \tilde{\mathcal{H}}_{\rm hop}^{l} + \oldjbm{u^{l}}\tilde{\mathcal{H}}_{\rm int}^{l}  \Big] = |t_1^{l}| \tilde{\mathcal{H}}_{}^{l},
\label{eq:htilde}
\end{equation}
in which $\tilde{\mathcal{H}}_{\rm hop}^{l} = \mathcal{H}_{\rm hop}^{l} / |t_1^{l}|$ and $\tilde{\mathcal{H}}_{\rm int}^{l} = \mathcal{H}_{\rm int}^{l} / U^{l}$ are the dimensionless hopping and interaction parts, expressed in units of their respective characteristic energies $|t_1^{l}|$ and $U^{l}$.
The full dimensionless intraplane LEH is $\tilde{\mathcal{H}}_{}^{l} = \mathcal{H}_{}^{l}/ |t_1^{l}|$\jbme{,} 
\jbme{and} the dimensionless ratio $u^{l}=U^{l}/|t_1^{l}|$ encodes the correlation strength.
\jbmd{As mentioned in Sec.~\ref{sec:introduction}, we also discuss the values of}
$|t_1^{\rm avg}|=(|t_1^{i}|+|t_1^{o}|)/2$ and $\oldjbm{u^{\rm avg}}=(u^{i}+u^{o})/2$.
Average values of other quantities with the superscript $l$ are defined similarly. 

\headline{
We compute the above LEH parameters $|t_1^{l}|$ and $U^{l}$ as follows.}
\oldjbm{We use the \texttt{RESPACK} code \cite{Nakamura2020,Moree2022}.} 
\oldmi{The standard calculation procedure} is presented in detail elsewhere \cite{Nakamura2020,Moree2022}.
First, we compute $t_1^{l}$ as
\begin{equation}
t_1^{l} = \int_{\jbmg{\Omega}} dr w^{*}_{l{\bf 0}}(r) h(r) w_{l{\bf R_1}}(r),
\end{equation}
in which $w_{l{\bf R}}$ is the \oldmi{Wannier} function of the AB orbital $(l{\bf R})$, \jbmg{$R_1=[100]$, $\Omega$ is the unit cell,
and $h$ is the one-particle part at the GGA level.}
Then, we compute $U^{l}$ as follows.
We compute the cRPA effective interaction $W_{\rm H}$, whose expression is \oldmi{found} in Appendix~\ref{app:crpa}\oldjbm{, Eq.~\eqref{eq:wh}}. 
We use a plane wave cutoff energy of $8$ Ry.
We deduce the onsite effective Coulomb interaction as 
\begin{equation}
U^{l} = \int_{\jbmg{\Omega}} dr \int_{\jbmg{\Omega}} dr' w_{l{\bf 0}}^{*}(r) w_{l{\bf 0}}^{*}(r') W_{\rm H}(r,r') w_{l{\bf 0}}^{}(r) w_{l{\bf 0}}^{}(r').
\label{eq:u}
\end{equation}
We also deduce the onsite bare Coulomb interaction $v^{l}$ by replacing $W_{\rm H}$ by the bare Coulomb interaction $v$ in Eq.~(\ref{eq:u}),
and the cRPA screening ratio $R^{l}=U^{l}/v^{l}$.
The obtained values of $|t_1^{l}|$, $U^{l}$, $v^{l}$ and $R^{l}$ are plotted in Fig.~\ref{fig:ratiop}.

\section{Pressure dependence of electronic structure at DFT level}
\label{sec:band}
   
\begin{figure*}[!htb]
\centering
\includegraphics[scale=0.38]{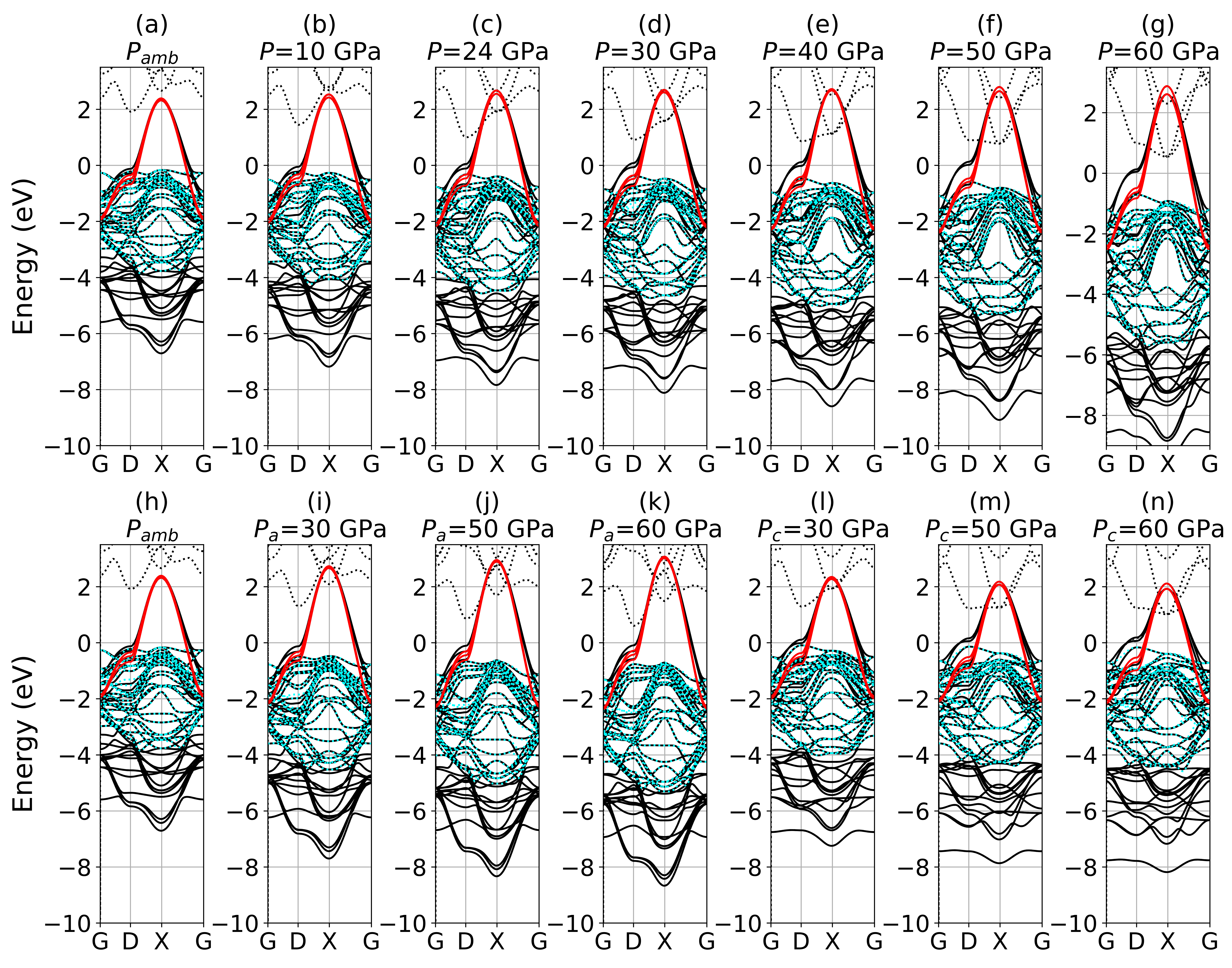} 
\includegraphics[scale=0.39]{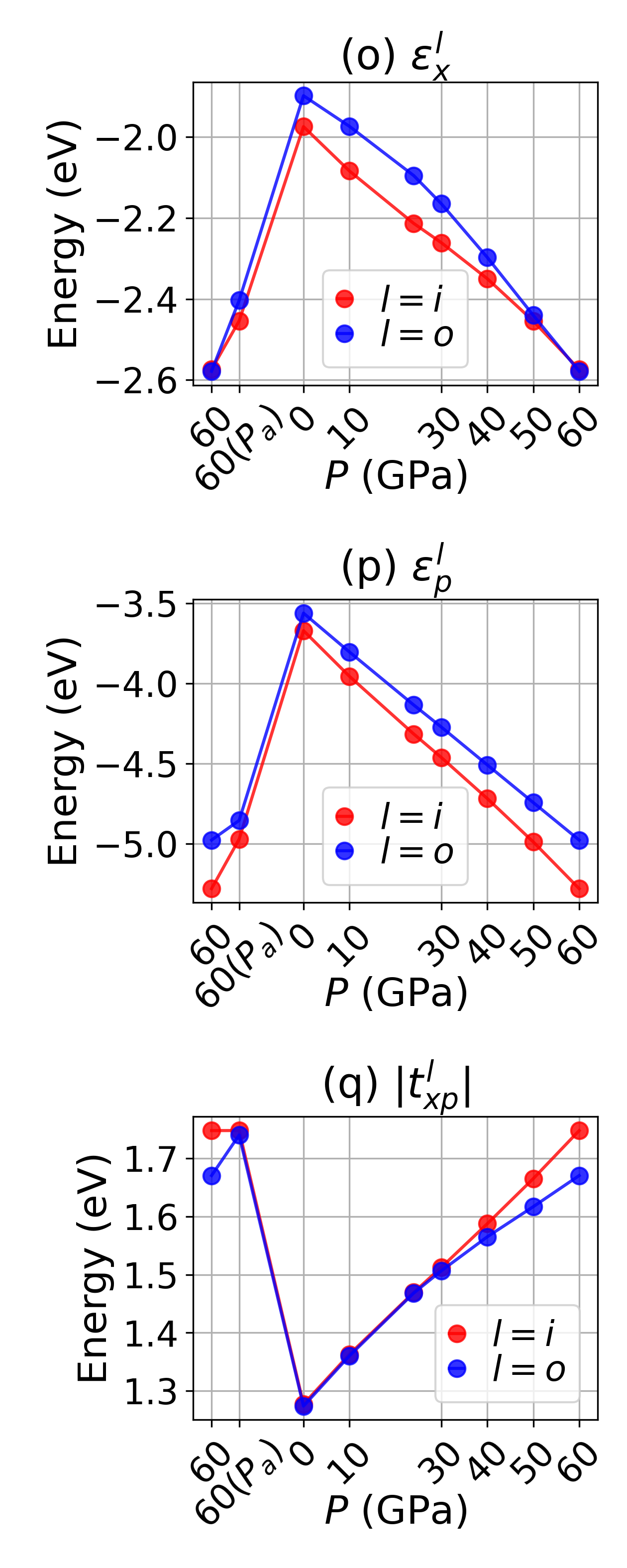} 
\caption{
\oldjbm{
Panels (a-g): Uniform pressure dependence of the GGA band structure.
We show the GGA bands outside (dashed black color) and inside (solid black color) the M space, 
the AB bands (red color) ,
and the 29 other bands in the band window, which are disentangled from the AB band (dashed cyan color).
High-symmetry points are, in coordinates of the reciprocal lattice: ${\rm G}=[0 \ 0 \ 0]$, ${\rm D}=[\yy{}{1/2} \ 0 \ 0]$, and ${\rm X}=[\yy{}{1/2} \ \yy{}{1/2} \ 0]$.
Panels (h-n): Uniaxial pressure dependence of the GGA band structure.
Panels (o-q): Uniform pressure dependence of 
the Cu$3d_{x^2-y^2}$ onsite energy $\epsilon_{x}^{l}$,
the in-plane O$2p_{\sigma}$ onsite energy $\epsilon_{p}^{l}$,
and the Cu$3d_{x^2-y^2}$/O$2p_{\sigma}$ hopping in the unit cell $t_{xp}^{l}$
in the IP ($l = i$) and OP ($l = o$). 
We also show the result at the uniaxial pressure $P_a = 60$ GPa [denoted as 60($P_a$)].
All quantities are obtained by using the optimized CP values.
}
}
\label{fig:bandp}
\end{figure*}

\headline{
Now, we show the result of GGA calculation as a function of $P$
and clarify what can be learned within the DFT level already.
}
The band dispersion is shown in Fig.~\ref{fig:bandp}(a-l).
We also show in Fig.~\ref{fig:bandp}(o-q) the onsite energy of the Cu$3d_{x^2-y^2}$ and in-plane O$2p_{\sigma}$ atomic-like Wannier orbital (ALWO).
(As explained in Appendix~\ref{app:comp_details}, we denote these ALWOs as M-ALWOs because they are in the M space.)
We also show the Cu$3d_{x^2-y^2}$/O$2p_{\sigma}$ hopping amplitude $|t_{xp}^{l}|=|t^{{\rm Cu(}l{\rm ),O(}l{\rm )}}_{x^2-y^2,p_{\sigma}}|$ in the unit cell.

\headline{We first elucidate main mechanisms of the following items [M$W$], [M$\epsilon$] and [M$t$] when $P$ increases:}

\noindent
\headline{[M$W$] Broadening of the M band dispersion in Fig.~\ref{fig:bandp}(a-g)}

\noindent
\headline{[M$\epsilon$] Decrease in onsite energies of M-ALWOs relative to the Fermi level in Fig.~\ref{fig:bandp}(o,p)}

\noindent
\headline{[M$t$] Increase in hoppings between M-ALWOs in Fig.~\ref{fig:bandp}(q).\\}

\headline{A simple interpretation of [M$W$], [M$\epsilon$] and [M$t$] is that $P$ works to reduce the interatomic distances.}
\oldjbm{
This causes two distinct effects: 
First, the electrons in the CuO$_2$ plane feel the stronger Madelung potential from ions in the crystal.
Indeed, the amplitude of the Madelung potential scales as $1/d$, where $d$ is the interatomic distance between the ion and the Cu or O atom in the CuO$_2$ plane.
The variation in Madelung potential modifies the M-ALWO onsite energies and causes [M$\epsilon$] (for details, see Appendix~\ref{app:dftband}).
Second, the overlap and hybridization between M-ALWOs increases, which causes [M$t$].
Both [M$\epsilon$] and [M$t$] increase the splitting of the B/NB (bonding/nonbonding) and AB bands, which causes [M$W$]:
The bandwidth $W$ of the M bands increases from $W \simeq 9$ eV at $P_{\rm amb}$ to $W \simeq 12$ eV at $P=60$ GPa [see Fig.~\ref{fig:bandp}(a-g)].
Simultaneously, the bandwidth $W_{\rm AB}$ of the AB band increases from $W_{\rm AB} \simeq 4$ eV at $P_{\rm amb}$ to $W_{\rm AB} \simeq 5.5$ eV at $P=60$ GPa,
which is caused by [M$t$].
Indeed, the increase in $|t_1^l|$ and thus $W_{\rm AB} \simeq 8|t_1|$ originates from the increase in $|t_{xp}^{l}|$, as discussed later in Sec.~\ref{sec:leht}.
}

\headline{
Effects of the uniaxial pressures $P_a$ and $P_c$ to [M$W$], [M$\epsilon$] and [M$t$] can also simply be accounted for when we consider the anisotropy of the overlap of the two M-ALWOs and the direction of the pressure.
}
\oldmi{For instance,} [M$W$] is caused by $P_a$ rather than $P_c$ [see Fig.~\ref{fig:bandp}(h-n)],
\oldmi{because the the AB bandwidth $W_{\rm AB}$ and $W$ are mainly determined by the 
overlap between Cu$3d_{x^2-y^2}$ and O$2p_{\sigma}$ ALWOs in a CuO$_2$ plane.} 
This increase in the bandwidth with $P_{a}$ was also mentioned in Ref.~\onlinecite{Sakakibara2012prboct} in the case of Hg1201.
On the other hand, the application of $P_c$ shifts a few specific bands: 
Hg$5d$-like bands are shifted from $-4/-5$ eV at $P_{\rm amb}$ to $-7$ eV at $P_c=30$ GPa.
However, $P_c$ does not modify $W_{\rm AB}$. 
Effects of uniaxial pressure on [M$\epsilon$] and [Mt] are also obviously and intuitively understood in a similar fashion:
We clearly see in Fig.~\ref{fig:bandp}(o-q) that [M$\epsilon$] and [Mt] are caused by $P_a$ rather than $P_c$.
For more details of the pressure effects, see Appendix~\ref{app:dftband}.

\section{Pressure dependence of AB effective Hamiltonian}
\label{sec:leh}
\begin{figure*}[!htb]
\centering
\includegraphics[scale=0.162]{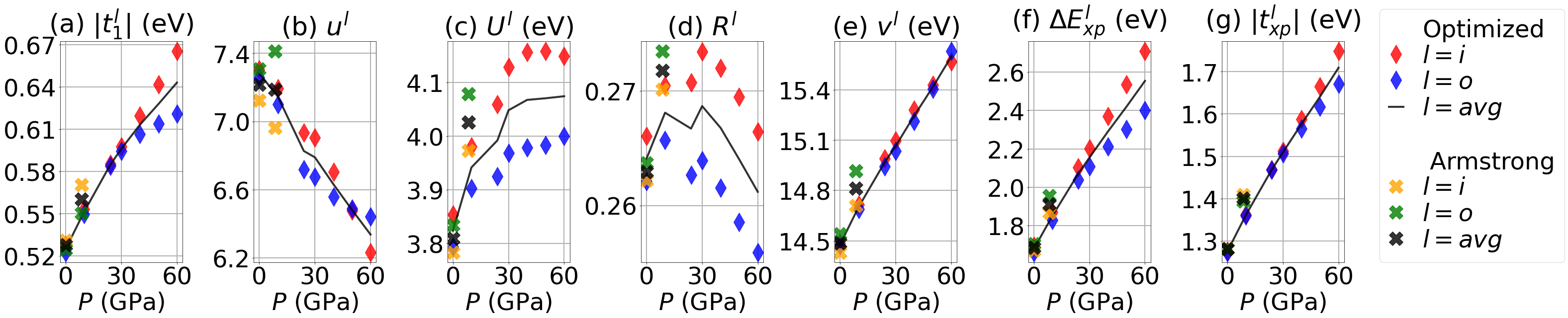}
\includegraphics[scale=0.162]{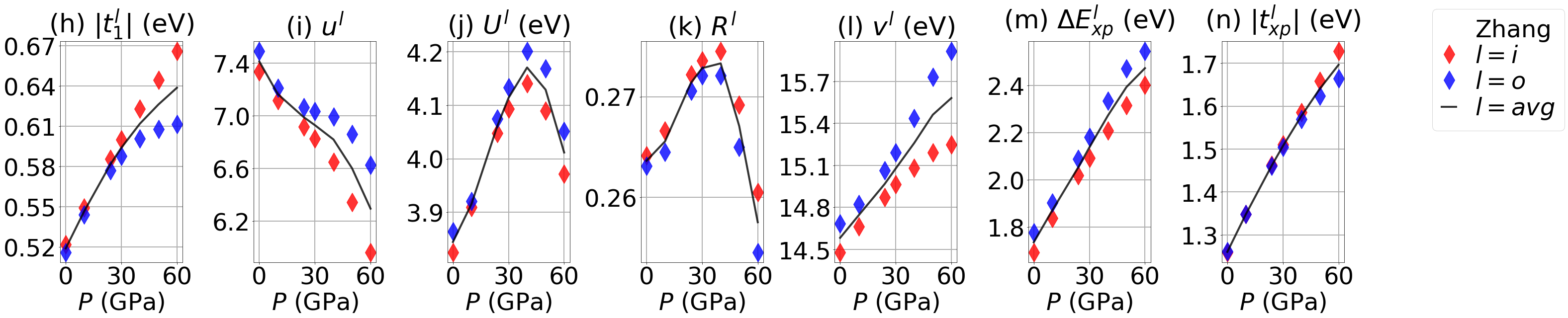}
\caption{
\oldjbm{
Uniform pressure dependence of the AB LEH parameter values in the IP ($l=i$), the OP ($l=o$), and the average value over the IP and OP ($l={\rm avg}$).
We show the basic energy unit $|t_1^{l}|$,
the correlation strength $u^{l}=U^{l}/|t_1^{l}|$,
the onsite effective Coulomb interaction $U^{l}$,
the screening ratio $R^{l}=U^{l}/v^{l}$,
and the onsite bare interaction $v^{l}$.
In addition, we show the charge transfer energy $\Delta E_{xp}^{l}$,
 and the amplitude of the hopping $t_{xp}^{l}$ between the Cu$3d_{x^2-y^2}$ and in-plane O$2p_{\sigma}$ ALWOs at the GGA level.
We show the quantities obtained by using the optimized CP values [panels (a-g)],
the experimental CP values from Armstrong \textit{et al.} \cite{Armstrong1995} at $P_{\rm amb}$ and $8.5$ GPa [crosses in the panels (a-g)],
and the CP values from Zhang \textit{et al.} [panels (h-n)].
}
}
\label{fig:ratiop}
\end{figure*}

\begin{figure*}[!htb]
\centering
\includegraphics[scale=0.162]{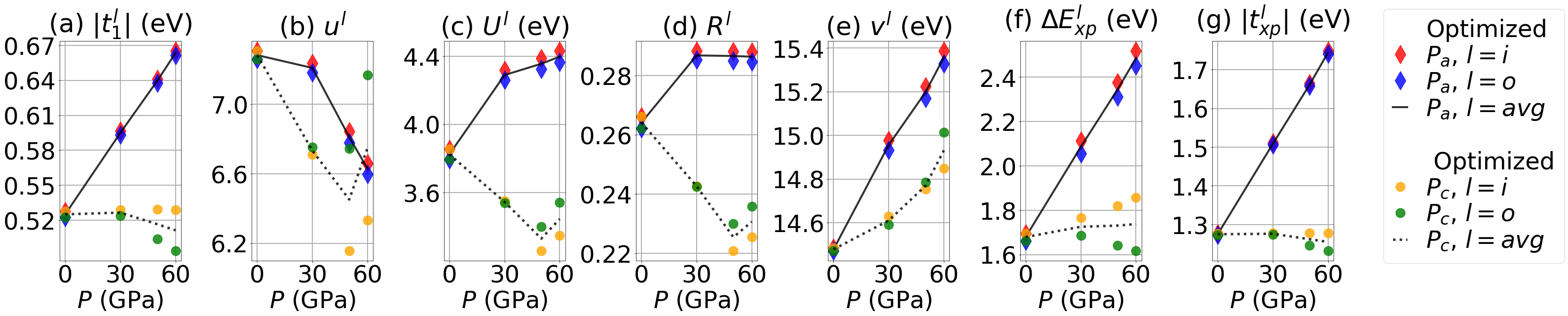}
\includegraphics[scale=0.162]{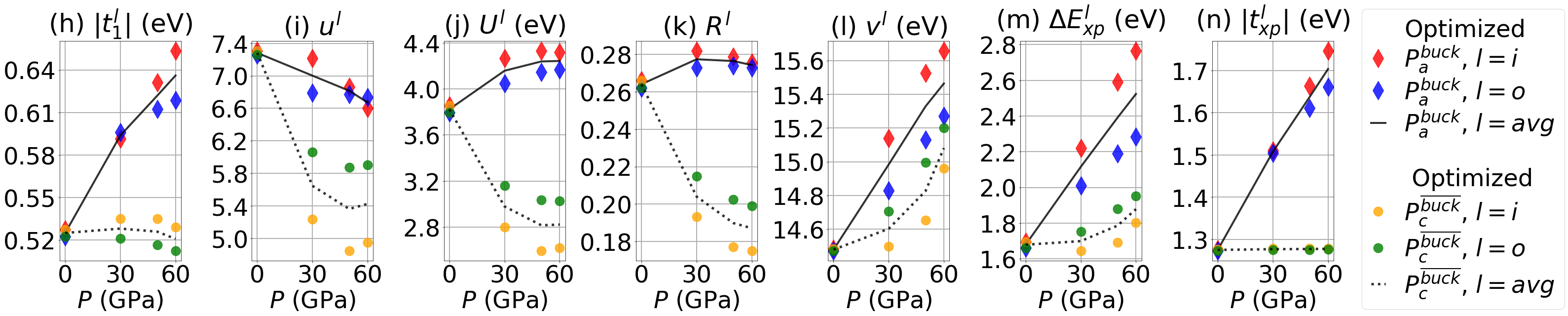}
\caption{
\oldjbm{
Uniaxial pressure dependence of the AB LEH parameter values.
Notations are the same as those in Fig.~\ref{fig:ratiop}.
All quantities are obtained by using the optimized CP values.
We show the $P_{a}$ and $P_{c}$ dependencies [panels (a-g)],
and the $P_{a}^{\rm buck}$ and $P_{c}^{\overline{\rm buck}}$ dependencies [panels (h-n)].
}
}
\label{fig:ratiopu}
\end{figure*}

\headline{
Now, we discuss the $P$ dependence of AB LEH parameters in Fig.~\ref{fig:ratiop}(a,b), in which the two main mechanisms (I,II) are visible:
(I) $|t_{1}^{l}|$ increases,
whereas (II) $u^{l}$ decreases.
}
In this section, we discuss the mechanisms of (I,II) that are summarized in Table~\ref{tab:summary}, 
and demonstrate that (I,II) are indeed physical and robust.
We discuss mainly \jbme{$|t_{1}^{\rm avg}|$ and $u^{\rm avg}$},
and discuss briefly the difference between values in the IP and OP.
\oldmi{A comparison with experiments will be made} separately in Sec.~\ref{sec:disc}.

\begin{table}[ht]
\caption{Summary of the variations in AB LEH parameters with $P < P_{\rm opt}$ and $P > P_{\rm opt}$ in Fig.~\ref{fig:ratiop}.
We use $\nearrow$, $\nearrow\nearrow$, $\simeq$, $\searrow$ or $\searrow\searrow$ if the quantity increases, strongly increases, remains static,
decreases or strongly decreases, respectively.
The variation in $u=U/|t_1|$ is controlled by that in $|t_1|$ and $U$.
The variation in $U=vR$ is controlled by that in the onsite bare interaction $v$ and the cRPA screening ratio $R$.
}
\label{tab:summary}
\begin{ruledtabular}
\begin{tabular}{cccccc}
& $|t_1|$ & $u=U/|t_1|$ & $U=vR$ & $v$ & $R$ \\
$P < P_{\rm opt}$  & $\nearrow\nearrow$ & $\searrow$ & $\nearrow$ & $\nearrow$ & $\nearrow$\\
$P > P_{\rm opt}$  & $\nearrow\nearrow$ & $\searrow\searrow$ & $\simeq$ /$\searrow$ & $\nearrow$ & $\searrow$/$\searrow\searrow$\\
\end{tabular}
\end{ruledtabular}
\end{table}


\subsection{Increase in $|t_{1}^{}|$ with $P$}
\label{sec:leht}

\headline{
The increase (I) in the $P$ dependence of $|t_{1}^{\rm avg}|$ [see Fig.~\ref{fig:ratiop}(a)] is purely caused by the reduction of cell parameter $a$ when the crystal is compressed along axis ${\bf a}$.
}
\oldjbm{Indeed, (I) is purely caused by the application of $P_a$ [see Fig.~\ref{fig:ratiopu}(a)], whose only effect is to reduce $a$.}
The underlying \oldmi{origin is simply} the increase in overlap between AB orbitals on neighboring Cu atoms due to the decrease in cell parameter $a$ when increasing $P_a$ \oldmi{as already discussed in Sec.~\ref{sec:band} at the DFT level}.
We note that $|t_{1}^{l}|$ has a similar $P$ dependence as that of $|t_{xp}^{l}|$ [see Fig.~\ref{fig:ratiop}(a,g)].
\oldmi{This is obvious because} the AB orbital is formed by the hybridization of Cu$3d_{x^2-y^2}$ and O$2p_{\sigma}$ M-ALWOs.

\headline{Note that, at $P > P_{\rm opt}$, $|t_{1}^{o}|$ is reduced with respect to $|t_{1}^{i}|$; this is because of the buckling of Cu-O-Cu bonds in the OP.}
\oldjbm{
Indeed, the decrease in $|t_{1}^{o}|-|t_{1}^{i}|$ and also $|t_{xp}^{o}|-|t_{xp}^{i}|$ occurs in the $P_{a}^{\rm buck}$ dependence [see Fig.~\ref{fig:ratiopu}(h)] but not in the $P_a$ dependence [see Fig.~\ref{fig:ratiopu}(a)],
and the value of $d^{z}_{\rm buck}$ is modified by the application of $P_{a}^{\rm buck}$ but not by the application of $P_a$.
Furthermore, the $P$ dependence of $|t_{1}^{o}|-|t_{1}^{i}|$ is consistent with that of $|d^{z}_{\rm buck}|$:
The decrease in $|t_{1}^{o}|-|t_{1}^{i}|$ starts at $P_{\rm opt}$ and is amplified at larger pressures [see Fig.~\ref{fig:ratiop}(a,g)], 
which is consistent with the increase in $|d^{z}_{\rm buck}|$ from $0.05$ \AA \ to $0.20$ \AA \ between $P_{\rm opt}$ and $60$ GPa (see Fig.~\ref{fig:cryst}).
The origin of the decrease in $|t_{1}^{o}|-|t_{1}^{i}|$ can be understood as follows:
When $|d^{z}_{\rm buck}|$ increases, the overlap between Cu$3d_{x^2-y^2}$ and O$2p_{\sigma}$ M-ALWOs in the OP is reduced.
Note that the buckling induced decrease in $|t_{1}^{}|$ has also been observed in the two-layer cuprate \jbmd{Bi2212} \cite{Moree2022}. 
}

\headline{
Comparison of results obtained from different CP values shows that (I) is physical and robust.
}
\oldjbm{
If we consider both (i) the optimized CP values and (ii) the CP values from Zhang \textit{et al.},
the $P$ dependencies of $|t_{1}^{l}|$ and $|t_{xp}^{l}|$ are very similar for (i) and (ii) [see Fig.~\ref{fig:ratiop}(a,g,h,n)].
This is intuitive since the $P$ dependence of $a$ is similar for (i) and (ii), 
and the $P$ dependence of $d^{z}_{\rm buck}$ at $P>P_{\rm opt}$ is also similar (see Fig.~\ref{fig:cryst}).
If we consider (iii) the experimental CP values from Armstrong \textit{et al.} \cite{Armstrong1995} at $P < 8.5$ GPa, 
the increase in $|t_1^{\rm avg}|$ and $|t_{xp}^{\rm avg}|$ is faster.
This is \oldmi{in accordance with} the faster decrease in $a$ for (iii) with respect to (i,ii) (see Fig.~\ref{fig:cryst}), \oldmi{and implies the uncertainty of the estimate} of $|t_1^{\rm avg}|$ at $P_{\rm opt}$, as discussed later in Section~\ref{sec:disc}.
}

\subsection{Decrease in $u^{}$ with $P< P_{\rm opt}$}
\label{sec:lehut}

\headline{At $P< P_{\rm opt}$, the decrease (II) in $u^{\rm avg}$ is largely induced by the increase (I) in $|t_{1}^{\rm avg}|$;
however, the increase in $U^{\rm avg}$ [see Fig.~\ref{fig:ratiop}(c)] partially cancels the decrease in $u^{\rm avg}$.}
\oldjbm{Thus, we discuss the $P$ dependence of $U^{\rm avg}$ \jbme{below}.}

\headline{The increase in $U^{\rm avg}$ is caused by two cooperative factors (i,ii) whose main origin is the reduction in $a$.}
\oldjbm{ These are
(i) the increase in onsite bare interaction $v^{\rm avg}$ [see Fig.~\ref{fig:ratiop}(c)], 
and (ii) the reduction in cRPA screening represented by the increase in the average value $R^{\rm avg}$ of the cRPA screening ratio $R^{l}=U^{l}/v^{l}$ [see Fig.~\ref{fig:ratiop}(d)]. 
In the following, we discuss the \oldmi{microscopic origins} of (i,ii).}
\\

\headline{
On (i), the increase in $v^{l}$ mainly originates from the increase in charge transfer energy $\Delta E^{l}_{xp}$ between Cu$3d_{x^2-y^2}$ and O$2p_\sigma$ M-ALWOs. 
}
\oldjbm{
This is because the increase in $\Delta E^{l}_{xp}$ reduces the importance of the Cu$3d_{x^2-y^2}$/O$2p_\sigma$ hybridization.
(The latter is roughly encoded in the ratio $O_{xp}^{l}=|t_{xp}^{l}|/\Delta E_{xp}^{l}$.)
The reduction in hybridization increases the Cu$3d_{x^2-y^2}$ atomic character and thus the localization of the AB orbital.
This is discussed and justified in the item (a) in Appendix~\ref{app:ct}.
This simple view is consistent with the systematic correlation between $v^{l}$ and $\Delta E_{xp}^{l}$
in this paper [see Fig.~\ref{fig:ratiop}(e,f,l,m) and also \jbmd{Appendix~\ref{app:unc}}], 
and also in the literature \cite{Hirayama2018,Moree2022}.
Still, note that the correlation between $v^{l}$ and $\Delta E_{xp}^{l}$ is slightly reduced at $P > P_{\rm opt}$ [see Fig.~\ref{fig:ratiop}(e,f,l,m) at $P > P_{\rm opt}$].
This is because 
\jbmc{$|t_{xp}^{o}|$ is reduced with respect to $|t_{xp}^{i}|$ at $P > P_{\rm opt}$ due to the nonzero $d^{z}_{\rm buck}$, which contributes to reduce $O_{xp}^{o}$ [see also the item (c) in Appendix~\ref{app:ct}].}
}

\headline{The increase in $\Delta E^{l}_{xp}$ mainly originates from the reduction in $a$.}
\oldjbm{
Indeed, the increase is mainly caused by $P_a$ [see Fig.~\ref{fig:ratiopu}(f)].
This is because the reduction in $a$ increases the energy of Cu$3d_{x^2-y^2}$ electrons with respect to that of O$2p_{\sigma}$ electrons (see Appendix~\ref{app:dftband}).
Although the reduction in $a$ is the main origin of the increase in the $P$ dependence of $\Delta E^{\rm avg}_{xp}$,
note that $\Delta E^{l}_{xp}$ depends not only on $a$ but also on other CPs (see \jbmd{Appendix~\ref{app:unc}}). %
}

\headline{
The concomitant increases in $v^{l}$ and $|t_1^{l}|$ seem counterintuitive, but can be explained as follows.
}
\oldjbm{The counterintuitive point is that} 
the increase in $v^{l}$ suggests a \oldmi{more localized} AB orbital whereas the increase in $|t_{1}^{l}|$ would be more consistent with a delocalization of the AB orbital.
Although the AB orbital is more localized, the increase in $|t_{1}^{l}|$ is explained by the increase in $|t^{l}_{xp}|$ with $P_a$ in Fig.~\ref{fig:ratiop}(\jbmb{g}).
This is discussed in detail in the item (b) in Appendix~\ref{app:ct}, which is summarized below.
We apply $P_a$ and examine the $a$ dependencies of $|t_1^{\rm avg}|$, $|t_{xp}^{\rm avg}|$ and $\Delta E_{xp}^{\rm avg}$,
and the average values $O_{xp}^{\rm avg}$ and $T_{xp}^{\rm avg}$ of $O_{xp}^{l}$ and $T_{xp}^{l}=|t_{xp}^{l}|^2/\Delta E_{xp}^{l}$.
The increase in $\Delta E_{xp}^{\rm avg}$ with $a$ is
faster than the increase in $|t_{xp}^{\rm avg}|$, but slower than the increase in $|t_{xp}^{\rm avg}|^2$.
As a result, when $a$ decreases,
$|t_1^{\rm avg}| \propto T_{xp}^{\rm avg} \propto 1/a^3$ increases.
On the other hand, $O_{xp}^{\rm avg} \propto a$ decreases, hence the increase in $v^{l}$.


\headline{On (ii), the decrease in cRPA screening [the increase in $R^{l}$ in Fig.~\ref{fig:ratiop}(d)]
is due to the broadening [M$W$] of the GGA band dispersion (whose origin is the reduction in $a$ as discussed in Sec.~\ref{sec:band}).}
Indeed, [M$W$] causes the increase in charge transfer energies between occupied bands and empty bands,
which reduces the amplitude of the cRPA polarization (see Appendix~\ref{app:screening} for details).
The increase in $R^{l}$ is monotonous, 
except for the small dip in the $P$ dependence of $R^{o}$ at $P \simeq 24$ GPa in Fig.~\ref{fig:ratiop}(d).
The dip may originate from the change in the sign of $d^{z}_{\rm buck}$ at $P \simeq 24$ GPa (see the next paragraph).
\\

\headline{
Comparison of results obtained from different CP values shows that (i,ii) are essentially correct, independently of the uncertainty on CP values.}
Let us consider the results obtained from the CP values from Zhang \textit{et al.} in Fig.~\ref{fig:ratiop}(h-n)
and compare them with the results obtained from the optimized CP values in Fig.~\ref{fig:ratiop}(a-g).
The increase in $v^{\rm avg}$ is well reproduced [see Fig.~\ref{fig:ratiop}(e,l)].
The increase in $R^{\rm avg}$ with $P$ is qualitatively reproduced [see Fig.~\ref{fig:ratiop}(d,k)];
however, the $P$ dependence of $R^{l}$ is not exactly the same and we discuss the difference below.\\

First, there is a small dip in the $P$ dependence of $R^{o}$ at $P \simeq 24$ GPa in Fig.~\ref{fig:ratiop}(d) (optimized CP values).
This dip is not observed in Fig.~\ref{fig:ratiop}(k) (CP values from Zhang \textit{et al.}).
This may be because the sign of $d^{z}_{\rm buck}$ does not change at $P \simeq 24$ GPa if we consider the CP values from Zhang \textit{et al.}, contrary to the optimized CP values (see the $P$ dependence of $d^{z}_{\rm buck}$ in Fig.~\ref{fig:cryst}).\\

Second, at $P_{\rm opt} = 30$ GPa, the value of $R^{i}$ is similar but the value of $R^{o}$ is larger in Fig.~\ref{fig:ratiop}(k) with respect to Fig.~\ref{fig:ratiop}(d).
This is because the values of both $d^{z}_{\rm Ca}$ and $d^{z}_{\rm Cu}$ are larger in Zhang \textit{et al.} with respect to the optimized CP value 
(the difference is $0.1$ \AA \ as seen in Fig.~\ref{fig:cryst}).
As shown in Appendix~\ref{app:unc},
the larger value of $d^{z}_{\rm Ca}$ increases $R^{o}$.
At the same time, the larger value of $d^{z}_{\rm Ca}$ ($d^{z}_{\rm Cu}$) decreases (increases) $R^{i}$. 
(Both effects cancel each other.)\\

Finally, if we consider the experimental CP values from Armstrong \textit{et al.}, the increases (i,ii) are faster [see Fig.~\ref{fig:ratiop}(d,e)].
This is consistent with the faster decrease in $a$ in Armstrong \textit{et al.} with respect to the optimized CP values
\jbme{and also those} 
from Zhang \textit{et al.} (see Fig.~\ref{fig:cryst}).
%

\subsection{Decrease in $u^{}$ with $P > P_{\rm opt}$}
\label{sec:lehuthighp}

\headline{
At $P > P_{\rm opt}$, the decrease in $u^{\rm avg}$ is faster because $R^{\rm avg}$ decreases.
}
Let us start from the $P$ dependence of $U^{\rm avg}$:
At $P > P_{\rm opt}$, $U^{\rm avg}$ ceases to increase [see Fig.~\ref{fig:ratiop}(c)]
and may even decrease if we consider the CP values from Zhang \textit{et al.} [see Fig.~\ref{fig:ratiop}(i)].
The origin is not the $P$ dependence of $v^{\rm avg}$, which increases monotonically [see Fig.~\ref{fig:ratiop}(e,l)],
but rather that of $R^{\rm avg}$, which shows a dome structure with a maximum at $P_{\rm scr} \simeq 30-40$ GPa and a decrease at $P > P_{\rm scr}$ [see Fig.~\ref{fig:ratiop}(d,k)].
The decrease in $R^{\rm avg}$ dominates the increase in $v^{\rm avg}$.

\headline{The decrease in $R^{\rm avg}$ looks physical, and robust with respect to the uncertainty on CP values.}
It is still observed if we consider the CP values from Zhang \textit{et al.} instead of the optimized CP values [see Fig.~\ref{fig:ratiopu}(k)],
even though the $P$ dependence of $R^{l}$ is modified.

\headline{
The decrease in $R^{\rm avg}$ is the result of a competition between $P_a$ and $P_c$. (The effect of $P_c$ is dominant at $P > P_{\rm opt}$.)
}
\oldjbm{
As seen in Fig.~\ref{fig:ratiopu}(d), 
applying only $P_a$ causes (i) the non-linear increase in $R^{\rm avg}$, 
which dominates at $P < P_{\rm opt}$ but saturates at $P > P_{\rm opt}$.
On the other hand, applying only $P_c$ causes (ii) the decrease in $R^{\rm avg}$, which becomes dominant at $P > P_{\rm opt}$.
\jbme{[(i,ii) are interpreted in terms of the cRPA polarization in Appendix~\ref{app:screening}.]}
The microscopic origin of (ii) is the decrease in both $d^{z}_{\rm Cu}$ and $d^{z}_{\rm O(ap)}$ when $P_c$ is applied (see Appendix~\ref{app:unc}).
}

\headline{
Note that, in the OP, the destructive effect of $P_c$ on $R^{o}$ and thus $u^{o}$ is cancelled by the buckling induced decrease in $|t_1^{o}|$.
}
\oldjbm{
Indeed, the $P_c$ dependence of $u^{o}$ in Fig.~\ref{fig:ratiopu}(b) shows a $6\%$ increase from $P_{\rm opt}$ to $60$ GPa.
This increase originates from the buckling of Cu-O-Cu bonds in the OP, 
because it does not appear in the $P_{c}^{\overline{\rm buck}}$ dependence of $u^{o}$ in Fig.~\ref{fig:ratiopu}(i),
and the value of $d^{z}_{\rm buck}$ is modified by applying $P_c$ but not by applying $P_{c}^{\overline{\rm buck}}$.
The buckling reduces $|t_1^{o}|$ as discussed in Sec.~\ref{sec:leht},
which is the main origin of the increase in $u^{o}$ from $P_{\rm opt}$ to $60$ GPa.
}


\section{Discussion}
\label{sec:disc}

\jbmd{Here, we discuss in detail how the experimental $P$ dependence of $T_{c}^{\rm opt}$
is predicted by considering (I,II) together with the assumptions \jbme{(A,B)} and the corrections \jbme{(C,D)} in Sec.~\ref{sec:introduction}.
\mib{We also discuss that (A) through (D) are all physically sound.}
}
\\

First, \mib{we emphasize that only by considering (A) and (B), the dome structure in the $P$ dependence of $T_{\rm c}^{\rm opt}$
is qualitatively understood.
Since (B) implies that $F_{\rm SC}$ stays at a plateau region around the peak of parabolic $P$ dependence between $P_{\rm amb}$ and $P_{\rm opt}$ as is seen in Fig.~\ref{fig:tcexp}(a). Then the dominant $P$ dependence of $F_{\rm SC}$ arises from $t_1$, which causes increase in $T_{\rm c}^{\rm est}$ in Eq.~(\ref{Tcscaling}). On the other hand, $F_{\rm SC}$ rather rapidly decreases with increasing $P$ above $P_{\rm opt}$, which dominates over the effect of increase in $|t_1|$.} 

\begin{figure}[ht]
\centering
\includegraphics[scale=0.43]{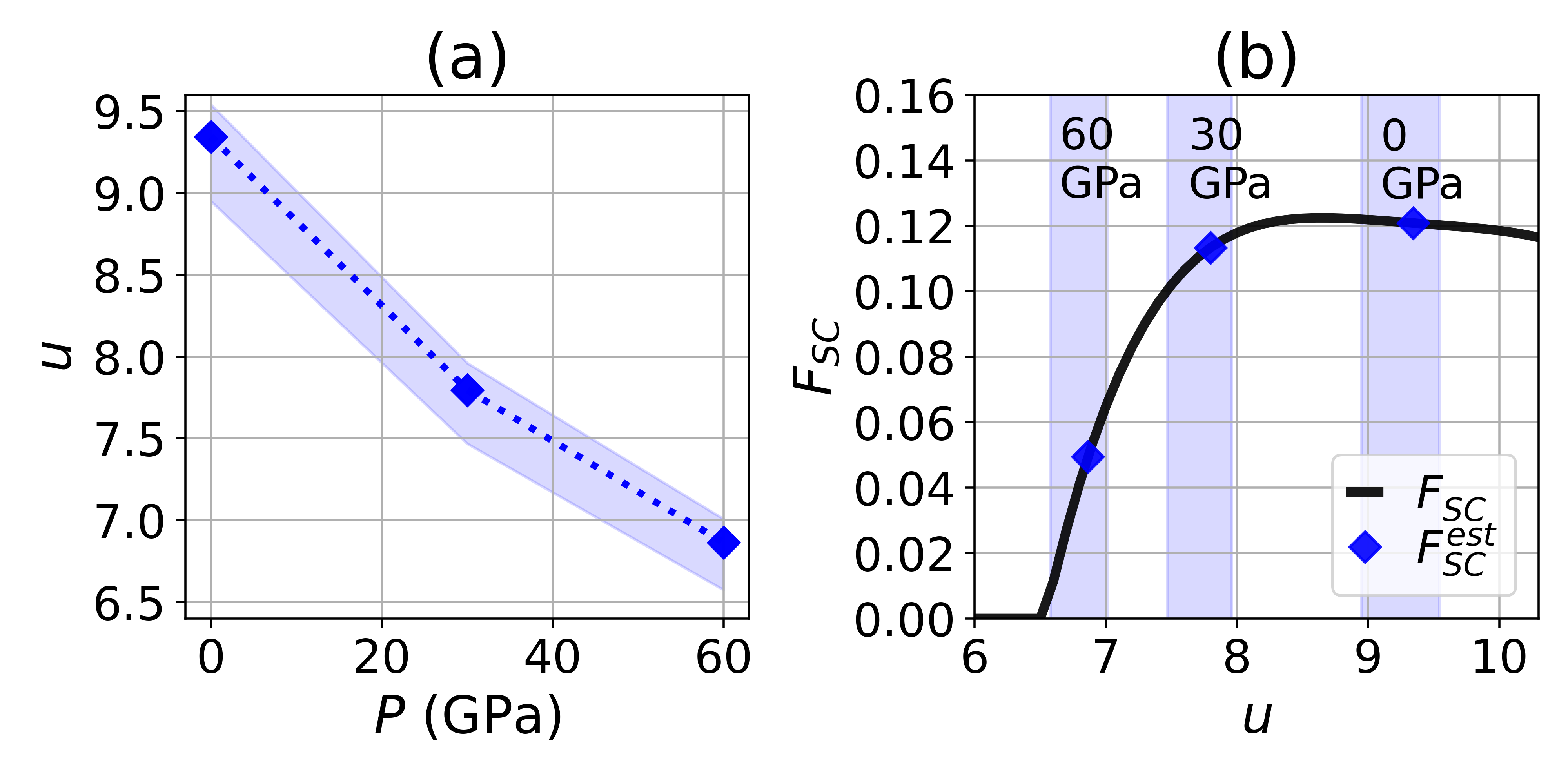}
\caption{
\jbmd{
Panel (a): $P$ dependence of \mib{estimated $u$ at the c$GW$-SIC+LRFB level denoted here as $u_{{\rm c}GW{\rm -SIC+``LRFB"}}$
The diamond symbols show $u_{{\rm c}GW{\rm -SIC+``LRFB"}}$ with the choice of $x_{\rm LRFB}=0.95$ correcting  
explicit calculations at the c$GW$-SIC level by employing Eq.(\ref{xLRFB}).
The dashed lines show linear interpolations between the diamond symbols.
The colored shaded area corresponds to the range $x_{\rm LRFB} = 0.91-0.97$.}
Panel (b): $u$ dependence of $F_{\rm SC}$ extracted from Schmid~\textit{et al.}~\cite{Schmid2023}, Fig.~10.
We also show 
\jbme{$F_{\rm SC}^{\rm est}$
at \jbme{$0$, $30$ and $60$ GPa}. 
}
}
}
\label{fig:tcexp2}
\end{figure}

\mib{The location of Hg1223 assumed in (B) is justified from \jbme{(C)}.}
\jbme{Without (C), we would have $u^{\rm avg} \simeq 7.2$ at $P_{\rm amb}$ and $\simeq 6.8$ at $P_{\rm opt}$:
Both values are below $u_{\rm opt} \simeq 8.0-8.5$, so that $F_{\rm SC}$ would quickly decrease with $P$, and (B) would not be valid.
On the other hand, if we apply (C), we have $u \simeq 9.3 \gtrsim u_{\rm opt}$ at $P_{\rm amb}$ and $\simeq 7.8 \simeq u_{\rm opt}$ at $P_{\rm opt}$ [see Fig.~\ref{fig:tcexp2}(a)], so that (B) becomes valid.
}

\jbme{\mib{Let us discuss more quantitative aspects. 
Although $F_{\rm SC }$ does not vary substantially with increasing $P$ below $P_{\rm opt}$,}
there is a small ($\simeq 5\%$) decrease in $F_{\rm SC}$
from $P_{\rm amb}$ to $P_{\rm opt}$ even after applying (C) [see Fig.~\ref{fig:tcexp}(a)].
If we apply (C) without (D),
the $\simeq 13-14\%$ increase in $|t_1^{\rm avg}|$ from $P_{\rm amb}$ to $P_{\rm opt}$ becomes the $\simeq 17\%$ increase in $|t_1|$.
However, the increase in 
\mib{$T_{\rm c}^{\rm est}$ estimated from Eq.(\ref{Tcscaling})
is only $\simeq 10\%$ due to the $\simeq 5\%$ decrease in $F_{\rm SC}$. 
If we apply (D) after (C), the increase in $|t_1|$ 
becomes $\simeq 22\%$,
so that the increase in $T_{\rm c}^{\rm est}$
becomes $\simeq 17\%$ and reproduces that in $T_{\rm c}^{\rm opt}$.
Note that the quantitative agreement between the increases in $T_{\rm c}^{\rm est}$
and 
$T_{\rm c}^{\rm opt}$ is very good at $x_{\rm LRFB}^{\rm est}=0.95$ at least for small $P$.
[see Fig.~\ref{fig:tcexp}(b)].
For completeness, note that (D) has a limitation: It relies on the $a$ dependence of $|t_1|$ at the GGA+cRPA level.
[For more details, see the last paragraph of Appendix~\ref{app:corr-d}.]
}}

\jbmd{Now, we \mib{argue that (A,B,C,D) are adequate from the physical point of view.}}
On (A), 
\mib{
it was shown that the scaling Eq.~(\ref{Tcscaling}) is equally satisfied for $N_{\ell}=1,2$ and $\infty$~\cite{Schmid2023}. This is on the one hand due to the fact that the interlayer coupling is small for all the cases and within a CuO$_2$ layer on the other hand, the superconductivity is mainly dependent on $t_1$ and $U$ only and the dependence on other parameters is weak within the realistic range. 
In the present case of Hg1223 with $N_{\ell} =3$, the interlayer coupling is again small. For instance, the ratio between the interlayer offsite Coulomb repulsion $V^{i,o}$ and $U^{\rm avg}$ is $V^{i,o}/U^{\rm avg} = 0.13$ at $P_{\rm amb}$ and the superconducting strength is expected to be governed by the single layer physics, which is the same as the  cases of $N_{\ell}=1,2$ and $\infty$.}

On (B), \mib{the statement that $u_{{\rm c}GW{\rm -SIC+``LRFB"}}$ at $P_{\rm amb}$ is above $u_{\rm opt} \jbmfmain{\sim}{ \ \simeq \ } 8.0$-8.5 is indeed satisfied in the \textit{ab initio} estimate by considering} \jbme{the correction (C).}
\jbme{As mentioned earlier, the GGA+cRPA estimate is $u^{\rm avg} \simeq 7.2 < u_{\rm opt}$ at $P_{\rm amb}$.
However, (C) yields $u_{{\rm c}GW{\rm -SIC+``LRFB"}} \simeq 9.3 \gtrsim u_{\rm opt}$ at $P_{\rm amb}$, 
and $u_{{\rm c}GW{\rm -SIC+``LRFB"}} \simeq 7.8 \simeq u_{\rm opt}$ at $P_{\rm opt}$.
[See Fig.~\ref{fig:tcexp2}(a).]}

\jbmd{On \jbme{(C)}, the 
calculation of $u_{{\rm c}GW{\rm -SIC+``LRFB"}}$ 
and $|t_1|_{{\rm c}GW{\rm -SIC+``LRFB"}}$ 
is detailed in Appendix~\ref{app:corr-c}.
}

\jbm{On \jbme{(D)}, 
it is plausible that our calculation overestimates $a$ by $\simeq 0.05$ \AA \ at \jbmc{$P > 8.5$ GPa,} 
because the same overestimation is already observed at $P=8.5$ GPa in Fig.~\ref{fig:cryst}.
\jbme{The \mib{derivation of improved $|t_1|_{{\rm c}GW{\rm -SIC+``LRFB"}}$} 
is detailed in Appendix~\ref{app:corr-d}.}
}

\mib{In Fig.~\ref{fig:tcexp}(b), although the pressure dependence of $T_{c}^{\rm opt}$ is nicely reproduced for $P<P_{\rm opt}$, the estimated $T_c^{\rm est}$ decreases more rapidly than  the experimental $T_c^{\rm opt}$ at $P>P_{\rm opt}$.
The origin of this discrepancy is not clear at the moment.
One possible origin is of course the uncertainty of the crystal parameters  
at high pressure because there exist no experimental data. Another origin would be the limitation of the inference for the LRFB correction taken simply by the constants $x_{\rm LRFB}$ and $y_{\rm LRFB}$. The third possibility is the possible inhomogeneity of the pressure in the experiments. The complete understanding of the origin of the discrepancy is an intriguing  future issue.}

\section{Summary and conclusion}
\label{sec:summary}

We have proposed the microscopic mechanism for the \mib{dome-like} $P$ dependence of $T_{c}^{\rm opt}$ in Hg1223
as the consequence of (I) and (II) obtained in this paper \jbmd{together with the assumptions (A,B) and the corrections (C,D)} mentioned in \jbmd{Sec.~\ref{Overview}} \jbm{and supported in Sec.~\ref{sec:disc}}.
We have also elucidated the microscopic origins of (I,II), which are summarized below.

\noindent
(I) The increase in $|t_1|$ is caused by the reduction in the cell parameter $a$ when the \jbme{crystal} is compressed along axis ${\bf a}$. 

\noindent
(II) The decrease in \jbme{$u$} 
is induced by (I), 
\jbmd{but is partially cancelled by the increase in $U$ at $P < P_{\rm opt}$.}
The increase in $U$ is caused by two cooperative factors:
\jbme{(i) The} increase in onsite bare interaction $v$, 
whose main origin is the reduction in Cu$3d_{x^2-y^2}$/O$2p_\sigma$ hybridization\jbme{, and}
(ii) the \jbmd{reduction in cRPA screening at $P < P_{\rm opt}$.}
Both (i) and (ii) originate from the reduction in $a$. 
At $P > P_{\rm opt}$, $U$ ceases to increase with increasing $P$, 
because the cRPA screening increases \jbme{due to} the compression along axis ${\bf c}$,
more precisely 
\jbme{
the reduction in distance $d^{z}_{\rm Cu}$ between the IP and OP [$d^{z}_{\rm O(ap)}$ between the OP and apical O],
which screens AB electrons in the IP (OP).
}

The elucidation of the above mechanisms offers a platform for future studies on cuprates under \jbme{$P$}
\jbme{and design of new compounds with even higher $T_{c}^{\rm opt}$:}
\jbmd{For instance, $T_{c}$ may be controlled by controlling $|t_1|$ via the cell parameter $a$. However, the increase in $|t_1|$ is a double-edged sword for the increase in $T_{c}$:}
\jbmd{O}n one hand, 
\jbmd{it} is the direct origin of the increase in 
\jbme{$T_{c}^{\rm opt} \propto |t_1|$}
at $P < P_{\rm opt}$ \jbmd{in Hg1223}.
On the other hand, it is a prominent cause of the decrease in \jbme{$u$ and thus $F_{\rm SC}$} 
\jbme{and $T_{c}^{\rm opt}$} at $P > P_{\rm opt}$.
Conversely, in the OP, the buckling of Cu-O-Cu bonds reduces $|t_1|$:
This reduces \jbme{$T_{c}^{\rm opt} \propto |t_1|$}, but this may also increase \jbme{$F_{\rm SC}$} and thus \jbme{$T_{c}^{\rm opt}$} if the value of $u$ is in the weak-coupling region [$u < \jbme{7.5}$ in Fig.~\jbmd{\ref{fig:tcexp2}(b)}].
For instance, the buckling may be identified as the main origin of the higher $T_{c}^{\rm opt}$ in Bi2212 ($T_{c}^{\rm opt} \simeq 84$ K~\cite{Torrance1988}) compared to Bi2201 ($T_{c}^{\rm opt} \simeq 6$ K~\cite{Torrance1988})\jbme{:}
The buckling reduces $|t_1|$ and thus increases $u$ in Bi2212 with respect to Bi2201~\cite{Moree2022},
so that Bi2212 is near the optimal region whereas Bi2201 is in the weak-coupling region~\cite{Schmid2023}.
This explains 
\jbme{the larger $|t_1|F_{\rm SC}$} in Bi2212~\cite{Schmid2023} despite the smaller $|t_1|$.



\section*{Acknowledgements}

We thank Michael Thobias Schmid for useful discussions.
This work was supported by MEXT as Program for Promoting Researches on the Supercomputer Fugaku (Basic Science for Emergence and Functionality in Quantum Matter ­Innovative Strongly-Correlated Electron Science by Integration of Fugaku and Frontier Experiments­, JPMXP1020200104 and JPMXP1020230411) and used computational resources of supercomputer Fugaku provided by the RIKEN Center for Computational Science (Project ID: hp200132, hp210163, hp220166 \oldjbm{and hp230169}).
We also acknowledge the financial support of JSPS Kakenhi Grant-in-Aid for Transformative Research
Areas, Grants Nos. JP22H05111 and  JP22H05114 (``Foundation of Machine Learning Physics"). 
Part of the results were obtained under the Special Postdoctoral Researcher Program at RIKEN.
The left panel of Fig.~ \ref{fig:cryst} was drawn by using software \texttt{VESTA} \cite{Momma2011}.

\renewcommand{\thesection}{A\arabic{section}}
\renewcommand{\theequation}{A\arabic{equation}}
\appendix

\begin{appendices}
\section{Method of MACE}
\label{app:method}

\headline{Here, as a complement to \jbmd{Sec.~\ref{sec:introduction}}, we summarize and comment the method of deriving the effective Hamiltonian, which consists of three steps.}

\noindent
\headline{
(i) First, starting from the crystal structure, 
the electronic structure of the material is calculated at the simplified Density Functional Theory (DFT) \cite{Hohenberg1964,Kohn1965} level.}
This framework uses the \jbmd{LDA or GGA} exchange-correlation functionals,
and a single-determinant wavefunction.
\jbmd{The electronic structure is either left at the LDA(GGA) level [in case the LDA(GGA)+cRPA is employed],
or preprocessed to the $GW$ level (if c$GW$-SIC is employed) supplemented with LRFB (if c$GW$-SIC+LRFB is employed), as explained in \jbmd{Sec.~\ref{sec:introduction}}.}

\noindent
\headline{
(ii) The description of the L space is improved by deriving a low-energy effective Hamiltonian (LEH) restricted to the L space.
}
In this LEH, the two-particle part is calculated at the constrained random phase approximation (cRPA) \cite{Aryasetiawan2004,Aryasetiawan2006} \jbmd{at} the \jbmc{GGA+cRPA} level.
\jbmd{At the c$GW$-SIC and c$GW$-SIC+LRFB levels}, the one-particle part of the LEH is also improved 
by removing the exchange-correlation double counting term \cite{Hirayama2013} and the self-interaction term~\cite{Hirayama2015} (see also \jbmd{Sec.~\ref{sec:introduction}}).
This properly describes high-energy (H) states such as core and semicore bands from closed shells,
but fails to describe many-body effects and strong electronic correlation in the low-energy (L) subspace near the Fermi level, even with the above preprocessing. 
In the case of cuprates, this L space is composed of the \jbme{AB} 
orbital centered on each Cu atom in the CuO$_2$ plane.
\jbmd{The correlation strength is quantified within the ratio $u$ whose value is typically above $7$ for the high-$T_{c}$ cuprates~\cite{Hirayama2018,Hirayama2019,Moree2022}.}

\noindent
\headline{
(iii) The LEH is solved by a many-body solver, e.g. many-variable Variational Monte-Carlo (mVMC) \cite{Misawa2019}.
\\
}

\headline{This three step MACE procedure allows to correctly describe the Mott physics in the mother compound and the SC phase in the carrier doped compound \cite{Ohgoe2020,Schmid2023}.}
In the mVMC solution, $F_{\rm SC}$ rapidly  increases with $u$ in the range $\jbme{7 \lesssim u \lesssim 8.5}$~\cite{Schmid2023}, 
which suggests an increase in $T_{c}$ with $u$ \cite{Schmid2023}, 
in agreement with the positive correlation between $u$ and $T_{c}^{\rm opt}$~\cite{Moree2022} in the same range of values of $u$.
This range corresponds to the weak-coupling and plateau regions [$\jbme{7 \lesssim u \lesssim 9}$ in the $u$ dependence of $F_{\rm SC}$ in Fig.~\jbmd{\ref{fig:tcexp2}(b)}].
These results led to the identification of the possibly universal scaling $T_c \simeq 0.16 |t_1| F_{\rm SC}$ in the solution of the AB LEH at the \jbmc{c$GW$-SIC+LRFB} level~\cite{Schmid2023}.
\\

\headline{
To predict the SC character of the material with the above MACE procedure,
insights may be obtained even prior to the computationally expensive solution (iii),
by examining intermediate quantities within the hierarchical structure of MACE.
}
Notably, the scaling $T_{c}^{\rm opt} \simeq 0.16 |t_1| F_{\rm SC}$ proposed in Ref.~\onlinecite{Schmid2023}
and the $u$ dependence of $F_{\rm SC}$ in Fig.~\jbmd{\ref{fig:tcexp2}(b)}
suggest it is possible to anticipate the crystal structure dependence of $T_{c}$ 
by studying the crystal structure dependence of LEH parameters (ii), particularly $|t_1|$ and $u$.
Following this idea, we tackle in this paper the derivation of the AB LEH (ii) for Hg1223 as a function of pressure, without performing explicitly the solution (iii) which is left for future studies.
Of course, the explicit many-body solution of the LEH (iii) is necessary to reach the final conclusion.

\headline{Furthermore, qualitative insights into the SC may be obtained by deriving the LEH parameters at the simple \jbmc{GGA+cRPA} level up to the process (ii), whereas \jbmc{c$GW$-SIC+LRFB} brings a mostly quantitative correction to the LEH parameters \cite{Moree2022}.
}
\jbme{N}ote that this quantitative correction by \jbmc{c$GW$-SIC+LRFB} is still important to stabilize the SC state with mVMC (iii) in practice:
The improvement \oldmi{by \jbmc{c$GW$-SIC+LRFB}} increases $U$ and thus $u$ by $10-15 \%$ in Bi2201 and Bi2212 \cite{Moree2022},
which allows \oldmi{quantitative estimate of the SC order in the mVMC solution.}
\oldmi{On the other hand}, at the simple \jbmc{GGA+cRPA} level, $u$ may be underestimated.
Nonetheless, \jbmc{GGA+cRPA} still reproduces the \oldmi{dependence of $u$ in the LEH parameters on the materials, 
and the CPs including pressure effects systematically in accordance with \jbmc{c$GW$-SIC+LRFB}~\cite{Moree2022}},
which allows to extract qualitatively correct trends in the LEH parameters by avoiding the large computational cost \cite{Moree2022} of \jbmc{c$GW$-SIC+LRFB}.
For instance, in the comparison 
between Bi2201 ($T_{c}^{\rm opt}\simeq 6$ K \cite{Torrance1988}) and Bi2212 ($T_{c}^{\rm opt}\simeq 84$ K \cite{Torrance1988}), 
$u$ is larger for Bi2212 at the \jbmc{c$GW$-SIC+LRFB} level, 
and this qualitative result is also reproduced at the \jbmc{GGA+cRPA} level in Ref.~\onlinecite{Moree2022}, Appendix C.

\headline{
Following the above idea, we \jbmd{mainly employ the GGA+cRPA scheme to derive the AB LEH (ii) for Hg1223.} 
}
\jbmd{We also employ the \jbme{c$GW$-SIC+LRFB scheme} in a limited case in Appendix~\ref{app:corr-c}, as explained in Sec.~\ref{sec:introduction}.}
\\

\section{Computational Details}
\label{app:comp_details}

\renewcommand{\theequation}{B\arabic{equation}}

\subsection{Choice of crystal parameter \oldjbm{values}}
\label{app:choicecryst}


\headline{
The CP values obtained from neutron diffraction powder and energy-dispersive synchrotron x-ray diffraction experiment \cite{Hunter1994,Eggert1994,Armstrong1995} are summarized in Fig.~\ref{fig:cryst}.
}

\headline{
There is an uncertainty on the CP values, especially after $P \simeq 9.2$ GPa.
}
\oldjbm{Indeed, the experimental $P$ dependence of the CP values varies between different works.
In addition, to our knowledge, the CP values at $P > 9.2$ GPa have not been completely determined in experiment.
Refs.~\onlinecite{Hunter1994,Armstrong1995} provide all \oldjbm{CP values}, but only up to $P \simeq 8.5-9.2$ GPa.
Ref.~\onlinecite{Eggert1994} provides the values of $a$ and $c$ up to $P \simeq 26$ GPa, but not the values of $d^{z}_{l}$.
Thus, the CP values are not available within the range $P_{\rm amb} < P < 45$ GPa that corresponds to 
the dome-like $P$ dependence of $T_{c}^{\rm opt}$ in Ref.~\onlinecite{Gao1994}.
}
\\

\headline{
To verify the robustness of our results with respect to the uncertainty on CP values,
we consider CP values obtained by two different theoretical calculations (i,ii), up to $60$ GPa.
}
\oldjbm{
We consider
(i) CP values obtained by a structural optimization (denoted as optimized CP values),
and (ii) CP values obtained in Zhang \textit{et al.} \cite{Zhang1997}.
We determine first (ii), then (i), as explained below.
}
\\

\headline{
On (ii), the values in Zhang \textit{et al.} have been obtained from a theoretical calculation, by the means of interatomic potentials. 
}
At $P < 9.2$ GPa, these values are in reasonable agreement with the different experimental values from Refs.~\onlinecite{Hunter1994,Armstrong1995}. Although the values of $a$ are overestimated with respect to Refs.~\onlinecite{Eggert1994,Hunter1994,Armstrong1995}, 
they are in good agreement with Ref.~\onlinecite{Eggert1994} at $24$ GPa.

\headline{However, the CP values from Zhang \textit{et al.} are available only up to $20$ GPa;
thus, we extrapolate their $P$ dependence up to $60$ GPa, as follows.
}
We fit \oldjbm{the $P$ dependence of $a$} 
by considering the Murnaghan equation of state
\begin{equation}
\frac{a(P)}{a(P_{\rm amb})} = \Bigg[ 1 + \frac{\kappa'}{\kappa} P \Bigg]^{-1/\kappa'},
\label{eq:murnaghan}
\end{equation}
as done in Ref.~\onlinecite{Eggert1994}\oldjbm{. We} deduce the values of the two parameters $\kappa$ and $\kappa'$, which are respectively the bulk modulus and its pressure derivative.
The same procedure is applied to $c$, $d^{z}_{\rm O(ap)}$, $d^{z}_{\rm Cu}$, $d^{z}_{{\rm Ba,O}(o)}=d^{z}_{\rm Ba}+d^{z}_{\rm buck}$, and $d^{z}_{\rm Ca,Ba}=d^{z}_{\rm Cu}-d^{z}_{\rm Ca}+d^{z}_{\rm Ba}$, whose values are extracted from Ref.~\onlinecite{Zhang1997}.
In the case of $d^{z}_{\rm buck}$, 
we fit the Cu($o$)-O($o$)-Cu($o$) bond angle as a function of $P$ in Ref.~\onlinecite{Zhang1997}, Fig.~5 with Eq.~(\ref{eq:murnaghan}).
Then, we deduce $d^{z}_{\rm Ba}$ from $d^{z}_{{\rm Ba,O}(o)}$ and $d^{z}_{\rm buck}$, and $d^{z}_{\rm Ca}$ from $d^{z}_{\rm Ca,Ba}$, $d^{z}_{\rm Cu}$ and $d^{z}_{\rm Ba}$.
We checked that values of $\kappa$ for these CPs from Ref.~\onlinecite{Zhang1997} are reproduced with a difference lower than $0.5\%$.
These values of $\kappa$ are $1.81\times10^{-3}$ GPa$^{-1}$ for $a$, $4.61\times 10^{-3}$ GPa$^{-1}$ for $c$, $7.01\times 10^{-3}$ GPa$^{-1}$ for $d^{z}_{\rm O(ap)}$, $2.94\times 10^{-3}$ GPa$^{-1}$ for $d^{z}_{\rm Cu}$, $0.64\times 10^{-3}$ GPa$^{-1}$ for $d^{z}_{{\rm Ba- O}(o)}$, and $1.535\times 10^{-3}$ GPa$^{-1}$ for $d^{z}_{\rm Ca- Ba}$.
\oldjbm{We obtain the CP values in Fig.~\ref{fig:cryst}.}
\\

\headline{
On (i), the optimized CP values are obtained by starting from (ii), 
and performing a structural optimization. 
}
\oldjbm{We impose the following constraint: The volume $V = a^2 c$ of the unit cell remains constant. 
This \jbmc{allows} to avoid the relaxation of the volume to its value at $P_{\rm amb}$. 
Other computational details are the same as those for the self-consistent calculation (see Appendix~\ref{app:dft}). 
Results are shown in Fig~\ref{fig:cryst}.
}
\\

\headline{
We deem (i) more reliable than (ii) because the structural optimization allows the rigorous minimization of the free energy of the crystal;
thus, we consider (i) in the main analyses of this paper, and (ii) as a complement.}
\oldjbm{Still, (ii) is useful to check the robustness of results obtained from (i):
We show that both (i) and (ii) yield the same qualitative $P$ dependence of AB LEH parameters (see Sec.~\ref{sec:leh}).
Of course, it would be desirable to determine accurately all CP values from $P_{\rm amb}$ to 60 GPa in future experimental works.
}

\headline{Note that, at $P = 60$ GPa, the negative value $d^{z}_{\rm buck} \simeq -0.2$ \AA \ obtained for both (i) and (ii) is physical, as discussed below.}
First, the $P$ dependence of $d^z_{\rm buck}$ at $P > P_{\rm opt}$ looks robust, because it is similar for (i) and (ii)  (see Fig.~\ref{fig:cryst}).
Second, the negative value of $d^{z}_{\rm buck}$ has a physical origin: The "collision" between the in-plane O in the OP and the Ca cation.
Indeed, when $P$ increases, 
the distance $d^{z}_{\rm Cu}-d^{z}_{\rm Ca}$ between the OP and Ca cation is reduced (see Fig.~\ref{fig:cryst}).
If we see the ions as rigid spheres, the Ca cation "collides" with the in-plane O in the OP,
so that the in-plane O is pushed outside of the OP.
This explains why $d^{z}_{\rm buck}$ becomes negative and $|d^{z}_{\rm buck}|$ increases.
In addition, the rigidity of Cu-O-Cu bonds may play a role in the increase in $|d^{z}_{\rm buck}|$:
When $a$ is decreased, $|d^{z}_{\rm buck}|$ is also increased to prevent the reduction in distance $d_{\rm Cu-O} = \sqrt{(a/2)^2 + (d^{z}_{\rm buck})^2}$ between Cu and in-plane O.

\subsection{Hole concentration}
\label{app:hole}

\headline{
Next, we take into account the experimental optimal value $p_{\rm opt}$ of the hole concentration $p$, 
which realizes $T_{c}^{\rm exp}$ \jbme{(the experimental value of $T_{c}^{}$)} close to $T_{c}^{\rm opt} \simeq 138$ K at $P_{\rm amb}$.
}
Experimentally, hole doping in the CuO$_2$ planes is realized by 
introduction of excess oxygen atoms
and/or partial substitution of atoms, e.g. Hg by Au,
so that the chemical formula of Hg1223 becomes Hg$_{1-x_{\rm s}}$Au$_{x_{\rm s}}$Ba$_2$Ca$_2$Cu$_3$O$_{8+\delta}$.
In that case, a rough estimate of the total hole concentration is 
$p_{\rm tot} = 2\delta + x_{\rm s}$,
which corresponds to the average hole concentration per CuO$_2$ plane
$p_{\rm av} = p_{\rm tot}/3 = (2\delta + x_{\rm s})/3$.

\headline{
At $P_{\rm amb}$, previous studies \cite{Gao1994,Bordet1996,Kotegawa2001,Yamamoto2015} suggest the optimal value of
$p_{\rm av}$ is $p_{\rm opt} \simeq \jbm{0.14-}0.20$. 
}
In Ref.~\onlinecite{Bordet1996}, the $x_s$ dependence of $T_{c}^{\rm exp}$ is explicitly studied: 
For $\delta=0.3$, we have $T_{c}^{\rm exp} \simeq 133$ K at $x_s=0$,
then $T_{c}^{\rm exp}$ decreases with $x_s$, 
so that the maximum value of $T_{c}^{\rm exp} \simeq 133$ K is reached at $p_{\rm av}= 2\delta/3 \simeq 0.2$. 
This value of $T_{c}^{\rm exp}$ corresponds to $T_{c}^{\rm opt} \simeq 138$ K~\cite{Gao1994}. 
Also, the value $p_{\rm opt} \simeq 0.2$ is consistent 
with Ref.~\onlinecite{Kotegawa2001} in which $T_{c}^{\rm exp} \simeq 115-133$ K at $p_{\rm av} \simeq 0.19 - 0.24$, 
and also with $p_{\rm opt} \simeq 0.19$ in Ref.~\onlinecite{Yamamoto2015}.
\jbm{
However, Ref.~\onlinecite{Gao1994} reports $p_{\rm opt} \simeq 0.14$ which corresponds to $T_{c}^{\rm opt}=138$ K. 
Thus, the maximal value of $T_{c}^{\rm exp} \simeq 133-138$ K is realized for experimental $p_{\rm opt} \simeq 0.14-0.20$ \cite{Gao1994,Bordet1996}.
We checked that the LEH parameters are insensitive to the variation in $p_{\rm av}$ in the range $\simeq 0.14-0.20$, as discussed below.
}

\headline{
Thus, in our calculations, we realize $p_{\rm av}=0.2$ by realizing $p_{\rm tot}=0.6$.
}
We do not consider excess oxygen, so that $\delta=0.0$; instead, we consider $x_{\rm s}=0.6$ to compensate the absence of excess oxygen,
and realize $p_{\rm tot}=0.6$.

\headline{
Also, we checked that our calculations correspond to optimal hole doping \jbm{$p_{\rm opt} \simeq 0.14-0.20$} 
not only at $P_{\rm amb}$ but also under pressure,
which allows a reliable comparison with the $P$ dependence of $T_{c}^{\rm opt}$~\cite{Gao1994}.
}
\jbm{
According to Ref.~\onlinecite{Yamamoto2015}, $p_{\rm opt}$ is reduced under pressure:
We have $p_{\rm opt} \simeq 0.19$ \jbm{($T_c^{\rm opt} \simeq 134$ K)} at $P_{\rm amb}$ 
but $p_{\rm opt} \simeq 0.163$ \jbm{($T_c^{\rm opt} \simeq 150$ K)} at $P=12$ GPa.
Linear extrapolation of the above pressure dependence of $p_{\rm opt}$ yields $p_{\rm opt} \simeq 0.12$ at $P_{\rm opt}=30$ GPa.
}
However, we have checked that this reduction in $p_{\rm opt}$ does not affect \jbm{substantially} the AB LEH parameters.
We consider $x_{\rm s}=0.4$ to realize $p_{\rm av} = 0.133$, and compare with results obtained at $p_{\rm av} = 0.2$.
The values of $|t_1^{l}|$ and $u^{l}$ at $P_{\rm opt}$ change by only $1\% - 2\%$ (see Table~\ref{tab:dop}). 
For completeness, we have also considered $p_{\rm av} = 0.133$ at $P_{\rm amb}$: 
In that case, the values of $|t_1^{l}|$ change by only $1\%$ and the values of $u^{l}$ increase by only $3\% - 6\%$ 
with respect to $p_{\rm av}=0.2$.
Thus, the $p_{\rm av}$ dependence of AB LEH parameters is weak,
and considering the same value of $p_{\rm av}=0.2$ at all pressures is acceptable. 

\begin{table}[!htb]
\caption{Values of $|t_1^{l}|$, $U^{l}$ and $u^{l}$ as a function of 
the average hole concentration per CuO$_2$ plane $p_{\rm av}$, 
at $P_{\rm amb}$ and $P_{\rm opt}$. 
We use the optimized CP values.
}
\label{tab:dop}
\begin{ruledtabular}
\begin{tabular}{ccccccccccc}
$P$ & $p_{\rm av}$ & $|t_1^{i}|$ & $|t_1^{o}|$ & $|t_1^{\rm avg}|$ &$U^{i}$ & $U^{o}$ & $U^{\rm avg}$ &$u^{i}$ & $u^{o}$ & $u^{\rm avg}$\\
$P_{\rm amb}$ & 0.133 & 0.526 & 0.519 & 0.522 & 3.97 & 3.98 & 3.98 & 7.55 & 7.68 & 7.61\\
$P_{\rm amb}$ & 0.2 & 0.528 & 0.523 & 0.525 & 3.85 & 3.79 & 3.82 & 7.31 & 7.26 & 7.28\\
$P_{\rm opt}$ & 0.133 & 0.596&  0.591&  0.594 & 4.10 & 3.87 & 3.99 & 6.88 & 6.54 & 6.71\\
$P_{\rm opt}$ & 0.2 & 0.598 & 0.594&  0.596&  4.13&  3.97 & 4.05 & 6.91 & 6.67 & 6.79\\
\end{tabular}
\end{ruledtabular}
\end{table}

\subsection{\oldjbm{DFT calculation}}
\label{app:dft}

\headline{
We perform the conventional DFT calculation as follows.
}
We use \texttt{Quantum ESPRESSO} \cite{QE-2009,QE-2017}, 
and optimized norm-conserving Vanderbilt pseudopotentials (PPs) \cite{Schlipf2015} \oldmi{by employing} the GGA-PBE functional \cite{Perdew1996} \oldmi{together with} the pseudopotentials \texttt{X\_ONCV\_PBE-1.0.upf} (\texttt{X} = Hg, Au, Ba, Ca, Cu and O) from the \texttt{http://www.quantum-espresso.org}. 
The substitution of Hg by Au is done by using the Virtual Crystal Approximation (VCA) \cite{nordheim1931electron}.
The Hg$_{1-x_{\rm s}}$Au$_{x_{\rm s}}$ fictitious atom 
is abbreviated as \yy{"}{``}Hg" \oldmi{from now on}. 
\oldjbm{We consider nonmagnetic calculations, a plane wave cutoff energy of $100$ Ry for wavefunctions, a Fermi-Dirac smearing of $0.0272$ eV, a $12 \times 12 \times 12$ $k$-point grid for the Brillouin zone sampling in the self-consistent calculation, and a $8 \times 8 \times 3$ k-point grid and 430 bands for the following non self-consistent calculation.}

\headline{
We obtain the GGA band dispersion in Fig.~\ref{fig:bandp}.}
In this band dispersion, the medium-energy (M) space \oldmi{near the Fermi level} is spanned by the 44 Cu$3d$, O$2p$ and Hg$5d$-like bands from $-10$ eV to $+3$ eV \oldmi{by defining the origin at} the Fermi level.

\headline{
First, we separate the M space from other bands as follows.
}
We compute the 44 atomic-like Wannier orbitals (ALWOs) spanning the M space (denoted as M-ALWOs), as \oldjbm{maximally localized Wannier orbitals} \cite{Marzari1997,Souza2001},
by using the RESPACK code \cite{Nakamura2020,Moree2022}.
The initial guesses are $d$, $p$ and $d$ \oldjbm{atomic orbitals} centered respectively at Cu($l$), O($l$) (with $l=i,o$ \oldmi{representing the inner and outer planes, respectively}) and at Hg atoms.
\oldjbm{\oldmi{44 ALWOs are constructed from the GGA band number from \#41 to \#87, which are numbered from the energy bottom of the GGA cutoff.}}
We preserve the band dispersion in the GGA by using the inner energy window from the bottom of the lowest band in the M space [the band in black color between $-7$ eV and $-10$ eV in Fig.~\ref{fig:bandp}(a-l)] to the bottom of the lowest empty band outside the M space 
[the dashed band in black color between the Fermi level and $+2$ eV in Fig.~\ref{fig:bandp}(a-l)].
Then, the three bands above the 44 M bands are disentangled \cite{Miyake2009} from the latter. 

\headline{
We obtain the M-ALWOs.
}
They are denoted as $(lj{\bf R})$, where ${\bf R}$ is the coordinate of the unit cell in the space [$xyz$] expanded in the (${\bf a},{\bf b},{\bf c}$) frame in Fig.~\ref{fig:cryst}, $j$ is the orbital index and $l$ is the index (defined in Table~\ref{tab:cryst}) giving the atom located in the cell at ${\bf R}$, on which $(lj{\bf R})$  is centered.
We then express the GGA one-particle part $h(r)$ in the M-ALWO basis, as 
\begin{equation}
h^{l,l'}_{j,j'}({\bf R}) = \int_{\jbmg{\Omega}} dr w_{lj{\bf 0}}^{*}(r) h(r) w_{l'j'{\bf R}}(r),
\label{eq:hopm}
\end{equation}
in which $w_{lj{\bf R}}$ is the one-particle wavefunction of $(lj{\bf R})$.
From Eq.~(\ref{eq:hopm}), we deduce the onsite energy $\epsilon^{l}_{l} = h^{l,l}_{j,j}({\bf 0})$ of the M-ALWO $(lj)$ at any ${\bf R}$,
and the hopping $t^{l,l'}_{j,j'}({\bf R}) = h^{l,l'}_{j,j'}({\bf R})$ between the M-ALWO $(lj{\bf 0})$ and the M-ALWO $(l'j'{\bf R})$.
In this paper, we discuss in particular
the Cu$3d_{x^2-y^2}$ and in-plane O$2p_{\sigma}$ onsite energies 
and the Cu$3d_{x^2-y^2}$/O$2p_{\sigma}$ hopping in the unit cell $t_{xp}^{l} = t_{x^2-y^2,p_\sigma}^{{\rm Cu}(l),{\rm O}(l)}$.
These quantities are given in Fig.~\ref{fig:bandp}(o-q).

\subsection{Low-energy subspace}
\label{sec:methleh}

\headline{
Then, we focus on the L space, which is spanned by the Cu$3d_{x^2-y^2}$/O$2p_{\sigma}$ AB band shown in red color in Fig.~\ref{fig:bandp}(a-g).
}
To construct the AB maximally localized Wannier orbitals, 
the initial guesses are the $d_{x^2-y^2}$ atomic orbitals centered on each of the three Cu($l$) atoms in the unit cell.
The band window is \oldmi{essentially the M space but we exclude the $N_{\rm excl}=14$ lowest bands from it}
to avoid catching the B/NB Cu$3d_{x^2-y^2}$/O$2p_{\sigma}$ character. 
Then, in the band window, we disentangle the 29 other bands from the AB band.

\subsection{Constrained polarization and effective interaction}
\label{app:crpa}

\headline{Then, we compute the cRPA polarization at zero frequency.
}
It is expressed as \cite{Nakamura2020}:
\begin{widetext}
\begin{equation}
\oldjbm{[\chi_{\rm H}]}_{GG'}^{}(\jbme{q})= -\frac{4}{N_k} \sum_{k} \sum_{n_u}^{\rm empty} \sum_{n_o}^{\rm occupied} (1-T_{n_o k} T_{n_u k+\jbme{q}}) 
\frac{M^{G}_{n_o, n_u}(k+\jbme{q},k)  [M^{G'}_{n_o, n_u}(k+\jbme{q},k) ]^{*}}{\Delta_{n_o, n_u}(k,\jbme{q})-i\eta},
\label{eq:chiou}
\end{equation}
\end{widetext} 
in which $\jbme{q}$ is a wavevector in the Brillouin zone, $G,G'$ are reciprocal lattice vectors, $nk$ is the Kohn-Sham one-particle state with energy $\epsilon_{nk}$ and wavefunction $\psi_{nk}$, and $T_{nk}=1$ if $nk$ belongs to the L space, and $T_{nk}=0$ else. 
The charge transfer energy 
\begin{equation}
\Delta_{n_o, n_u}(k,\jbme{q}) = \epsilon_{n_u k+\jbme{q}} - \epsilon_{n_o k}
\label{eq:cte}
\end{equation}
encodes the difference in onsite energies of $n_u k+\jbme{q}$ and $n_o k$,
and the interstate matrix element
\begin{equation}
M^{G}_{n_o, n_u}(k+\jbme{q},k) = \int_{\Omega} dr \psi^{*}_{n_u k +\jbme{q}}(r) e^{i(\jbme{q}+G)r} \psi_{n_o k}(r)
\label{eq:m}
\end{equation} 
encodes the wavefunctions $\psi_{nk}$, and also encodes the overlap between ALWOs since the latter are constructed from $\psi_{nk}$.
We deduce the cRPA effective interaction as
\begin{equation}
W_{\rm H} = \Big(1 - v\chi_{\rm H}\Big)^{-1} v,
\label{eq:wh}
\end{equation}
in which $v$ is the bare Coulomb interaction.
We deduce the onsite Coulomb repulsion in Eq.~\eqref{eq:u}.


\section{\jbme{Correction of $u$ and $|t_1|$: Improvement from the GGA+cRPA level to the c$GW$-SIC+\jbmf{LRFB} level}}
\label{app:corr-c}

\renewcommand{\theequation}{C\arabic{equation}}

\jbmf{Here, we give details on the calculation of $x_{\rm LRFB}^{}$ and $y_{\rm LRFB}^{}$ in Eqs.~\eqref{xLRFB} and~\eqref{yLRFB}
which allows to deduce $u_{{\rm c}GW{\rm -SIC+``LRFB"}}$ and $|t_1|_{{\rm c}GW{\rm -SIC+``LRFB"}}$ in Hg1223.
[This corresponds to the correction (C) mentioned in Sec.~\ref{Overview}.]}

\jbme{First, we \mib{address again the computational load
of the direct} c$GW$-SIC+LRFB calculation for Hg1223.
This 
calculation requires 
the LRFB preprocessing, whose extension to the cuprates with $N_{\ell}=3$ is 
\mib{computationally demanding, because one needs}
to solve the three-orbital Hamiltonian \mib{consisting of three CuO$_2$ planes in total by an accurate quantum many-body solver} (see Ref.~\onlinecite{Moree2022} for details) 
by taking into account the inter-CuO$_2$ plane hopping and interaction parameters. 
\mib{We leave} such an extension 
for future studies.}
\mib{Instead we employ the procedure (C1) and (C2)}
mentioned in Sec.~\ref{Overview}, \mib{because it already allows us to reach physically transparent understanding}.

\jbme{\mib{In the procedure (C1), we improve 
the AB LEH} from the GGA+cRPA level to the c$GW$-SIC level. 
Since the ratios $u_{{\rm c}GW{\rm -SIC}}/u^{\rm avg}$ 
and $|t_1|_{{\rm c}GW{\rm -SIC}}/|t_1^{\rm avg}|$ 
may have strong materials dependence and also pressure dependence,
\mib{due to the diversity of the global band structure outside of the AB band, we need to perform this procedure with respect to each material and pressure separately.} 
For instance, in Hg1223, we have $u_{{\rm c}GW{\rm -SIC}}/u^{\rm avg} \simeq 1.36$ at $P_{\rm amb}$ and $\simeq 1.21$ at 30 GPa.
The \mib{calculated c$GW$-SIC level of the parameters is} shown in Table~\ref{tab:appmace};
computational details of the c$GW$-SIC calculation are given at the end of this Appendix.
}

\jbme{\mib{To perform (C2), we employ the material independent constants $x_{\rm LRFB}$ and $y_{\rm LRFB}$ to correct the c$GW$-SIC results obtained in (C1), because this procedure is only to readjust mainly the onsite Coulomb interaction $U$ and this correction is materials insensitive. This readjustment arises from the correction of the relative chemical potential between the AB and B/NB bands to keep the electron fillings of the Cu3$d$ and O2$p$ orbitals, while the band structure of AB and B/NB bands by readjusting their chemical potentials and this chemical potential shift are indeed material insensitive in the known four compounds~\cite{Moree2022} because of the similar AB and B/NB band structures of the cuprates in general.}
In fact, our explicit calculations of $x_{\rm LRFB}$ and $y_{\rm LRFB}$ for several other cuprates (Hg1201, CaCuO$_2$, Bi2201, and Bi2212) show that, near optimal hole doping, $x_{\rm LRFB} \simeq 0.91-0.97$ and $y_{\rm LRFB} \simeq 0.99-1.06$ are rather universal and almost independent of the material.
Thus, it may be \mib{reasonable} to assume that Hg1223 near the optimal hole doping has similar values of $x_{\rm LRFB}$ of $y_{\rm LRFB}$,
and the narrow range of uncertainty allows accurate estimation of the Hamiltonian parameters.}

\begin{figure}[!ht]
\includegraphics[scale=0.5]{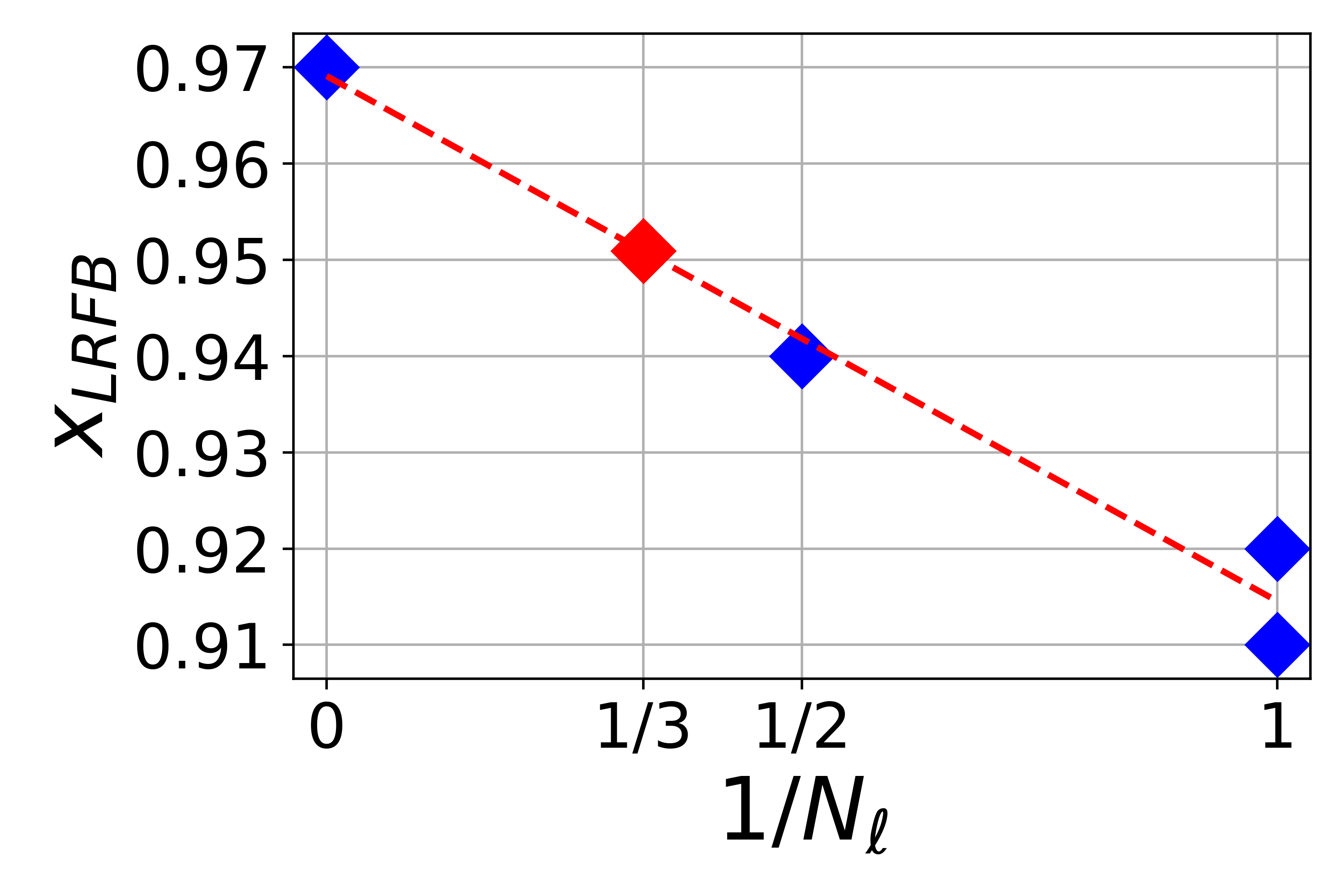}
\caption{\jbmf{$1/N_{\ell}$ dependence of calculated $x_{\rm LRFB}$ (blue symbols) 
for Hg1201, CaCuO$_2$, Bi2201 and Bi2212 listed in Table~\ref{tab:appmace} \mib{and their linear fitting (red dashed line).
Red diamond is the estimate for Hg1223 ($x_{\rm LRFB}^{\rm est} = 0.951$)  obtained from the interpolation at $1/N_{\ell}=1/3$.}}}
\label{fig:xlrfbnell}
\end{figure}

Still, the small uncertainty on $x_{\rm LRFB} \simeq 0.91-0.97$ causes a \mib{possible quantitative error} on the $P$ dependence of $T_{\rm c}^{\rm est}$ [see Fig.~\ref{fig:tcexp}(b)], even though the qualitative dome structure is \mib{robust}.
We \mib{thus narrow down} the estimate of $x_{\rm LRFB}$ as follows.
In 
\jbmf{Fig.~\ref{fig:xlrfbnell}}, we see a small \mib{but systematic} linear dependence of $x_{\rm LRFB}$ on $1/N_{\ell}$.
Linear \mib{interpolation} of the $1/N_{\ell}$ dependence of $x_{\rm LRFB}$ yields $x_{\rm LRFB}^{\rm est} \jbmf{ \ =0.951} \simeq 0.95$ at $N_{\ell}=3$. 
Thus, we assume $x_{\rm LRFB} = 0.95$ in Hg1223; 
for completeness, we also admit the range of uncertainty $x_{\rm LRFB} \simeq 0.91-0.97$.
On $y_{\rm LRFB}$, there is no clear $1/N_{\ell}$ dependence of $y_{\rm LRFB}$, so that we simply assume $y_{\rm LRFB}=1.0$.
(Note that the results shown in Fig.~\ref{fig:tcexp} and Fig.~\ref{fig:tcexp2} do not depend on the value of $y_{\rm LRFB}$.)
\jbmf{We deduce the values of $u_{{\rm c}GW{\rm -SIC+``LRFB"}}$ and $|t_1|_{{\rm c}GW{\rm -SIC+``LRFB"}}$ that are shown in Table~\ref{tab:appmace}.}

\jbme{The universality of calculated $x_{\rm LRFB}$ and $y_{\rm LRFB}$ may be understood as follows.
The LRFB corrects the value of $\Delta E_{xp}$ by an amount $\Delta \mu$ whose value is similar for all optimally doped compounds (we obtain $\Delta \mu \simeq 1.1-1.4$ eV in Ref.~\onlinecite{Moree2022}, Table IV).
This universality in $\Delta \mu$ is consistent with the universality in $x_{\rm LRFB}$ and $y_{\rm LRFB}$.
}

\jbme{Note that \jbmf{$u_{{\rm c}GW{\rm -SIC+``LRFB"}}$ and $|t_1|_{{\rm c}GW{\rm -SIC+``LRFB"}}$ are rough estimates} of the actual c$GW$-SIC+LRFB result.
In the actual c$GW$-SIC+LRFB calculation, more complex factors such as the self-doping of the IP and OP \cite{Kotegawa2001} and the Coulomb interaction between the IP and OP may affect the result of the LRFB calculation.
(Clarification of these factors is left for future studies.)
Nonetheless, the simple above estimate supports the assumption (B) in Sec.~\ref{sec:introduction}.
}
\\

\begin{table*}
\caption{
\jbme{
Values of $u$ and $|t_1|$ calculated for Hg1201, CaCuO$_2$, Bi2201, Bi2212 and Hg1223,
at the average hole concentration per CuO$_2$ plane $p_{\rm av}$ close to the optimal hole concentration. 
$N_{\ell}$ is the number of adjacent CuO$_2$ layers sandwiched between block layers.
On Hg1201, CaCuO$_2$, Bi2201 and Bi2212,
we show the values of $u_{{\rm c}GW{\rm -SIC+LRFB}}$ and $|t_1|_{{\rm c}GW{\rm -SIC+LRFB}}$ at $P_{\rm amb}$ taken from Ref.~\onlinecite{Moree2022}, and
\jbmf{the values of $u_{{\rm c}GW{\rm -SIC}}$ and $|t_1|_{{\rm c}GW{\rm -SIC}}$ at $P_{\rm amb}$ calculated in this paper. The values of $x_{\rm LRFB}$ and $y_{\rm LRFB}$ at $P_{\rm amb}$ \mib{are calculated} from Eqs.~\eqref{xLRFB} and~\eqref{yLRFB}.}
On Hg1223, we show the values of $u^{\rm avg}$ and $|t_1^{\rm avg}|$ at the GGA+cRPA level, and the values of $u_{{\rm c}GW{\rm -SIC}}$ and $|t_1|_{{\rm c}GW{\rm -SIC}}$ calculated in this paper at $P_{\rm amb}$, $30$ GPa and $60$ GPa.
We also show the values of \mib{$u_{{\rm c}GW{\rm -SIC+``LRFB"}}$ and $|t_1|_{{\rm c}GW{\rm -SIC+``LRFB"}}$ estimated from Eqs.~\eqref{xLRFB} and~\eqref{yLRFB} with the choices of $\jbme{x_{\rm LRFB}} = 0.91, 0.95$ and $0.97$ and $y_{\rm LRFB} = 1.0$
inferred  by analyzing other compounds} (see the main text).
We also show the values of \jbmf{$|t_1|_{{\rm c}GW{\rm -SIC+``LRFB"}}$ obtained after} applying the correction (D).
}
}
\label{tab:appmace}

\begin{ruledtabular}
\begin{tabular}{llllllll}
							  & Hg1201  & CaCuO$_2$ & Bi2201 & Bi2212 & Hg1223 & Hg1223 & Hg1223 \\
$P$	(GPa)			&   0 & 0 & 0 & 0 & 0 & 30 & 60\\
$p_{\rm av}$   &                     0.1 & 0.1 & 0.2 & 0.2 &  0.2 &  0.2 &  0.2 \\
\hline
\jbme{$N_{\ell}$} &   1 & $\infty$ & 1 & 2 & 3 & 3 & 3\\
\jbme{$1/N_{\ell}$} &   1 & 0 & 1 & 0.5 & 0.333 & 0.333 & 0.333\\
\hline
$u^{\rm avg}$             &      --- & --- & --- & ---  & 7.22 & 6.80 & 6.35 \\             
$u_{{\rm c}GW{\rm -SIC}}$              & 8.06 & 8.39 & 9.06 & 9.97 &  9.83 & 8.21 & 7.23\\
$u_{{\rm c}GW{\rm -SIC+LRFB}}$    & 7.35 & 8.10 & 8.34 & 9.37 & --- & --- & ---\\
$x_{\rm LRFB}$                                                       & 0.91 & 0.97 & 0.92 & 0.94 & --- & --- & ---\\
estimated $x_{\rm LRFB}$                                   & --- & --- & --- & ---  & 0.91/0.95/0.97 & 0.91/0.95/0.97 & 0.91/0.95/0.97\\
\jbmf{$u_{{\rm c}GW{\rm -SIC+``LRFB"}}$}    & --- & --- & --- & ---  &  8.95/9.34/9.54 & 7.47/7.80/7.96 & 6.58/6.87/7.01\\
\hline
$|t_1^{\rm avg}|$             &      --- & --- & --- & ---  & 0.528 & 0.596 & 0.643  \\        
$|t_1|_{{\rm c}GW{\rm -SIC}}$             & 0.526 & 0.526 & 0.498 & 0.436 & 0.485 & 0.569 & 0.615\\
$|t_1|_{{\rm c}GW{\rm -SIC+LRFB}}$    & 0.544 & 0.521 & 0.527 & 0.451 & --- & --- & ---\\
$y_{\rm LRFB}$                                                       & 1.034 & 0.990 & 1.058 & 1.034 & --- & --- & ---\\
estimated $y_{\rm LRFB}$                                   & --- & --- & --- & ---  & 1.0 & 1.0 & 1.0\\
\jbmf{$|t_1|_{{\rm c}GW{\rm -SIC+``LRFB"}}$}             & --- & --- & --- & --- & 0.485 & 0.569 & 0.615\\
\jbmf{$|t_1|_{{\rm c}GW{\rm -SIC+``LRFB"}}$ [after (D)]}              & --- & --- & --- & --- & 0.485 & 0.593 & 0.642\\
\end{tabular}
\end{ruledtabular}
\end{table*}

\paragraph*{\jbm{Computational details of the \jbmc{c$GW$-SIC} scheme ---}}
\headline{\jbmc{We apply the c$GW$-SIC scheme to Hg1201, Bi2201, Bi2212\jbmd{, CaCuO$_2$} and Hg1223 as follows.}}
\jbmc{On Hg1201, Bi2201\jbmd{, Bi2212 and CaCuO$_2$}, we consider the same computational conditions and hole concentration 
as in Ref. \onlinecite{Moree2022}.}
On Hg1223, we first preprocess the 44 bands within the M space from the GGA level to the $GW$ level.
(The $GW$ preprocessing is presented in detail in Ref.~\onlinecite{Moree2022}, Appendix A2.)
The random phase approximation (RPA) polarization is calculated by using $100$ real frequencies and $30$ imaginary frequencies; the maximum modulus of the frequency is $19.8$ Ha.
The exchange-correlation potential is sampled in the real space by using a $120 \times 120 \times 540$ grid to sample the unit cell.
In the calculation of the $GW$ self-energy, we reduce the computational cost by employing the scheme sketched in Ref.~\onlinecite{Moree2022}, Appendix E, with the cutoff energy $\epsilon = 0.01$ eV.
Other computational details are the same as those in Appendix~\ref{app:comp_details}.
We obtain the $GW$ electronic structure, in which the M bands are preprocessed at the $GW$ level and the other bands are left at the GGA level.
Then, we derive the AB LEH.
We start from the $GW$ electronic structure, and construct the AB MLWO.
The band window is the M space but we exclude the \jbme{$N_{\rm excl}$ lowest bands from it.
(We use $N_{\rm excl}=9$ at $P_{\rm amb}$, and $N_{\rm excl}=10$ at 30 GPa and 60 GPa.)}
Then, we use the cRPA to calculate the two-particle part and $U$.
We also use the \jbmd{c$GW$} to calculate the one-particle part and $|t_1|$.
(Details about the \jbmd{c$GW$} scheme can be found in Ref.~\onlinecite{Moree2022}, Appendix A5.)


\section{\jbme{Correction of $|t_1^{}|$ by correcting the cell parameter $a$}}
\label{app:corr-d}

\renewcommand{\theequation}{D\arabic{equation}}

\jbme{
Here, we give details about the correction (D) mentioned in Sec.~\ref{sec:introduction}\jbmf{.}
To correct the $P$ dependence of $|t_1|$, we correct
(i) the $P$ dependence of $a$ in Fig.~\ref{fig:cryst},
then combine the corrected (i) with 
(ii) the $a$ dependence of $|t_1|$ estimated in Appendix~\ref{app:ct}, Eq.~\eqref{adept1}.
}

\jbme{
On (i), the $P$ dependence of $a$ is shown in Fig.~\ref{fig:cryst}.
The experimental values of $a$ are available at $P_{\rm amb}$ \cite{Armstrong1995} and $P=8.5$ GPa, but not at $P > 8.5$ GPa.
At $P_{\rm amb}$, the experimental $a$ and optimized $a$ are in very good agreement (the difference is $\simeq 0.004$ \AA).
However, at $P=8.5$ GPa, the optimized $a$ overestimates the experimental $a$ by $\simeq 0.05$ \AA. 
We assume that such an overestimation also happens at $P > 8.5$ GPa, and we correct the $P$ dependence of optimized $a$ accordingly.
The values of the $P$ dependent correction $\Delta a(P)$ are
$\Delta a(P) = 0$ \AA \ if $P=P_{\rm amb}$ and $\Delta a(P) = \Delta a = -0.05$ \AA \ if $P > P_{\rm amb}$,
and the $P$ dependent corrected $a$ is denoted as
\begin{equation}
\tilde{a}(P) = a(P) + \Delta a(P).
\label{eq:deltaa}
\end{equation}
}

\jbme{On (ii), Eq.~\eqref{adept1} gives $|t_1|(a) \propto 1/a^3$.
Combination with Eq.~\eqref{eq:deltaa} yields:
\begin{equation}
|t_1|[\tilde{a}(P)] =  \frac{|t_1|[a(P)]}{1 + 3\frac{\Delta a(P)}{a(P)} + 3 \Big[ \frac{\Delta a(P)}{a(P)} \Big]^2 + \Big[ \frac{\Delta a(P)}{a(P)} \Big]^3},
\label{eq:t1deltaa}
\end{equation}
which allows to determine $|t_1|[\tilde{a}(P)]$ as a function of $|t_1|[a(P)]$.
\jbmf{The last two terms in the denominator of Eq.~\eqref{eq:t1deltaa} are negligible because $|\Delta a(P)/a(P)| \simeq 0.014 \ll 1$, 
so that we have:}
\begin{equation}
\jbmf{|t_1|[\tilde{a}(P)]} = \frac{ \jbmf{|t_1|[a(P)]} }{1 + 3\frac{\Delta a(P)}{a(P)}}.
\label{eq:t1deltaa2}
\end{equation}
\jbmf{[Note that $|t_1|[\tilde{a}(P)] \geq |t_1|[a(P)]$ because $\Delta a(P) \leq 0$.]}
\jbmf{We use Eq.~\eqref{eq:t1deltaa2} to correct the $P$ dependence of $|t_1|_{{\rm c}GW{\rm -SIC+``LRFB"}}$.}
The values of \jbmf{$|t_1|_{{\rm c}GW{\rm -SIC+``LRFB"}}$ that are obtained after applying (D) are shown in Table~\ref{tab:appmace} at $P_{\rm amb}$, $30$ GPa and $60$ GPa.}
}
\\

\jbme{For completeness, we mention a limitation of the correction (D):
It relies on the dependencies (i) and (ii) mentioned above, 
and (ii) is determined at the GGA+cRPA level.
The only way to improve slightly the approximation in (D) and Eq.~\eqref{eq:t1deltaa2} would be to 
take the optimized CPs at $P=30$ GPa and reduce the cell parameter $a$ by $0.05$ \AA \ 
(the estimated difference between optimized $a$ and experimental $a$), 
then perform explicitly the c$GW$-SIC calculation from the CP with the reduced $a$, 
then deduce \jbmf{$|t_1|_{{\rm c}GW{\rm -SIC+``LRFB"}}$ and $u_{{\rm c}GW{\rm -SIC+``LRFB"}}$.}
However, this improvement is computationally expensive, 
and we do not expect it to change the results significantly.
Thus, we do not consider it here.
}

\section{Pressure dependence of intermediate quantities}
\label{app:pdepinterm}

\renewcommand{\theequation}{E\arabic{equation}}

\subsection{Pressure dependence of the DFT band structure and Madelung potential}
\label{app:dftband}

\begin{figure*}[!htb]
\centering
\includegraphics[scale=0.395]{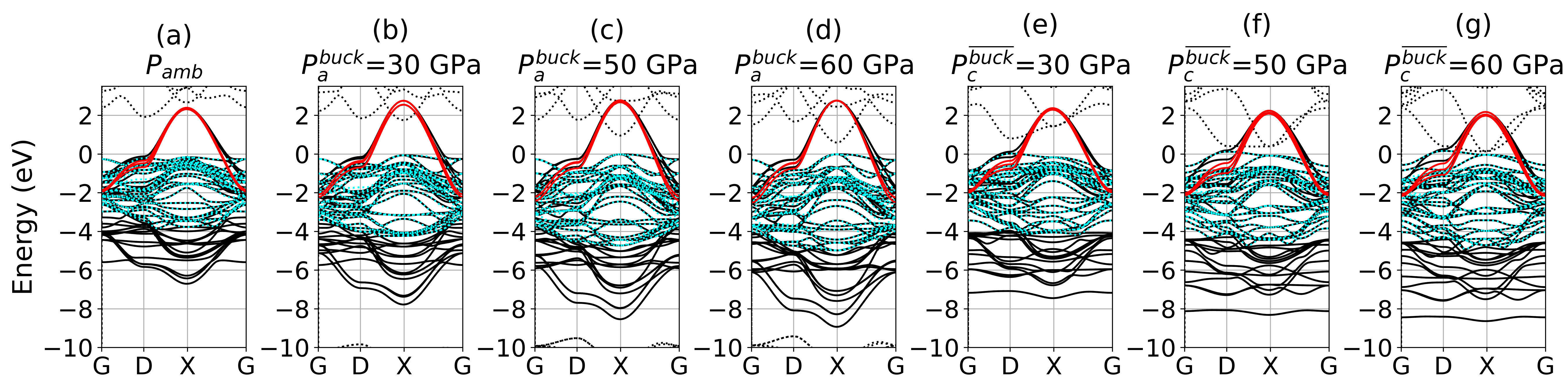} 
\caption{
$P_{a}^{\rm buck}$ and $P_{c}^{\overline{\rm buck}}$ dependence of the GGA band structure obtained by using the optimized CP values. 
Notations are the same as those in Fig.~\ref{fig:bandp}.
}
\label{fig:bandpc}
\end{figure*}

\begin{figure*}[!htb]
\centering
\includegraphics[scale=0.395]{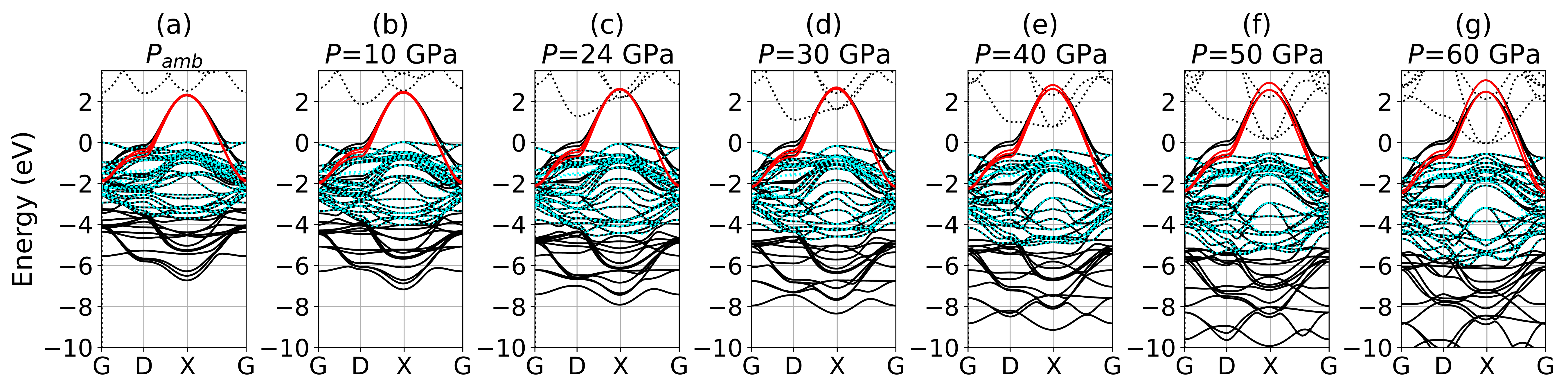} 
\caption{
$P$ dependence of the GGA band structure obtained by using the CP values from Zhang \textit{et al.}.
Notations are the same as those in Fig.~\ref{fig:bandp}.
}
\label{fig:bandpzhang}
\end{figure*}

\headline{
Here, as a complement to Sec.~\ref{sec:band}, we show that [M$W$] is robust with respect to the definition of uniaxial pressure and with respect to the uncertainty on CP values.
}
First [M$W$] is caused by $P_{a}^{\rm buck}$ rather than $P_{c}^{\overline{\rm buck}}$ (see Fig.~\ref{fig:bandpc}),
which is consistent with Fig.~\ref{fig:bandp} in which [M$W$] is caused by $P_{a}^{}$ rather than $P_{c}^{}$:
The main origin of [M$W$] is indeed the reduction in $a$, and the variation in $d^{z}_{\rm buck}$ with $P_{a}^{\rm buck}$ does not affect this result.
Also, if we use the CP values from Zhang \textit{et al.} instead of the optimized CP values, 
[M$W$] is well reproduced (see Fig.~\ref{fig:bandpzhang}).
\\

\headline{
In addition, we discuss the mechanisms of [M$\epsilon$] and [M$W$] in terms of Madelung potential created by ions in the crystal.
}
\oldjbm{As shown in Sec.~\ref{sec:band}, [M$\epsilon$] and [M$W$] are mainly caused by the reduction in $a$.
This may be understood as follows.
The main contribution of the Madelung potential felt by electrons in the CuO$_2$ plane is from the positive Cu and negative O ion within the plane.
Then, the energy of an electron at the Cu$3d$ orbital gets higher when the surrounding O ions become closer to the Cu site, namely, if $a$ is reduced.
On the contrary, an electron at the O$2p_{\sigma}$ orbital feels opposite for the reduced $a$.
This makes the difference of the electronic levels for the Cu$3d_{x^2-y^2}$ and O$2p_{\sigma}$ larger. 
More precise calculation including long-range Coulomb potential by DFT supports this simple view is essentially correct.
}

\begin{figure}
\includegraphics[scale=0.5]{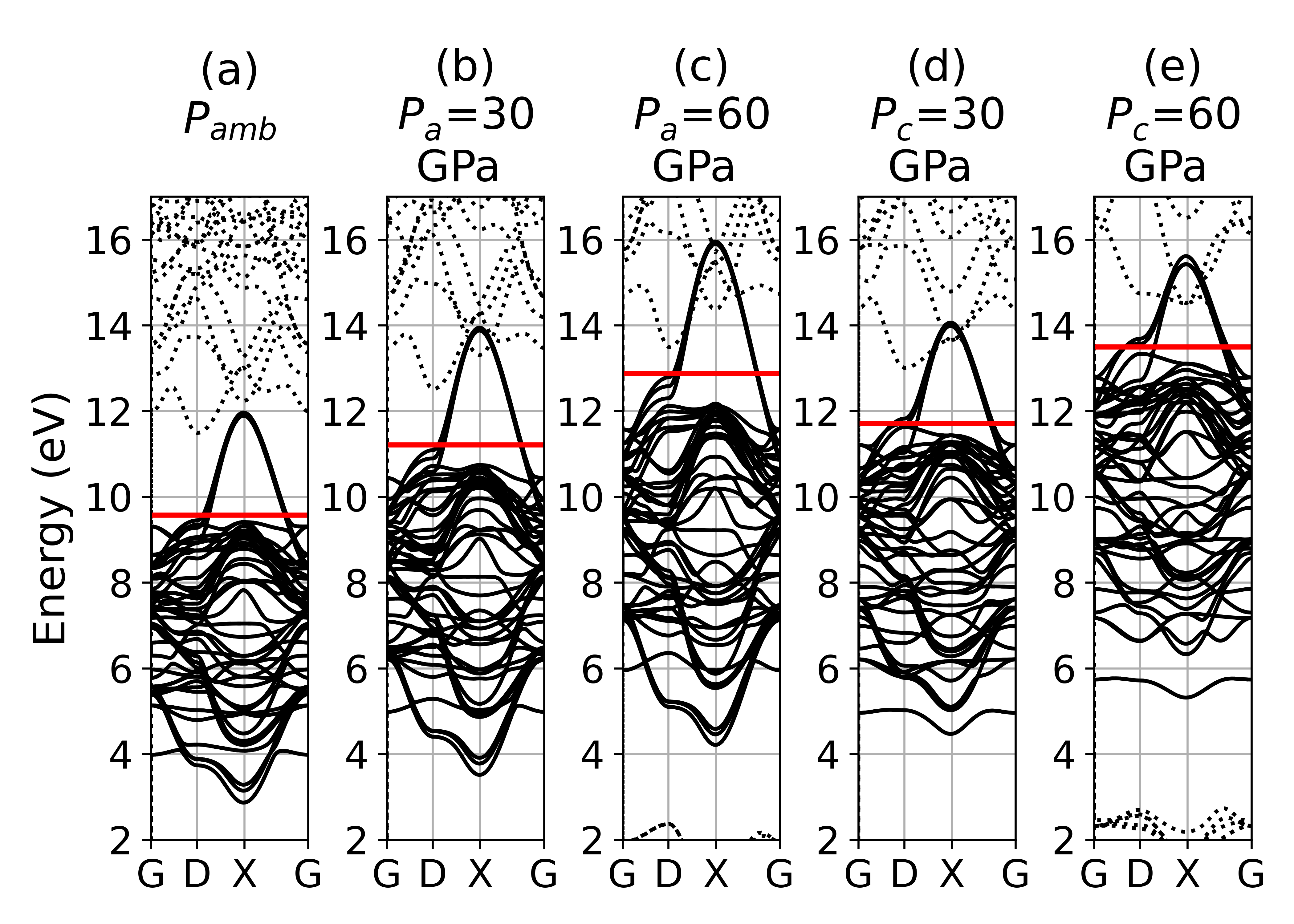}
\caption{$P_a$ and $P_c$ dependencies of the band structure and Fermi energy.
We show the bands inside the M space (solid black color) and outside M space (dotted black color).  
These band structures correspond to those in Fig.~\ref{fig:bandp}(h,i,k,l,n),
except that the band energies are not renormalized with respect to the Fermi energy. 
The latter is given by the horizontal line in red color.
}
\label{fig:bandpabs}
\end{figure}

\headline{The $P_a$ induced increase in energy of Cu$3d$ bands is illustrated in Fig.~\ref{fig:bandpabs}(a,b,c).}
The application of $P_a$ increases the absolute energy of Cu$3d$ bands.
(The absolute energy is defined as the energy without renormalization with respect to the Fermi level.)
Note that examining the pressure dependence of absolute energies does make sense, 
because the chemical composition of the crystal is not modified by the application of pressure.

\headline{The application of $P_c$ increases the energy of not only Cu$3d$ bands but also O$2p$ bands in Fig.~\ref{fig:bandpabs}(a,d,e),
so that [M$W$] does not occur.}
This can also be understood in terms of Madelung potential from in-plane O anions.
When $P_c$ is applied, the distance $d^{z}_{\rm Cu}$ between the IP and OP is reduced.
This reduces not only (i) the interatomic distance between the O anion in the OP (IP) and the Cu in the IP (OP),
but also (ii) the interatomic distance between the O anion in the OP (IP) and the O in the IP (OP),
The concomitant reduction in (i) and (ii) causes the concomitant increase in Cu$3d$ and O$2p$ electronic levels.

\subsection{Pressure dependence of the onsite bare interaction and Cu$3d_{x^2-y^2}$/O$2p_{\sigma}$ charge transfer energy}
\label{app:ct}

\headline{
Here, as a complement to Sec.~\ref{sec:lehut}, we discuss the following points:\\
\noindent
(a) The increase in onsite bare interaction $v$ is caused by the reduction in Cu$3d_{x^2-y^2}$/O$2p_{\sigma}$ hybridization when $\Delta E_{xp}$ increases.\\
\noindent
(b) The concomitant increases in $|t_1|$ and $v$ when $a$ decreases can be understood by further analysis of the $a$ dependence of quantities.\\
\noindent
(c) The reduction in the correlation between $v^{l}$ and $\Delta E^{l}_{xp}$ at $P > P_{\rm opt}$ originates from the non-equivalence of the IP and OP,
especially the buckling of Cu-O-Cu bonds in the OP.
\\
}

\begin{figure}[!htb]
\centering
\includegraphics[scale=0.5]{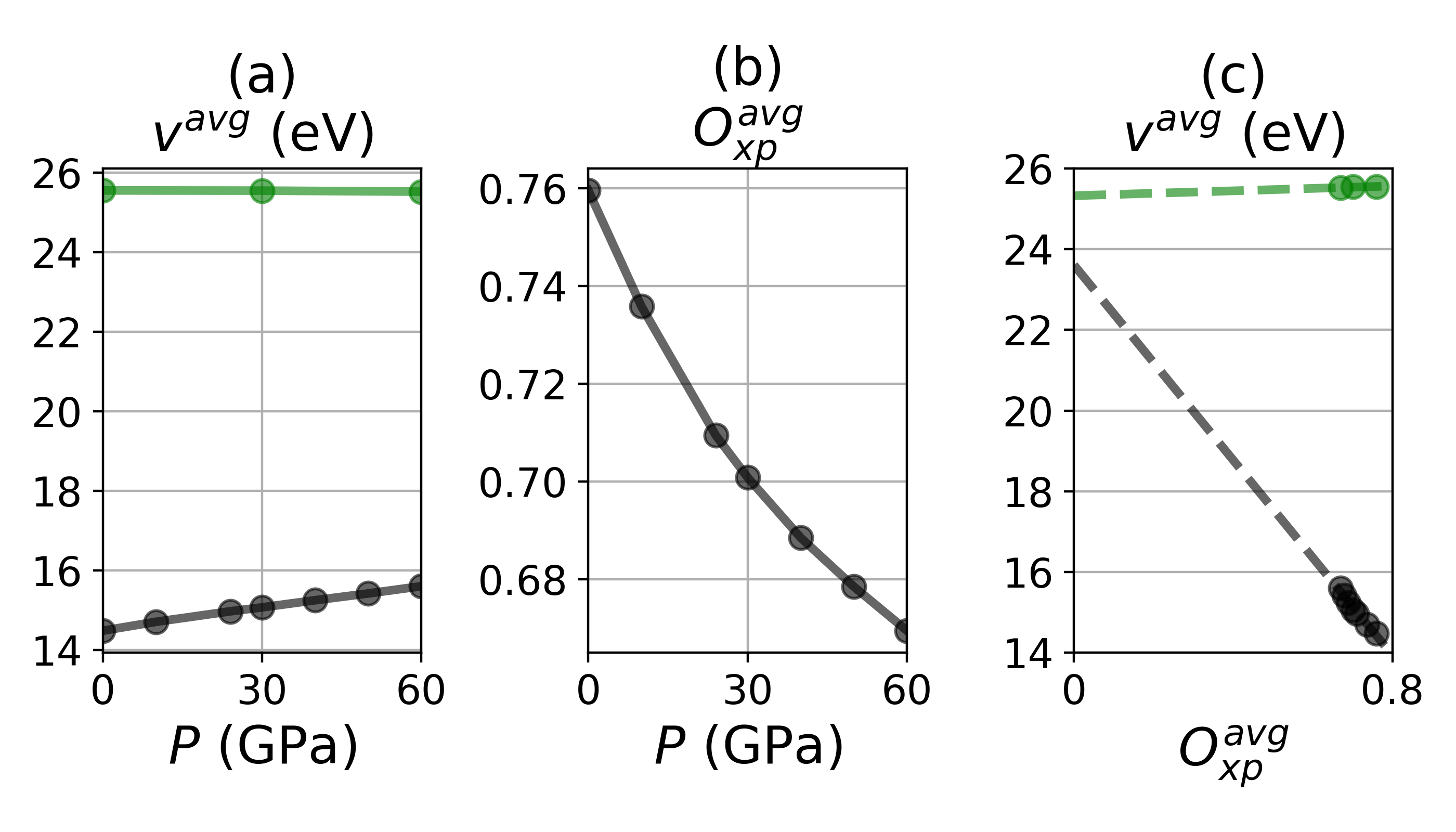}
\caption{
\oldjbm{
Panel (a): $P$ dependence of the average value $v^{\rm avg}$ over the IP and OP of the onsite bare interaction.
We show $v^{\rm avg}$ in the AB (black dots) and Cu$3d_{x^2-y^2}$ (green dots) Wannier orbitals.
Panel (b): $P$ dependence of the average value $O_{xp}^{\rm avg}$ of the ratio $O_{xp}^{l}=|t_{xp}^{l}|/\Delta E_{xp}^{l}$,
which encodes the Cu$3d_{x^2-y^2}$/O$2p_{\sigma}$ hybridization.
Panel (c): $O_{xp}^{\rm avg}$ dependence of $v^{\rm avg}$ in the AB (black dots) and Cu$3d_{x^2-y^2}$ (green dots) Wannier orbitals.
The dashed lines show the linear extrapolation to $O_{xp}^{\rm avg}=0$ (the limit in which the Cu$3d_{x^2-y^2}$/O$2p_{\sigma}$ hybridization becomes negligible).
}
}
\label{fig:vrxp}
\end{figure}

\headline{
On (a), a first remark is that the Cu$3d_{x^2-y^2}$/O$2p_{\sigma}$ hybridization reduces $v$ by reducing the atomic Cu$3d_{x^2-y^2}$ character of the AB orbital.
}
In the AB orbital, the onsite bare interaction is $v^{\rm avg} \simeq 14.5-15.5$ eV,
but in the Cu$3d_{x^2-y^2}$ M-ALWO, the onsite bare interaction $v^{\rm avg}_{x} \simeq 25.5$ eV is larger [see Fig.~\ref{fig:vrxp}(a)].
This is because the Cu$3d_{x^2-y^2}$ M-ALWO has atomic character, and is more localized than the AB orbital.
In the limit of zero hybridization, the AB orbital is equivalent to the Cu$3d_{x^2-y^2}$ M-ALWO if we neglect the effect of other orbitals:
In that case, $v^{\rm avg}_{} = v^{\rm avg}_{x}$.
However, the hybridization is always non-zero in the realistic cuprate, 
so that the atomic Cu$3d_{x^2-y^2}$ character of the AB orbital is reduced.

\headline{
Second, the importance of the Cu$3d_{x^2-y^2}$/O$2p_{\sigma}$ hybridization decreases with $P$.
}
The importance of the hybridization is roughly encoded in the ratio $O_{xp}=|t_{xp}|/\Delta E_{xp}$ between the Cu$3d_{x^2-y^2}$/O$2p_{\sigma}$ hopping amplitude and Cu$3d_{x^2-y^2}$/O$2p_{\sigma}$ charge transfer energy.
$O_{xp}$ decreases when the hybridization is reduced, and becomes zero when the hybridization is negligible.
And, $O_{xp}^{\rm avg}$ decreases with $P$ [see Fig.~\ref{fig:vrxp}(b)].

\headline{
Thus, the atomic Cu$3d_{x^2-y^2}$ character of the AB orbital increases with $P$:
We interpret this as the origin of the increase in $v$.
}
To confirm this, we show explicitly that $v^{\rm avg}_{} \simeq v^{\rm avg}_{x}$ in the limit of zero hybridization.
Let us consider the $O_{xp}^{\rm avg}$ dependence of $v^{\rm avg}$ in Fig.~\ref{fig:vrxp}(c), 
which is obtained by combining the $P$ dependencies of $v^{\rm avg}$ and $O_{xp}^{\rm avg}$ in Fig.~\ref{fig:vrxp}(a,b).
In the \textit{ab initio} calculation, we have $O_{xp}^{\rm avg} \simeq 0.7-0.8$,
so that the $O_{xp}$ dependence of $v^{\rm avg}$ can be explicitly obtained only within this range.
However, the value of $v^{\rm avg}$ in the limit of zero hybridization may be estimated by performing a linear extrapolation of the \textit{ab initio} $O_{xp}^{\rm avg}$ dependence of $v^{\rm avg}$.
The extrapolation yields $v^{\rm avg} \simeq 24$ eV at $O_{xp}^{\rm avg}=0$ [see Fig.~\ref{fig:vrxp}(c)], 
which is similar to $v^{\rm avg}_{x} \simeq 25.5$ eV.
This suggests the AB orbital becomes the Cu$3d_{x^2-y^2}$ M-ALWO in the limit of zero hybridization, as mentioned above.
\\

\headline{
On (b), we analyze the $a$ dependence of $|t_1^{\rm avg}|$, $\Delta E_{xp}^{\rm avg}$ and $|t_{xp}^{\rm avg}|$,
as well as the average values $O_{xp}^{\rm avg}$ and $T_{xp}^{\rm avg}$ of $O_{xp}^{l}=|t_{xp}^{l}|/\Delta E_{xp}^{l}$ and $T_{xp}^{l}=|t_{xp}^{l}|^2/\Delta E_{xp}^{l}$; the key point is that, when $a$ decreases, $O_{xp}^{\rm avg}$ decreases whereas $|t_1^{\rm avg}| \propto T_{xp}^{\rm avg}$ increases.
}
To obtain the $a$ dependence of the above quantities,
we take the values of $a$ as a function of pressure in Fig.~\ref{fig:cryst}
and combine them with the $P_a$ dependence of $|t_1^{l}|$, $\Delta E_{xp}^{l}$ and $|t_{xp}^{l}|$ in Fig.~\ref{fig:ratiopu}(a,f,g).
Note that we consider the $P_a$ dependence instead of the $P$ dependence.
This is because the application of $P_a$ modifies only the value of $a$:
This allows to extract accurately the $a$ dependence while avoiding the $d^{z}_{l}$ dependence of quantities.

\begin{figure}[!htb]
\centering
\includegraphics[scale=0.45]{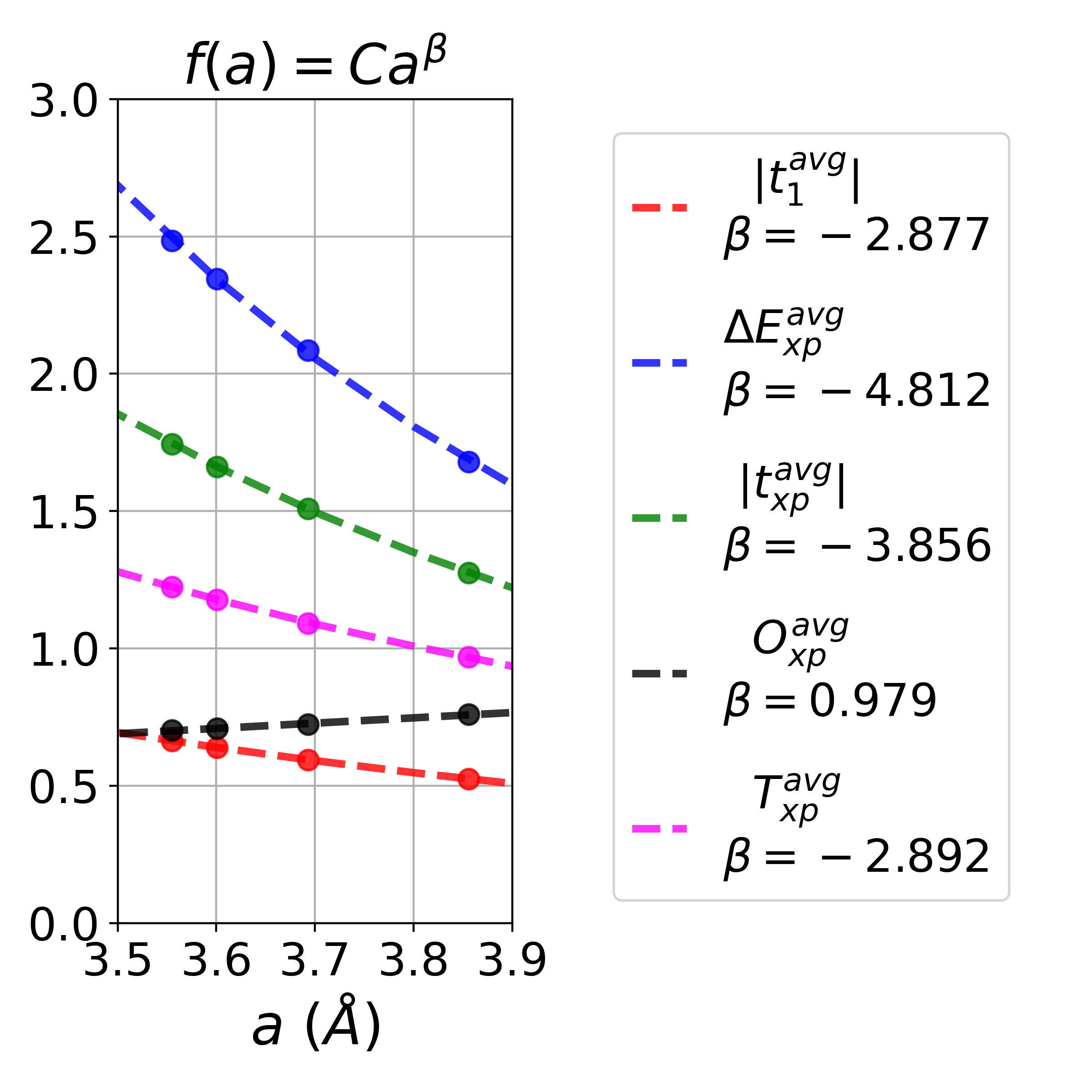}
\caption{
Cell parameter $a$ dependence of $|t_1^{\rm avg}|$, $\Delta E_{xp}^{\rm avg}$, $|t_{xp}^{\rm avg}|$, and the average values $O_{xp}^{\rm avg}$ and $T_{xp}^{\rm avg}$ of $O_{xp}^{l}=|t_{xp}^{l}|/\Delta E_{xp}^{l}$ and $T_{xp}^{l}=|t_{xp}^{l}|^2/\Delta E_{xp}^{l}$ (dots).
We \jbmd{start from the optimized CP values at $P_{\rm amb}$, and apply $P_a$ to modify only the value of $a$.}
The dashed curves show the interpolation of the $a$ dependence
by the function $f(a)= C a^{\beta}$, where $\beta$ and $C$ are the fitting parameters.
The legend shows the obtained values of $\beta$.
}
\label{fig:vrxp2}
\end{figure}

\headline{
We interpolate the $a$ dependencies of the above quantities,
by using the fitting function $f(a)= C a^{\beta}$, 
where $\beta$ and $C$ are the fitting parameters.} 
We examine the values of $\beta$, which encode the speed of variation in quantities with $a$.
The obtained values of $\beta$ are shown in Fig.~\ref{fig:vrxp2}.

\headline{First, we have:
\begin{equation}
|t_{1}^{\rm avg}| \propto T_{xp}^{\rm avg} \propto 1/a^3.
\label{adept1}
\end{equation}
}
Indeed, the value of $\beta$ for $|t_1^{\rm avg}|$ is $\beta(|t_1^{\rm avg}|) \simeq -2.88$, which is very close to $-3$.
This is consistent with the $1/r^3$ decay of the density-density correlation function \cite{Misawa2007}.
Also, $\beta(T_{xp}^{\rm avg}) \simeq -2.89$ is almost identical to $\beta(|t_1^{\rm avg}|)$.

\headline{Second, we have:
\begin{equation}
O_{xp}^{\rm avg} \propto a.
\label{adepv}
\end{equation}
}
Indeed, the $O_{xp}^{\rm avg}$ dependence of $a$ is almost linear: $\beta(O_{xp}^{\rm avg})=0.98$ is very close to $1$.


\headline{The above equations \eqref{adept1} and \eqref{adepv}
show that both $|t_{1}^{i}|$ and $v^{i}$ increase when $a$ decreases.
}
Indeed, in the item (a), we have clarified that $v^{\rm avg}$ increases when $O_{xp}^{\rm avg}$ decreases,
and Eq.~\eqref{adepv} shows that $O_{xp}^{\rm avg}$ decreases when $a$ decreases.
Note that $\beta(\Delta E_{xp}^{\rm avg}) < \beta(|t_{xp}^{\rm avg}|) <0$:
This is the origin of the positive value of $\beta(O_{xp}^{\rm avg})$.
On the other hand, $2 \beta(|t_{xp}^{\rm avg}|) < \beta(\Delta E_{xp}^{\rm avg}) <0$:
This is the origin of the negative value of $\beta(|t_1^{\rm avg}|) \simeq \beta(T_{xp}^{\rm avg})$.
\\

\headline{
On (c), the non-equivalence of the IP and OP causes a slight difference in the $\Delta E^{l}_{xp}$ dependence of $v^{l}$ for $l=i$ and $l=o$
[see Fig.~\ref{fig:ratiop}(e,f,l,m) at $P > P_{\rm opt}$].
}
\oldjbm{
\jbmc{This is simply because} $|t_{xp}^{o}|$ is reduced at $P>P_{\rm opt}$ due to the increase in $|d^{z}_{\rm buck}|$ [see Fig.~\ref{fig:ratiop}(g)].
This contributes to reduce $O_{xp}^{o}=|t_{xp}^{o}|/\Delta E_{xp}^{o}$, which increases $v^{o}$ [see the item (a)].
This \jbmc{explains} why, at $P > P_{\rm opt}$, $v^{o} \simeq v^{i}$ even though $\Delta E_{xp}^{o} < \Delta E_{xp}^{i}$
[see Fig.~\ref{fig:ratiop}(e,f)].
On the other hand, if we apply \jbmc{only} $P_a$ (which modifies only $a$ without modifying $d^{z}_{\rm buck}$),
$|t_{xp}^{o}|$ is not reduced with respect to $|t_{xp}^{i}|$ [see Fig.~\ref{fig:ratiop}(g)],
and the $\Delta E_{xp}^{l}$ dependence of $v^{l}$ is very similar for $l=i$ and $l=o$ [see Fig.~\ref{fig:ratiop}(e,f)].
}

\headline{The non-equivalence of the IP and OP also causes a slight difference in the $P$ dependence of $\Delta E^{l}_{xp}$ for $l=i$ and $l=o$ in Fig.~\ref{fig:ratiop}(f,m).
}
\oldjbm{
\jbmc{This is because $\Delta E^{l}_{xp}$ depends not only on $a$, but also on $d^{z}_{l}$} [see Fig.~\ref{fig:uncertainty}(f) in Appendix~\ref{app:unc}].
For instance, $\Delta E^{i}_{xp}$ ($\Delta E^{o}_{xp}$) increases (decreases) when $d^{z}_{\rm Ca}$ decreases.
And, $d^{z}_{\rm Ca} \simeq 1.48$ \AA \ in the optimized CP values is smaller than $d^{z}_{\rm Ca} \simeq 1.59$ \AA \ in the CP values from Zhang \textit{et al.}.
This is why $\Delta E^{i}_{xp} > \Delta E^{o}_{xp}$ for the optimized CP values,
but $\Delta E^{i}_{xp} < \Delta E^{o}_{xp}$ for the values from Zhang \textit{et al.} [see Fig.~\ref{fig:ratiop}(f,m)].
}

\subsection{Pressure dependence of the screening}
\label{app:screening}

\headline{
Here, as a complement to Sec.~\ref{sec:lehuthighp},
we discuss the uniaxial pressure dependence of $R^{\rm avg}$,
from which the dome structure in the uniform pressure dependence of $R^{\rm avg}$ originates.
\\
}

\headline{First, we discuss the $P_a$ dependence of $R^{\rm avg}$ in Fig.~\ref{fig:ratiopu}(d).}

\noindent
\headline{(i) At $P_a < P_{\rm opt}$, the increase in $R^{\rm avg}$ is explained by the broadening [M$W$] of the GGA band dispersion.}
More precisely, the origin is the increase in charge transfer energies (\ref{eq:cte}) (schematically denoted as $\Delta$ in this Appendix) 
between occupied and empty bands, due to [M$W$] discussed in Sec.~\ref{sec:band}.
The increase in $\Delta$ participates in the decrease of the amplitude of the cRPA polarization (\ref{eq:chiou}),
schematically denoted as $|\chi| \propto 1/\Delta$.
This reduces the cRPA screening and thus increases $R^{\rm avg}$.

\noindent
\headline{(ii) At $P_a > P_{\rm opt}$, the increase in $R^{\rm avg}$ ceases.}
This is because the effect of [M$W$] is progressively reduced:
We have $\partial |\chi_{}^{}| / \partial \Delta_{} \propto -1/\Delta_{}^2$, 
so that the larger $P_a$ and thus $\Delta_{}$, 
the smaller the decrease in $|\chi_{}^{}|$ when $\Delta_{}$ is further increased, and the less important the effect of [M$W$].
In addition, when $P_a$ increases, 
the charge transfer energy $\Delta_{\rm M-empty}$ between the M bands and empty bands outside M space is reduced,
because the energy of the Cu$3d$ bands increases [see Fig.~\ref{fig:bandpabs}(a,b,c)].
This may contribute to increase $|\chi_{}^{}| \propto 1/\Delta_{\rm M-empty}$ and cancel the effect of  [M$W$] at high pressure.
\\

\headline{Second, we discuss the decrease in the $P_c$ dependence of $R^{\rm avg}$ in Fig.~\ref{fig:ratiopu}(d).}
This is because [M$W$] does not occur when $P_c$ is applied, contrary to $P_a$.
Thus, $\Delta$ does not increase.
On the other hand, $\Delta_{\rm M-empty}$ is reduced because the energy of Cu$3d$ bands increases [see Fig.~\ref{fig:bandpabs}(a,d,e)].
As a result, $|\chi_{}^{}| \propto 1/\Delta_{\rm M-empty}$ increases.

\section{\oldjbm{Crystal parameter dependence of effective Hamiltonian parameters at optimal pressure}}
\label{app:unc}

\renewcommand{\theequation}{F\arabic{equation}}

\begin{figure*}[!htb]
\centering
\includegraphics[scale=0.38]{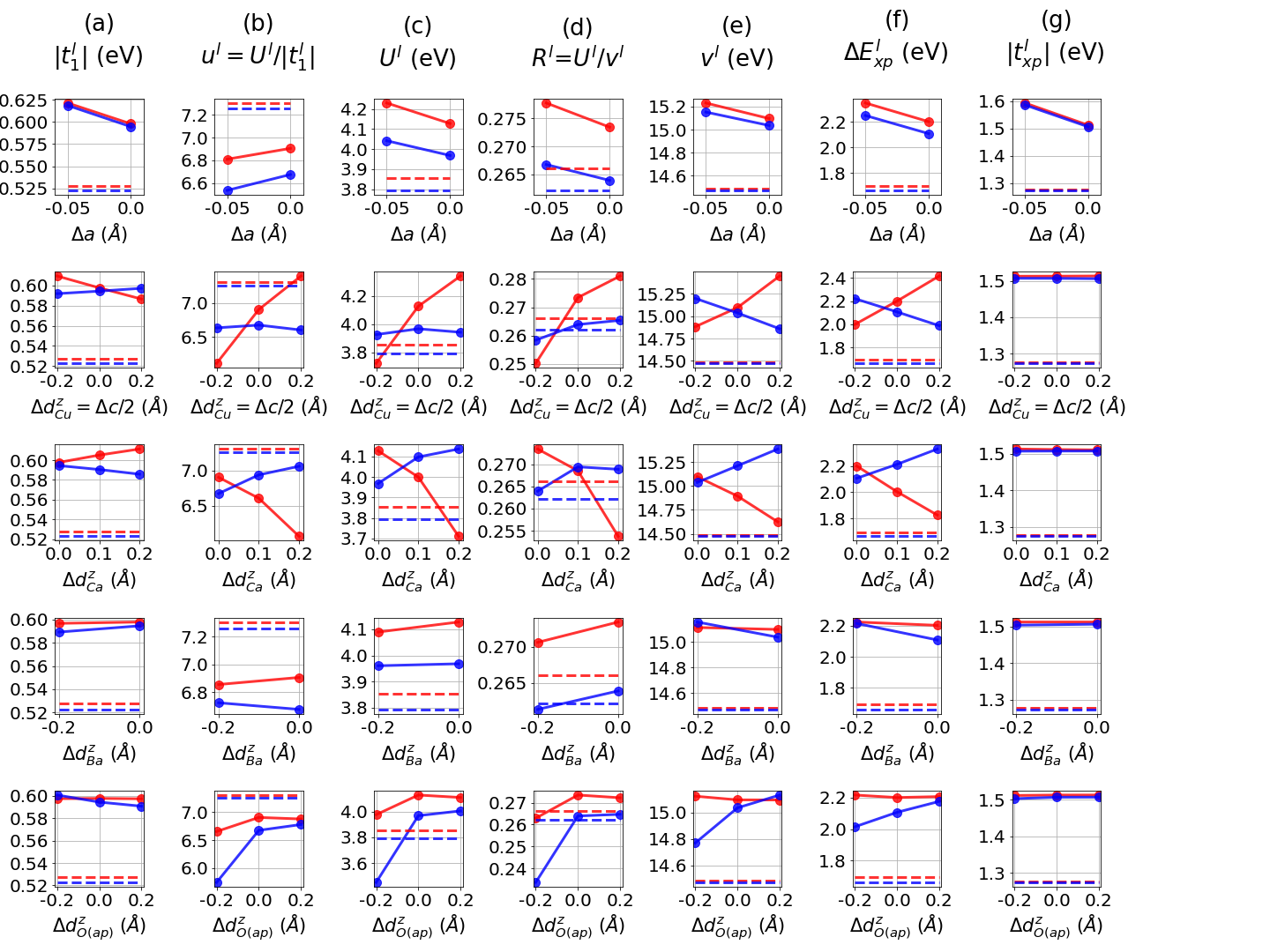}
\caption{
CP dependencies of $|t_1^{l}|$, $u^{l}$, $U^{l}$, $R^{l}=U^{l}/v^{l}$, $v^{l}$, $\Delta E_{xp}^{l}$ and $|t_{xp}^{l}|$.
We show quantities in the IP ($l=i$) and OP ($l=o$) in red color and blue color, respectively.
The quantities are obtained by using the optimized CP values at $P_{\rm opt}=30$ GPa, 
and modifying the values of $a$, $c$ and $d^{z}_{l}$ by $\Delta a$, $\Delta c$, and $\Delta d^{z}_{l}$, respectively.
Note that when $\Delta d^{z}_{\rm Cu}$ is applied, $\Delta c=2\Delta d^{z}_{\rm Cu}$ is also applied so that all interatomic distances in the block layer remain unchanged.
The horizontal dashed lines represent the values at $P_{\rm amb}$, for comparison.
}
\label{fig:uncertainty}
\end{figure*}

\begin{figure}
\centering
\includegraphics[scale=0.4]{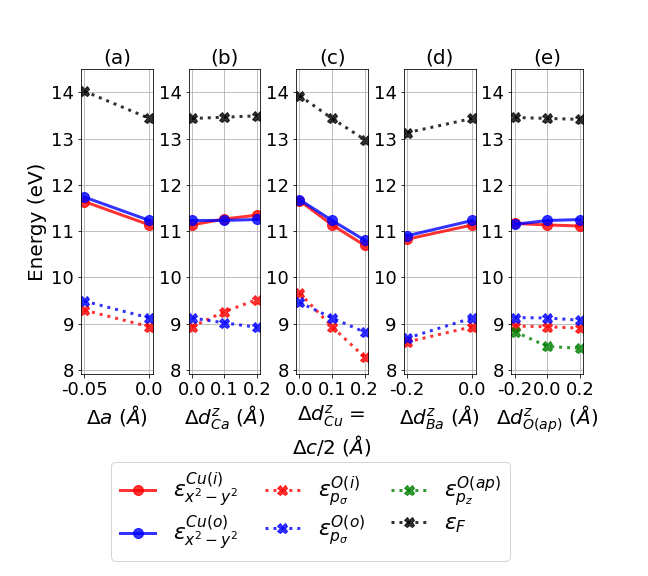}
\caption{\oldjbm{CP dependencies of the 
Fermi energy $\epsilon_{\rm F}^{}$ 
and onsite energies $\epsilon_{i}^{l}$ of the Cu$3d_{x^2-y^2}$ and O$2p_{\sigma}$ ALWOs.
The quantities are obtained by using the optimized CP values at $P_{\rm opt}=30$ GPa, 
and modifying the values of $a$, $c$ and $d^{z}_{l}$ by $\Delta a$, $\Delta c$, and $\Delta d^{z}_{l}$, respectively.
Note that when $\Delta d^{z}_{\rm Cu}$ is applied, $\Delta c=2\Delta d^{z}_{\rm Cu}$ is also applied so that all interatomic distances in the block layer remain unchanged.
In the panel (e), we also show the $d^{z}_{\rm O(ap)}$ dependence of the onsite energy $\epsilon^{\rm O(ap)}_{p_z}$ of the apical O$2p_{z}$ orbital.
}
}
\label{fig:unconsite}
\end{figure}

Here, as a complement to Sec.~\ref{sec:leh}, we analyze the CP dependencies of AB LEH parameters around $P_{\rm opt}$.
We start from the optimized CP values at $P_{\rm opt}$ and modify separately the values of each CP. 
The modified values are given in Fig.~\ref{fig:cryst} (open squares).
The CP dependencies of AB LEH parameters are shown in Fig.~\ref{fig:uncertainty}. 

We summarize the main results below: 

\noindent
(i) As for $|t_1^{l}|$, the $a$ dependence is the strongest.

\noindent
(ii) As for $u^{i}$, the $d^{z}_{\rm Ca}$ and $d^{z}_{\rm Cu}$ dependencies are the strongest.

\noindent
(iii) As for $u^{o}$, the $d^{z}_{\rm Ca}$ and $d^{z}_{\rm O(ap)}$ dependencies are the strongest.

Also, (ii,iii) suggest the origin of the decrease in $R^{l}$ at $P > P_{\rm scr}$ in Sec.~\ref{sec:lehuthighp}, Fig.~\ref{fig:ratiop}(d,k):
The decreases in $R^i$ and $R^o$ are caused respectively by the decreases in $d^{z}_{\rm Cu}$ and $d^{z}_{\rm O(ap)}$.
\\

\paragraph*{$a$ dependence of AB LEH parameters ---}

At $P_{\rm opt}$, the optimized value of $a \simeq 3.69$ \AA \ is the same as that from Zhang \textit{et al.}. 
Still, this value might be overestimated. 
Indeed, the $P$ dependence of experimental values \cite{Hunter1994,Armstrong1995} shows the faster decrease at lower pressures (see Fig.~\ref{fig:cryst}). 
Thus, we consider the modification $\Delta a$ of $a$ at $P_{\rm opt}$, such that $-0.05$ \AA \ $\leq \Delta a \leq 0 $ \AA \ at $P_{\rm opt}$.

\headline{The $a$ dependence of $|t_1|$ is strong [see Fig.~\ref{fig:uncertainty}(a)], as discussed in Sec.~\ref{sec:leht}.}
We note that the $15\%$ increase in $|t_1|$ from $P_{\rm amb}$ to $P_{\rm opt}$ becomes \jbmc{$18-19\%$} if $\Delta a = -0.05$ \AA.
Thus, the $3\%$ difference between the increase in $|t_1|$ and that in $T_{c}^{\rm opt}$ may be understood by admitting the above uncertainty on $a$ at $P_{\rm opt}$ (see the discussion in Sec.~\ref{sec:disc}).
\\

\paragraph*{$d^{z}_{\rm Ca}$ dependence of AB LEH parameters ---}

The optimized value $d^{z}_{\rm Ca} \simeq 1.48$ \AA \ is lower than that from Zhang \textit{et al.} ($d^{z}_{\rm Ca} \simeq 1.59$ \AA). 
Thus, we consider  $0.0$ \AA \ $\leq \Delta d^{z}_{\rm Ca} \leq +0.2$ \AA \ to examine the $d^{z}_{\rm Ca}$ dependence of AB LEH parameters.

\headline{Increasing $d^{z}_{\rm Ca}$ causes the rapid decrease in $u^{i}$ and increase in $u^{o}$ [see Fig.~\ref{fig:uncertainty}(b)], 
due to the decrease in $\Delta E_{xp}^{i}$ and increase in $\Delta E_{xp}^{o}$ [see Fig.~\ref{fig:uncertainty}(f)].
}
\oldjbm{
Indeed, $v^{l}$ and $R^{l}$ are correlated with $\Delta E_{xp}^{l}$.
The correlation between $v^{l}$ and $\Delta E_{xp}^{l}$ has been discussed in Appendix~\ref{app:ct}, 
and the increase in $\Delta E_{xp}^{l}$ also contributes to increase $R^{l}$ by reducing the cRPA screening between Cu$3d_{x^2-y^2}$/O$2p_{\sigma}$ B/NB and AB bands.
}

\headline{
The increase (decrease) of $\Delta E_{xp}^{l}$ originates from the positive Madelung potential created by the Ca cation, which stabilizes electrons in the vicinity of the Ca cation.
}
\oldjbm{
When $d^{z}_{\rm Ca}$ increases, the Ca cation becomes closer to (farther from) the O atoms in the OP (IP).
Thus, the O$2p_{\sigma}$ orbitals in the IP (OP) are destabilized (stabilized) [see Fig.~ \ref{fig:unconsite}(b)].
The Cu$3d_{x^2-y^2}$ orbitals are also destabilized, but less than O$2p_{\sigma}$ orbitals because Cu atoms are farther from Ca compared to in-plane O.
The above simple view is supported by the fact that 
the variation in $\epsilon_{p_{\sigma}}^{l}$ with $d^{z}_{\rm Ca}$
and also the variation in LEH parameters with $d^{z}_{\rm Ca}$
are twice faster in the IP compared to the OP [see Fig.~ \ref{fig:unconsite}(b) and Fig.~\ref{fig:uncertainty}].
This is because the IP is surrounded by twice more Ca cations than the OP (see Fig.~\ref{fig:cryst}).
However, note that the average values of LEH parameters do not vary substantially, because the $\Delta d^{z}_{\rm Ca}$ dependencies of LEH parameters in the IP and OP compensate each other.
This explains why the increase in $\Delta E_{xp}^{\rm avg}$ from $P_{\rm amb}$ to $P_{\rm opt}$ originates from $P_a$ rather than $P_c$ (see Sec.~\ref{sec:lehut}).
}
\\

\paragraph*{$d^{z}_{\rm Cu}$ dependence of AB LEH parameters ---}

The optimized value $d^{z}_{\rm Cu} \simeq 2.82$ \AA \ is lower than that from Zhang \textit{et al.} ($d^{z}_{\rm Cu} \simeq 2.91$ \AA). 
Thus, we consider  $0.0$ \AA \ $\leq \Delta d^{z}_{\rm Cu} \leq +0.2$ \AA \ to examine the $d^{z}_{\rm Cu}$ dependence of AB LEH parameters.

\headline{Increasing $d^{z}_{\rm Cu}$ causes the rapid increase in $u^{i}$ [see Fig.~\ref{fig:uncertainty}(b)],
 due to the decrease in both $v^{i}$ and $R^{i}$ [see Fig.~\ref{fig:uncertainty}(d,e)].}
 
\headline{The decrease in $v^{i}$ is caused by the decrease in $\Delta E_{xp}^{i}$ [see Fig.~\ref{fig:uncertainty}(e,f)].}
\oldjbm{$\Delta E_{xp}^{i}$ decreases because the in-plane O anions in the OP become farther from those in the IP.
As a result, the O$2p_{\sigma}$ electrons in the IP are stabilized [see Fig.~ \ref{fig:unconsite}(c)], 
because the Madelung potential from O anions in the OP is weaker.
}

\headline{However, the decrease in $\Delta E_{xp}^{i}$ may not be sufficient to explain the decrease in $R^{l}$.}
We see that from $\Delta d^{z}_{\rm Cu} = 0.0$ \AA \ to $\Delta d^{z}_{\rm Cu} = -0.2$ \AA, 
$R^{o}$ slightly decreases and $R^{i}$ sharply decreases [see Fig.~\ref{fig:uncertainty}(d)].
The decrease in $R^{o}$ is not consistent with the increase in $\Delta E_{xp}^{o}$ which contributes to increase $R^{o}$; 
also,  the decrease in $R^{i}$ is very sharp compared to the smooth decrease in $\Delta E_{xp}^{i}$.

\headline{Instead, the decrease in $R^{l}$ may be caused by an increase in cRPA screening between adjacent CuO$_2$ planes.}
This is intuitive because $\Delta d^{z}_{\rm Cu} = -0.2$ \AA \ reduces the distance between the CuO$_2$ planes in the real space.
This increases the overlap and hybridization between M-ALWOs in the IP and OP, which may increase the cRPA screening (see also the discussion about the $d^{z}_{\rm O(ap)}$ dependence of the screening below).
The interplane cRPA screening particularly affects the IP, because the IP is adjacent to two OPs whereas the OP is adjacent to only the IP;
this explains the sharp decrease in $R^{i}$.
\\

\paragraph*{$d^{z}_{\rm Ba}$ dependence of AB LEH parameters ---}

The optimized value $d^{z}_{\rm Ba} \simeq 1.96$ \AA \ is similar to that from Zhang \textit{et al.} ($d^{z}_{\rm Ba} \simeq 1.98$ \AA).
Still, for completeness, we consider $-0.2$ \AA \ $\leq \Delta d^{z}_{\rm Ba} \leq 0.0$ \AA \ in order to examine the $d^{z}_{\rm Ba}$ dependence of AB LEH parameters.

\headline{Decreasing $d^{z}_{\rm Ba}$ does not cause a significant variation in $u^{l}$ [see Fig.~\ref{fig:uncertainty}(b)].}
Still, we note that $v^{o}$ and $\Delta E_{xp}^{o}$ slightly increase [see Fig.~\ref{fig:uncertainty}(e,f)].
This is because the positive Madelung potential from Ba cation felt by the OP is stronger (see the above discussion on the $d^{z}_{\rm Ca}$ dependence of $\Delta E_{xp}^{l}$).
Note that the positive Madelung potential from Ba cation does not affect the IP, because the IP is separated from the Ba cation by the OP (see Fig.~\ref{fig:cryst}).
\\

\paragraph*{$d^{z}_{\rm O(ap)}$ dependence of AB LEH parameters ---}

The optimized value $d^{z}_{\rm O(ap)} \simeq 2.22$ \AA \ is slightly lower than that from Zhang \textit{et al.} ($d^{z}_{\rm O(ap)} \simeq 2.32$ \AA). 
Thus, we consider $0.0$ \AA $\leq \Delta d^{z}_{\rm O(ap)} \leq +0.2$ \AA \ to examine the $d^{z}_{\rm O(ap)}$ dependence of AB LEH parameters.
In addition, we consider $-0.2$ \AA \ $\leq \Delta d^{z}_{\rm O(ap)} \leq$ $0.0$ \AA \ to probe the effect of apical O displacement at higher pressures.

\headline{In the $d^{z}_{\rm O(ap)}$ dependence of $u^{o}$ [see Fig.~\ref{fig:uncertainty}(b)],
there is a 
sharp decrease in $u^{o}$ when $d^{z}_{\rm O(ap)}$ decreases.}
\oldjbm{
This decrease has also been observed in the case of Bi2201 and Bi2212~\cite{Moree2022}.
It has two origins:
(i) the decrease in $v^{o}$ due to the decrease in $\Delta E_{xp}^{o}$ [see Fig.~\ref{fig:uncertainty}(e,f)],
and more prominently (ii) the decrease in $R^{o}$ [see Fig.~\ref{fig:uncertainty}(d)].
(ii) is due to the cRPA screening of AB electrons by the apical O,
and this screening increases when $d^{z}_{\rm O(ap)}$ decreases as in Bi2201 and Bi2212~\cite{Moree2022}.
Note that, contrary to $R^{o}$, $R^{i}$ does not decrease significantly when $d^{z}_{\rm O(ap)}$ decreases:
This is because the IP is protected from the cRPA screening from apical O by the OP, which separates the IP from the apical O (see Fig.~\ref{fig:cryst}).
}

\begin{figure}
\includegraphics[scale=0.36]{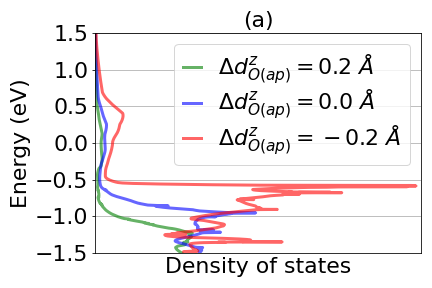}
\includegraphics[scale=0.33]{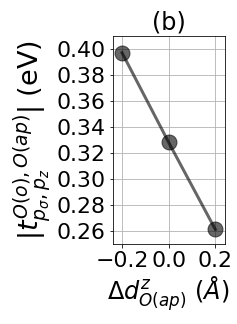}
\caption{Panel (a): $\Delta d^{z}_{\rm O(ap)}$ dependence of the partial density of states of the apical O$2p_{z}$ M-ALWO.
Panel (b): $\Delta d^{z}_{\rm O(ap)}$ dependence of the amplitude $|t^{{\rm O}(o),{\rm O(ap)}(o)}_{p_\sigma,p_z}|$ of the apical O $2p_{z}$/in-plane O$2p_{\sigma}$ hopping.
The quantities are obtained by using the optimized CP values at $P_{\rm opt}=30$ GPa, 
and modifying the value of $d^{z}_{\rm O(ap)}$ by $\Delta d^{z}_{\rm O(ap)}$.
}
\label{fig:dosoap}
\end{figure}

\headline{
A possible origin of (ii) is the increase in hybridization between the  apical O $2p_{z}$ orbital and AB orbital in the OP.}
\oldjbm{
We show in Fig.~\ref{fig:dosoap}(a) the partial density of states of the apical O$2p_{z}$ M-ALWO.
We see that bands at the Fermi level have slight apical O$2p_{z}$ character, in addition to the dominant AB character.
This originates from the hybridization between the AB orbital and the apical O$2p_{z}$ orbital.
The apical O$2p_{z}$ partial density of states at Fermi level increases when $d^{z}_{\rm O(ap)}$ decreases,
which suggests the increase in hybridization between apical O $2p_{z}$ and AB orbitals.
This is further supported by the increase in amplitude $|t^{{\rm O}(o),{\rm O(ap)}(o)}_{p_\sigma,p_z}|$ of the apical O $2p_{z}$/in-plane O$2p_{\sigma}$ hopping when $d^{z}_{\rm O(ap)}$ decreases [see Fig.~\ref{fig:dosoap}(b)], because the AB orbital is partly constructed from the in-plane O$2p_{\sigma}$ orbital.
}

\end{appendices}

%

\end{document}